\documentclass{pasj01}

\usepackage{comment}
\usepackage{url}
\usepackage{color}

\Received{2020 May 29}
\Accepted{2020 December 15}
\Published{$\langle$publication date$\rangle$}
\SetRunningHead{Astronomical Society of Japan}{Usage of \texttt{pasj00.cls}}

\begin{document}
\title{CO Multi-line Imaging of Nearby Galaxies (COMING). IX. $^{\bf 12}$CO({\bf \textit{J}}\,=\,2--1)/$^{\bf 12}$CO({\bf \textit{J}}\,=\,1--0) line ratio on kiloparsec scales}
\author{
	Yoshiyuki \textsc{Yajima}\altaffilmark{1,*},
	Kazuo \textsc{Sorai}\altaffilmark{1,2,3,4},
	Yusuke \textsc{Miyamoto}\altaffilmark{5},\\
	Kazuyuki \textsc{Muraoka}\altaffilmark{6},
	Nario \textsc{Kuno}\altaffilmark{3,4,7},
	Hiroyuki \textsc{Kaneko}\altaffilmark{8,5},\\
	Tsutomu T. \textsc{Takeuchi}\altaffilmark{9,10},
	Atsushi \textsc{Yasuda}\altaffilmark{3},
	Takahiro \textsc{Tanaka}\altaffilmark{3},\\
	Kana \textsc{Morokuma-Matsui}\altaffilmark{11} and
	Masato I. N. \textsc{Kobayashi}\altaffilmark{12,13}
}
\altaffiltext{1}{Department of Cosmosciences, Graduate School of Science, Hokkaido University, N10 W8, Kita-ku, Sapporo, Hokkaido 060-0810, Japan}
\altaffiltext{2}{Department of Physics, Faculty of Science, Hokkaido University, N10 W8, Kita-ku, Sapporo, Hokkaido 060-0810, Japan}
\altaffiltext{3}{Graduate School of Pure and Applied Sciences, University of Tsukuba, 1-1-1 Tennodai, Tsukuba, Ibaraki 305-8571, Japan}
\altaffiltext{4}{Tomonaga Center for the History of the Universe, University of Tsukuba, 1-1-1 Tennodai, Tsukuba, Ibaraki 305-8571, Japan}
\altaffiltext{5}{National Astronomical Observatory of Japan, 2-21-1 Osawa, Mitaka, Tokyo 181-8588, Japan}
\altaffiltext{6}{Department of Physical Science, Osaka Prefecture University, 1-1 Gakuen, Sakai, Osaka 599-8531, Japan}
\altaffiltext{7}{Department of Physics, School of Science and Technology, Kwansei Gakuin University, 2-1 Gakuen, Sanda, Hyogo 669-1337, Japan}
\altaffiltext{8}{Graduate School of Education, Joetsu University of Education, 1, Yamayashiki-machi, Joetsu, Niigata 943-8512, Japan}
\altaffiltext{9}{Division of Particle and Astrophysical Science, Nagoya University, Furo-cho, Chikusa-ku, Nagoya, Aichi 464-8602, Japan}
\altaffiltext{10}{The Research Center for Statistical Machine Learning, The Institute of Statistical Mathematics, 10-3 Midori-cho, Tachikawa, Tokyo 190-8562, Japan}
\altaffiltext{11}{Institute of Astronomy, The University of Tokyo, 2-21-1 Osawa, Mitaka, Tokyo 181-8588, Japan}
\altaffiltext{12}{Department of Earth and Space Science, Graduate School of Science, Osaka University, 1-1 Machikaneyama-cho, Toyonaka, Osaka 560-0043, Japan}
\altaffiltext{13}{Astronomical Institute, Graduate School of Science, Tohoku University, Aoba-ku, Sendai, Miyagi 980-8578, Japan}
\email{yajima@astro1.sci.hokudai.ac.jp}
\maketitle

\begin{abstract}
While molecular gas mass is usually derived from $^{12}$CO($J$\,=\,1--0) --- the most fundamental line to explore molecular gas --- it is often derived from $^{12}$CO($J$\,=\,2--1) assuming a constant $^{12}$CO($J$\,=\,2--1)/$^{12}$CO($J$\,=\,1--0) line ratio ($R_{2/1}$).
We present variations of $R_{2/1}$ and effects of the assumption that $R_{2/1}$ is a constant in 24 nearby galaxies using $^{12}$CO data obtained with the Nobeyama 45-m radio telescope and IRAM 30-m telescope.
The median of $R_{2/1}$ for all galaxies is 0.61, and the weighted mean of $R_{2/1}$ by $^{12}$CO($J$\,=\,1--0) integrated-intensity is 0.66 with a standard deviation of 0.19.
The radial variation of $R_{2/1}$ shows that it is high ($\sim$0.8) in the inner $\sim$1 kpc while its median in disks is nearly constant at 0.60 when all galaxies are compiled.
In the case that the constant $R_{2/1}$ of 0.7 is adopted, we found that the total molecular gas mass derived from $^{12}$CO($J$\,=\,2--1) is underestimated/overestimated by $\sim$20\%, and at most by 35\%.
The scatter of a molecular gas surface density within each galaxy becomes larger by $\sim$30\%, and at most by 120\%.
Indices of the spatially resolved Kennicutt--Schmidt relation by $^{12}$CO($J$\,=\,2--1) are underestimated by 10--20\%, at most 39\% in 17 out of 24 galaxies.
$R_{2/1}$ has good positive correlations with star-formation rate and infrared color, and a negative correlation with molecular gas depletion time.
There is a clear tendency of increasing $R_{2/1}$ with increasing kinetic temperature ($T_{\rm kin}$).
Further, we found that not only $T_{\rm kin}$ but also pressure of molecular gas is important to understand variations of $R_{2/1}$.
Special considerations should be made when discussing molecular gas mass and molecular gas properties inferred from $^{12}$CO($J$\,=\,2--1) instead of $^{12}$CO($J$\,=\,1--0).
\end{abstract}

\section{Introduction}
Molecular gas is a crucial component in the interstellar medium (ISM) because star formation is the main physical process in the universe, and stars form from cold molecular gas.
Therefore, it is essential to understand the properties of molecular gas to investigate galaxies and their evolution.
$^{12}$CO($J$\,=\,1--0) line has been used as a tracer for cold molecular gas (e.g., \cite{Solomon87}; \cite{Young91}) because H$_2$ molecule cannot be directly observed in cold environments due to no electric dipole moment.
Since the $J$\,=\,1 energy level of a $^{12}$CO molecule is low ($\Delta E/k_{\rm B} \sim 5.5$ K, where $\Delta E$ is energy gap and $k_{\rm B}$ is the Boltzmann constant), $^{12}$CO is easily excited to the $J$\,=\,1 level and emit $^{12}$CO($J$\,=\,1--0) line even in cold conditions.
In addition, the critical density of $^{12}$CO($J$\,=\,1--0) is a few hundred cubic centimeters in an optical thick region.
This value is quite lower than that of other molecular gas tracers.
Furthermore, $^{12}$CO is an abundant molecule in the ISM except hydrogen and helium, has large dissociation energy, and its abundance ratio is nearly uniform in molecular clouds.
Thus, $^{12}$CO($J$\,=\,1--0) is the most useful line to study the bulk of cold molecular gas.
Molecular gas surface density ($\Sigma_{\rm mol}$) in galactic disks is usually derived from the following equation with the integrated intensity of $^{12}$CO($J$\,=\,1--0), $I_{^{12}\mathrm{CO}(1-0)}$,
\begin{eqnarray}
\left( \frac{\Sigma_{\mathrm{mol}}}{M_{\odot}\>\, \mathrm{pc^{-2}}} \right) &=&
1.36 \times 3.20\> \cos i \left[ \frac{I_{^{12}\mathrm{CO}(1-0)}}{\mathrm{K\>\, km\>\, s^{-1}}} \right] \nonumber \\
&& \left[ \frac{X_{\mathrm{CO}}}{2.0 \times 10^{20}\>\, {\rm cm^{-2}\>\, (K\>\, km\>\, s^{-1})^{-1}}} \right],
\end{eqnarray}
where $i$ is the inclination angle of the galactic disk, $X_{\rm CO}$ is the CO-to-H$_2$ conversion factor that converts $I_{^{12}\mathrm{CO}(1-0)}$ into column density of H$_2$.
Helium in molecular gas contributes the factor of 1.36 and the other factor 3.20 represents the unit conversion of K km s$^{-1}$ into $M_{\odot}$ pc$^{-2}$.
The product of $1.36 \times 3.20 = 4.35$ in units of $M_{\odot}$ pc$^{-2}$ (K km s$^{-1}$)$^{-1}$ corresponding to $X_{\rm CO} = 2.0 \times 10^{20}$ cm$^{-2}$ (K km s$^{-1}$)$^{-1}$ is also widely used as the CO-to-H$_2$ conversion factor that converts $I_{^{12}\mathrm{CO}(1-0)}$ into $\Sigma_{\rm mol}$ [or luminosity of $^{12}$CO($J$\,=\,1--0) into molecular gas mass] usually denoted as ``$\alpha_{\rm CO}$'' (e.g., \cite{Bolatto13}; \cite{Leroy13}; \cite{Schruba19}).

When the atmosphere is extremely dry, which is common at sites of short-millimeter and submillimeter telescopes, the observation efficiency of $^{12}$CO($J$\,=\,1--0) for the local universe is usually lower than that with $^{12}$CO($J$\,=\,2--1).
This is because the rest frequency of $^{12}$CO($J$\,=\,1--0), 115 GHz, is close to the O$_2$ absorption band at 118 GHz, and the $^{12}$CO($J$\,=\,2--1) attenuation caused by water vapor is small at its rest frequency of 230 GHz in such sites.
Thus, $^{12}$CO($J$\,=\,2--1) is often used to observe molecular gas instead of $^{12}$CO($J$\,=\,1--0) (e.g., \cite{Leroy09}, hereafter L09; \cite{Druard14}; \cite{Sun18}).
In this case, $\Sigma_{\rm mol}$ is derived from the following equation instead of equation 1 by assuming the integrated-intensity ratio of $^{12}$CO($J$\,=\,2--1)/$^{12}$CO($J$\,=\,1--0) (hereafter, $R_{2/1}$),
\begin{eqnarray}
\left( \frac{\Sigma_{\mathrm{mol}}}{M_{\odot}\> \mathrm{pc^{-2}}} \right) &=&
1.36 \times 3.20\> \cos i \>\, R_{2/1}^{-1}\,\, \left[ \frac{I_{^{12}\mathrm{CO}(2-1)}}{\mathrm{K\>\, km\>\, s^{-1}}} \right] \nonumber \\
&& \left[ \frac{X_{\mathrm{CO}}}{2.0 \times 10^{20}\>\, {\rm cm^{-2}\>\, (K\>\, km\>\, s^{-1})^{-1}}} \right],
\end{eqnarray}
where $I_{^{12}\mathrm{CO}(2-1)}$ is the integrated intensity of $^{12}$CO($J$\,=\,2--1).
The value of 0.7--0.8 is usually assumed to be a constant $R_{2/1}$ (e.g., L09; \cite{Leroy13}).

Although $^{12}$CO($J$\,=\,2--1) is used as a proxy of $^{12}$CO($J$\,=\,1--0), the energy of $^{12}$CO $J$\,=\,2 level ($\Delta E/k_{\rm B} \sim 16.5$ K) is higher than that in the typical cold molecular clouds ($\sim 10$ K).
Therefore, it is possible that $^{12}$CO molecules are not well excited to reach the $J$ = 2 level based on molecular gas properties (e.g., \cite{Penaloza17}).
Molecular gas traced by $^{12}$CO($J$\,=\,2--1) is warmer and/or denser than that traced using $^{12}$CO($J$\,=\,1--0).
$R_{2/1}$ is influenced by the physical conditions of molecular gas.

The systematic variations of $R_{2/1}$ have been reported in previous observations of molecular gas in the Milky Way.
For nearby giant molecular clouds (GMCs), {while $R_{2/1}$ is medium ($\sim 0.7$--0.8) in the ridges of GMCs, it exceeds unity in interfaces of H\emissiontype{II} regions and OB associations, and it is relatively low in the peripheries of GMCs ($\sim 0.5$)} (\cite{Sakamoto94}; \cite{Nishimura15}).
GMCs with active star formation tend to show high $R_{2/1}$ while those with quiescent star formation show low ratios.
In large scales of the Milky Way, $R_{2/1}$ decreases from 0.75 at 4 kpc to 0.5 at 8 kpc as a function of the Galactocentric radius (\cite{Sakamoto95}; \cite{Sakamoto97}).
Further, a $R_{2/1}$ gradient appears across spiral arms; it gradually increases toward the downstream of arms.
In addition, $R_{2/1}$ shows higher values that are close to unity in the Galactic center compared with the Galactic disk (\cite{Oka98}; \cite{Sawada01}).

For external galaxies, a pioneering study by \citet{Braine92} reported that there is a moderate positive correlation between $R_{2/1}$ and infrared (IR) color, and they attempted to infer molecular gas properties from $R_{2/1}$ around the center of nearby galaxies.
High $R_{2/1}$ (0.9 with the scatter of $\sim 0.3$) was reported in the Large Magellanic Cloud (LMC), and positions where $R_{2/1}$ is high do not always coincide with massive star-formation regions (\cite{Sorai01}).
It was argued that it reflects dense molecular gas, which is ready to form stars and is not due to warmed gas by radiations from massive stars, whereas the low metallicity environment in LMC may also influence $R_{2/1}$.
In several nearby galaxies, the median of $R_{2/1}$ in disks is approximately constant ($\sim 0.8$) and $R_{2/1}$ increases ($> 1$) in the center similar to that in the Milky Way (L09), while some galaxies show a nearly constant $R_{2/1}$ in the entire galaxy including the center (e.g., M\,33; \cite{Druard14}).
\citet{Leroy13} reported the median of 0.66 with standard deviations of $\sim 0.3$ for approximately 30 nearby galaxies.
Based on this result, some studies using $^{12}$CO($J$\,=\,2--1) data assumed the constant $R_{2/1} = 0.7$ and converted the intensity of the line into that of $^{12}$CO($J$\,=\,1--0) to derive molecular gas mass and its surface density.

A detailed study to understand $R_{2/1}$ as a probe of molecular gas properties in the extragalactic field reported that while typical $R_{2/1}$ is $\sim 0.7$, it is relatively high (0.8--0.9) in the leading side (the downstream) of the spiral arms and low (0.4--0.6) in the inter-arms for M\,51 as well as the Milky Way (\cite{Koda12}).
In addition, they found that $R_{2/1}$ increases as surface density of star-formation rate ($\Sigma_{\rm SFR}$) and star-formation efficiency (SFE) in this galaxy.
They also suggested that high $R_{2/1}$ is relevant to warm molecular gas by active star formation and compressed molecular gas before star formation.
With regard to $R_{2/1}$ and star formation activity, it was found that the correlation of $R_{2/1}$ with IR color is better than that of $R_{2/1}$ with far-ultraviolet (FUV), and far IR intensities in M\,83 (\cite{Koda20}).
It was argued that these are attributed to warm molecular gas heated by dust, UV photons, and cosmic ray from supernovae, considering high $R_{2/1}$ in the downstream of arms where many massive stars are seen.

As described above, $R_{2/1}$ has systematic variations within and among galaxies, and it reflects molecular gas conditions such as temperature.
Hence, it would be possible that molecular gas mass, its related quantities and relations derived from $^{12}$CO($J$\,=\,2--1) with assumed constant $R_{2/1}$ are misled (e.g., \cite{Momose13}) although there remains the uncertainty of the CO-to-H$_2$ conversion factor.
It is better to test the validity of the assumption that $R_{2/1}$ is constant, and its effect on derived quantities and relations which are relevant to molecular gas, especially for many types of galaxies.
Furthermore, the causes of $R_{2/1}$ variations should be investigated with physical properties of molecular gas.

The largest CO-mapping survey for nearby galaxies in the world, CO Multi-line Imaging of Nearby Galaxies (COMING; \cite{Sorai19}\footnote{See also $\langle$\url{https://astro3.sci.hokudai.ac.jp/~radio/coming/}$\rangle$}, hereafter S19) was carried out using the receiver FOur-beam REceiver System on the 45-m Telescope (FOREST; \cite{Minamidani16}) installed on the Nobeyama 45-m telescope.
COMING mapped 147 nearby galaxies in $^{12}$CO, $^{13}$CO, and C$^{18}$O $J$\,=\,1--0 lines.
The addition of other CO-mapping surveys such as \citet{Kuno07} (hereafter, K07) and L09 enable us to investigate spatial $R_{2/1}$ variations in many types of nearby galaxies and the effects of the assumption for a constant $R_{2/1}$ in a large area of galaxies.
With those CO data, we verify the assumption of fixed $R_{2/1}$.
The key questions in this paper are as follows: (i) How does the assumption of a constant $R_{2/1}$ affect molecular gas mass itself and its relevant relations?
Is the assumption that $R_{2/1}$ is a constant reasonable?;
(ii) What changes $R_{2/1}$? How does $R_{2/1}$ vary depending on molecular gas properties?

The remainder of the paper is organized as follows.
In section 2, the samples of this study, analysis of CO cubes, intensity accuracy of CO considering calibration and pointing, and ancillary data sets are explained.
The spatial distribution, statistics, and radial distribution of $R_{2/1}$ are described in the first half of section 3.
In the latter half of the section, we report effects of the constant $R_{2/1}$ on molecular gas mass, its scatter within a galaxy, and the $\Sigma_{\rm SFR}$--$\Sigma_{\rm mol}$ relation (i.e., the molecular gas Kennicutt--Schmidt relation).
To investigate the relation of $R_{2/1}$ and molecular gas properties, the correlations of representative quantities and $R_{2/1}$ are examined in the first half of section 4.
We attempted to derive the intrinsic properties of molecular gas, number density, and kinetic temperature from $^{12}$CO($J$\,=\,1--0), $^{12}$CO($J$\,=\,2--1), and $^{13}$CO($J$\,=\,1--0) for COMING galaxies and compared them with $R_{2/1}$.
These discussions and implications are described in the latter half of section 4.
Finally, the conclusions of this paper are provided in section 5.

\section{Data sets}
\begin{table}[t!]
  \begin{center}
  \tbl{Samples in this study.}{
  \begin{tabular}{lccccc} \hline
galaxy & $D$ & $i$ & P.A. & Res. & $^{12}$CO(1--0) ref. \\
 & [Mpc] & [deg] & [deg] & [kpc] &  \\
 & (1) & (2) & (3) & (4) & (5) \\
\hline
NGC\,337 & 18.9 & 44.5 & 119.6 & 1.6 & S19 \\
NGC\,628 & 9.02 & 7 & 20 & 0.74 & S19 \\
NGC\,2146 & 27.7 & 62 & $-$43.5 & 2.3 & S19 \\
NGC\,2798 & 29.6 & 60.7 & 158.7 & 2.4 & S19 \\
NGC\,2841 & 14.60 & 73.7 & 152.6 & 1.2 & S19 \\
NGC\,2903 & 9.46 & 67 & $-$155 & 0.78 & S19 \\
NGC\,2976 & 3.63 & 64.5 & $-$25.5 & 0.30 & S19 \\
NGC\,3034 & 3.53 & 81 & 68 & 0.34 & S19 \\
NGC\,3077 & 3.81 & 38.9 & 63.8 & 0.31 & S19 \\
NGC\,3184 & 8.7 & 21 & $-$174 & 0.72 & K07 \\
NGC\,3198 & 13.40 & 71.5 & $-$145.0 & 1.1 & S19 \\
NGC\,3351 & 10.7 & 41 & $-$168 & 0.88 & K07 \\
NGC\,3521 & 14.20 & 63 & $-$19 & 1.2 & S19 \\
NGC\,3627 & 9.04 & 52 & 176 & 0.75 & S19 \\
NGC\,3938 & 17.9 & 20.9 & $-$154.0 & 1.5 & S19 \\
NGC\,4214 & 2.93 & 30 & 65 & 0.24 & S19 \\
NGC\,4254 & 16.5 & 42 & 66 & 1.4 & K07 \\
NGC\,4321 & 16.5 & 27 & 138 & 1.4 & K07 \\
NGC\,4536 & 16.5 & 64.2 & $-$54.5 & 1.4 & S19 \\
NGC\,4559 & 7.31 & 63.1 & $-$36.8 & 0.60 & S19 \\
NGC\,4569 & 16.5 & 64 & 22 & 1.4 & K07 \\
NGC\,4579 & 16.5 & 41.7 & 92.1 & 1.4 & S19 \\
NGC\,4736 & 4.3 & 40 & $-$61 & 0.4 & K07 \\
NGC\,5055 & 9.04 & 61 & 98 & 0.75 & S19 \\
NGC\,5194 & 7.7 & 20 & 176 & 0.63 & K07 \\
NGC\,5457 & 7.2 & 18 & 42 & 0.59 & K07 \\
NGC\,5713 & 19.5 & 33 & $-$157 & 1.6 & S19 \\
NGC\,6946 & 5.5 & 40 & $-$118 & 0.5 & K07 \\
NGC\,7331 & 13.90 & 75.8 & 167.7 & 1.1 & S19 \\
  \hline
  \end{tabular}}
  \label{table:sample}
  \begin{tabnote}
    (1) Adopted distance.
    (2) Inclination angle of the disk.
    (3) Position angle of the major axis of the disk for the redshifted side (against the north direction, the positive values correspond to the counterclockwise direction).
    (4) The linear scale corresponded to the angular resolution of \timeform{17''} along the major axis of the disk.
    (5) References of $^{12}$CO($J$\,=\,1--0) data.
    References of $D$, $i$, P.A. are the same as in S19 or K07.
  \end{tabnote}
  \end{center}
\end{table}

\subsection{CO data and their analysis}
$^{12}$CO($J$\,=\,1--0) data used in this paper were taken from COMING firstly.
Refer to sections 3 and 4 in S19 for details about the settings of observations, calibration, and data reduction for COMING.
In addition, Nobeyama CO Atlas of Nearby Spiral Galaxies (K07) is used as the second reference of $^{12}$CO($J$\,=\,1--0) data.
Since the samples of K07 includes galaxies with large appearances, it makes it possible to discuss variations of $R_{2/1}$ in large area of galaxies.
For the $^{12}$CO($J$\,=\,2--1) reference, we use HERA CO-Line Extragalactic Survey (HERACLES; L09) carried out with the IRAM 30-m telescope.

The $^{12}$CO($J$\,=\,1--0) data obtained by K07 and $^{12}$CO($J$\,=\,2--1) data obtained by L09 were convolved to match the angular resolution of \timeform{17''}, which is the original resolution of COMING.
Before measurements of $R_{2/1}$, the coordinates for all CO cubes were matched and regridded so that the grid size becomes \timeform{8''}.
Then, velocity channels were binned so that the velocity resolution is 20 km s$^{-1}$. 
These processes were performed to improve signal-to-noise ratio ($S/N$).
After that, the baseline was subtracted.

The method of baseline subtraction is the same as COMING Auto-Reduction Tool (COMING ART; see section 4.2 in S19 for detail), however, signal channels are defined by $^{12}$CO($J$\,=\,2--1), not by $^{12}$CO($J$\,=\,1--0) because $^{12}$CO($J$\,=\,2--1) data achieves a much better sensitivity than that of $^{12}$CO($J$\,=\,1--0).
That is, the result that whether the channel is a `signal' or `noise' for each channel of $^{12}$CO($J$\,=\,2--1) in a position was applied to that of $^{12}$CO($J$\,=\,1--0) at the channel of the same velocity in the position.
This methodology does not induce any biases for $^{12}$CO($J$\,=\,1--0) and $^{12}$CO($J$\,=\,2--1) integrated-intensities because the sensitivities of the $^{12}$CO($J$\,=\,2--1) data are much better than those of $^{12}$CO($J$\,=\,1--0).
The threshold to evaluate each channel as a `signal' or a `noise' is set as 3\,$\sigma${, where $\sigma$ is the root mean square (R.M.S.) measured in the defined emission-free velocity range beforehand.

The samples in this study consist of 29 galaxies that are overlapped between S19 $+$ K07, and L09.
They are listed in table \ref{table:sample}.
For the overlapped galaxies between S19 and K07 (NGC\,2903, NGC\,3351, NGC\,3521, NGC\,3627, NGC\,5055), we used S19 data except for NGC\,3351 because the baseline of the S19 data for this galaxy is heavily undulated.

\subsection{Intensity accuracy of CO data regarding calibration and pointing error}
\begin{figure*}[t!]
 \begin{center}
 \includegraphics[width=16cm]{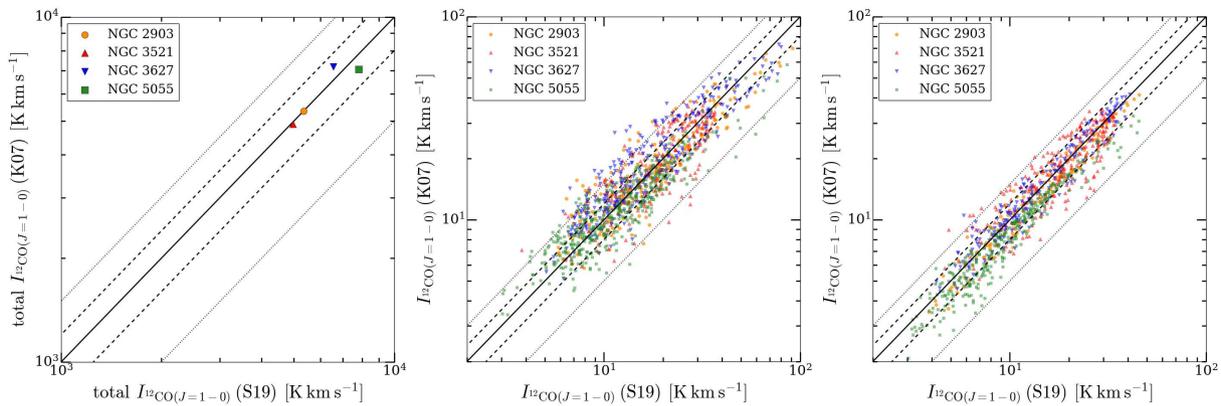}
 \end{center}
 \caption{
 Consistency of integrated intensity for NGC\,2903, NGC\,3521, NGC\,3627, and NGC\,5055 between S19 and K07 summing up within the whole disk, which corresponds to total flux (left), at the original resolution and grid size (center) and convolved data so that the  resolution and grid size are fixed to 1.5 kpc and 650 pc (right), respectively.
 Pixels below 5\,$\sigma$ are masked in the center and right panel.
 The solid, dashed, and dotted lines in each panel indicate inconsistency of 0\%, $\pm 20$\%, and $\pm 50$\%, respectively.
  }
 \label{fig:mom0_correl}
\end{figure*}

Intensity accuracy of CO data is important to study $R_{2/1}$ because its contrast is relatively small (usually $R_{2/1}=0.4$--1.0) as argued in Koda et al. (\yearcite{Koda12}, \yearcite{Koda20}).
Therefore, calibration and pointing error is an issue in this paper.
We demonstrated how intensity of $^{12}$CO($J$\,=\,1--0) data is accurate with overlapped 4 galaxies (NGC\,2903, NGC\,3521, NGC\,3627, NGC\,5055) between S19 and K07.
Although NGC\,3351 is an overlapped galaxy, the baseline of this galaxy is heavily undulated and thus, we excluded this galaxy to test intensity consistency.
Both S19 and K07 data was regridded and the baseline was subtracted in the same way described in the last subsection.

We firstly tested consistency of total integrated-intensity within the whole disk (i.e., total flux) for S19 and K07 data and integrated intensity at position-to-position in the original resolution (17$''$) and grid size (8$''$).
The left and center panel of figure \ref{fig:mom0_correl} shows correlations of total integrated-intensities and integrated intensities at each pixel, respectively.
The threshold was set to be 5\,$\sigma$ in the central panel.
The difference of total integrated-intensity is at most $\sim 10\%$.
The scatter of intensity at position-to-position is $\sim 25\%$ in the R.M.S. level within each galaxy.
Hence, calibration error of $^{12}$CO($J$\,=\,1--0) used in this paper is up to 10\% and intensity accuracy including pointing error and calibration error is 25\% at the original resolution.

In section 3.3 and 4.1, we use convolved CO data so that the spatial resolution and grid size are fixed to 1.5 kpc and 650 pc, respectively.
In addition, we stacked spectra within concentric annuli in disks and galactic structures in analyses of section 4.2.
Thus, we next tested consistency of convolved S19 and K07 data to match the resolution of 1.5 kpc and grid size of 650 pc.
The right panel of figure \ref{fig:mom0_correl} shows consistency of integrated intensity between S19 and K07 for convolved data.
The scatter at position-to-position in this case is $\sim 15\%$ in each galaxy.
Therefore, intensity error combining calibration and pointing error in section 3.3 and 4.1 is 15\%.
Since in stacking analysis, spectra are stacked in a large area (galactic components and concentric annuli whose width is $r_{25}/8$ where $r_{25}$ is $B$-band isophotal radius at 25 mag arcsec$^{-2}$), intensity error due to pointing error will be negligible.

We would like to note that intensity accuracy of $^{13}$CO($J$\,=\,1--0) against $^{12}$CO($J$\,=\,1--0) is relatively good because $^{13}$CO($J$\,=\,1--0) data was taken  simultaneously with $^{12}$CO($J$\,=\,1--0) in COMING observations.
In addition, since COMING applied the On-The-Fly (OTF) mode in their observation, while observation in K07 applied the Position-Switch,  calibration and pointing accuracy of S19 would be better than that of K07.

We could not test how calibration/pointing error of $^{12}$CO($J$\,=\,2--1) data is in the same way of S19 and K07 because there is no available archival data of the line.
However, L09 mentioned that daily variations of $^{12}$CO($J$\,=\,2--1) intensity at high $S/N$ regions is 20\%.
Therefore, intensity error of $^{12}$CO($J$\,=\,2--1) combining calibration and pointing error is 20\%, which is quite better than that of $^{12}$CO($J$\,=\,1--0).
Assuming the fraction of contribution by pointing error is the same as the $^{12}$CO($J$\,=\,1--0) case, calibration error of $^{12}$CO($J$\,=\,2--1) will be $\sim 8$\%.
Hence, there will be also calibration error of $\sim 10$\% in $^{12}$CO($J$\,=\,2--1) data, which is common in short millimeter-wave observations.
Similarly, intensity error due to calibration/pointing error in convolved $^{12}$CO($J$\,=\,2--1) data will be $\sim 12$\%.
To summarize above descriptions, $R_{2/1}$ error due to calibration error is 13\%, due to calibration$+$pointing error at original angular resolution is 32\%, and at fixed spatial resolution and grid size is 19\%, respectively.

\subsection{Data in other wavelengths}
Star-formation rate (SFR) was derived from FUV and 22-$\micron$ band intensity based on the following equation by \citet{Casasola17} and \citet{Leroy08} that adopted the Kroupa initial mass function (IMF; \cite{Kroupa01}) as with the method in other COMING papers (e.g., \cite{Muraoka19}; Takeuchi et al. 2020 in prep.):
\begin{eqnarray}
\left( \frac{{\Sigma_{\rm SFR}}}{M_{\odot}\>\, {\rm yr}^{-1}\>\, {\rm kpc}^{-2}} \right) =
&1.59&\>\, \cos i \, \bigg[ 3.2 \times 10^{-3} \left( \frac{I_{22 \micron}}{{\rm MJy\>\, sr}^{-1}} \right) \nonumber \\
&+& 8.1 \times 10^{-2} \left( \frac{I_{\rm FUV}}{{\rm MJy\>\, sr}^{-1}} \right) \bigg],
\end{eqnarray}
where $I_{22 \micron}$ and $I_{\rm{FUV}}$ are intensities of the 22-$\micron$ and FUV band.
The FUV maps were obtained with the GALEX Ultraviolet Atlas of Nearby Galaxies (\cite{GildePaz07}) and were retrieved from the NASA/IPAC Extragalactic Database (NED).
The 22-$\micron$ maps obtained with the WISE band 4 were retrieved from the NASA/IPAC Infrared Science Archive.

In addition, the IR color that represents dust temperature is used as an indicator of ISM conditions.
To measure the IR color in many galaxies as much as possible, we used the intensity ratio of the 70-$\micron$ to 160-$\micron$ band obtained with Herschel/PACS.
This filter selection reduces the sample number of galaxies because of a lack of samples in the far-infrared range.


\begin{figure*}[t!]
 \begin{center}
  \includegraphics[width=16cm]{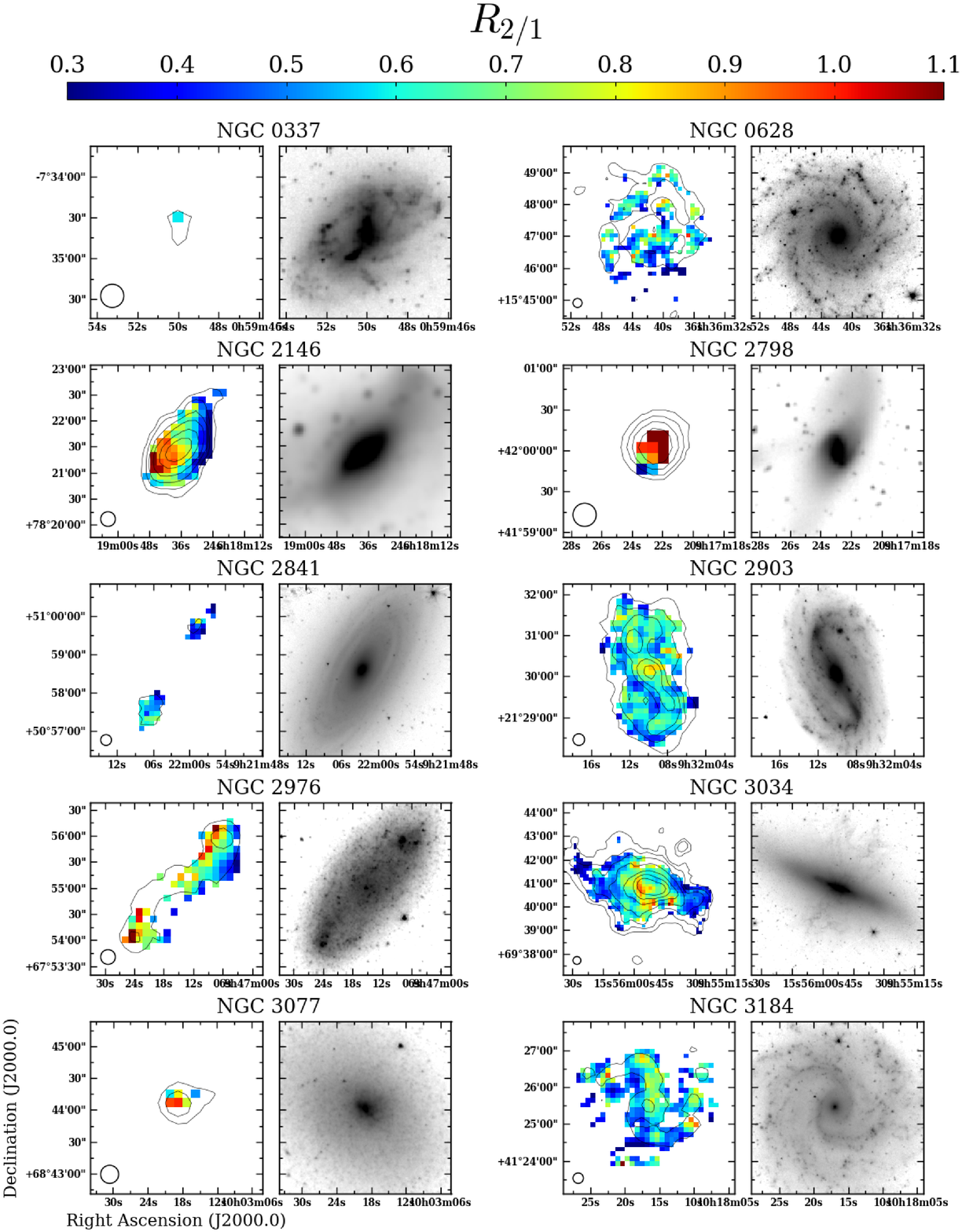}
 \end{center}
 \caption{$R_{2/1}$ maps (left) and near-infrared images in gray scale (right) of all galaxies in the samples.
 In $R_{2/1}$ maps, pixels are masked when either integrated intensities of $^{12}$CO($J$\,=\,1--0) or $^{12}$CO($J$\,=\,2--1) at the pixel do not reach 4.5\,$\sigma$.
 The middle of the color bar corresponds to $R_{2/1} = 0.7$, which is the usually assumed value for a constant $R_{2/1}$.
 Open circles in the bottom left corner in each panel indicate the angular resolution of \timeform{17''}.
 Black contours indicate the integrated intensity of $^{12}$CO($J$\,=\,2--1) at levels 2, 5, 10, 20, 50, 100, and 200 K km s$^{-1}$.
 The reference of near-infrared data is Spitzer/IRAC 3.6-$\micron$ images obtained by the S$^4$G survey \citep{Sheth10} except for NGC\,2146.
 For NGC\,2146, the WISE 3.4-$\micron$ image is used alternatively.
  }
 \label{fig:ratio_map}
\end{figure*}
\setcounter{figure}{1}
\begin{figure*}[t!]
 \begin{center}
  \includegraphics[width=16cm]{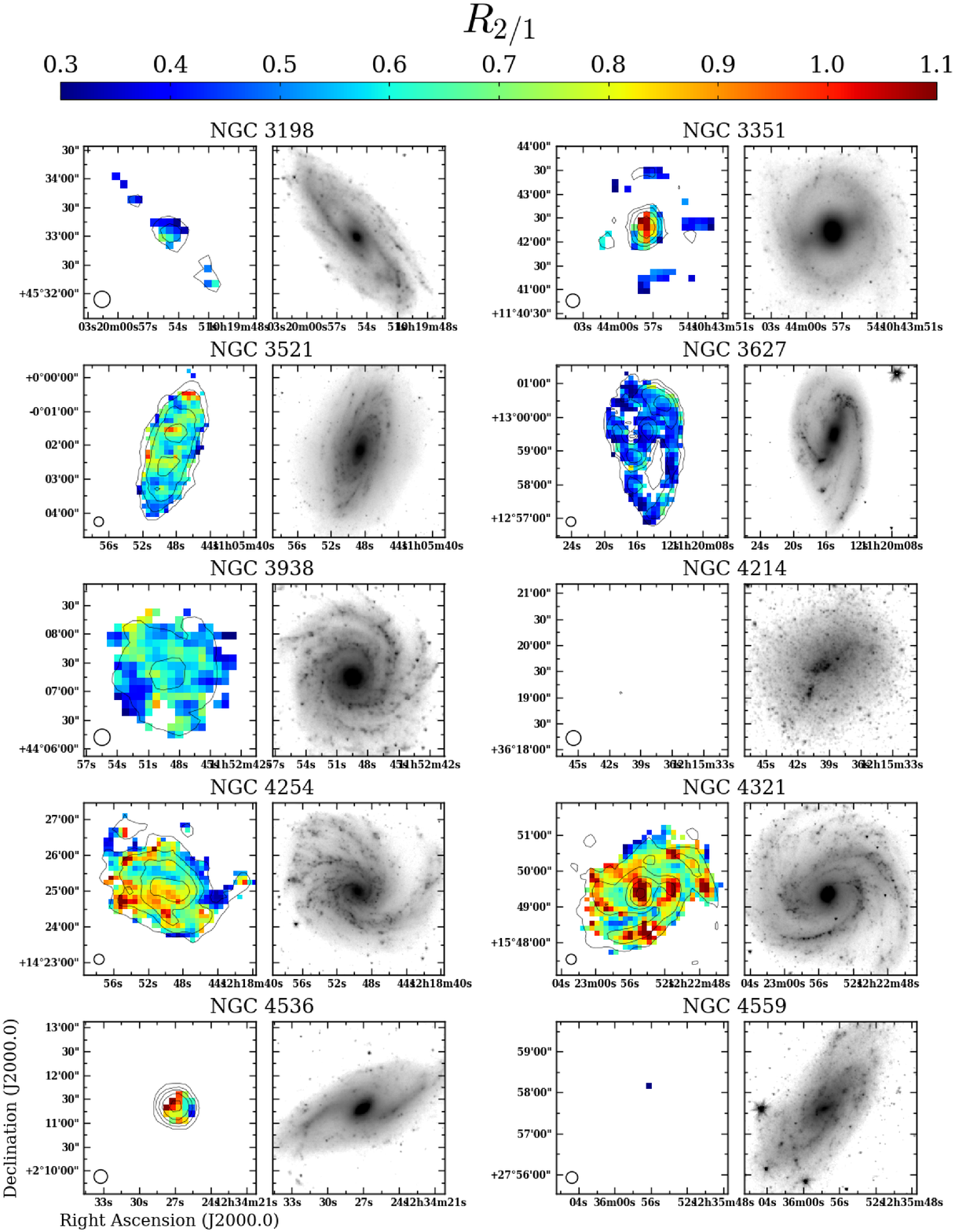}
 \end{center}
 \caption{(Continued.)
 }
 \label{fig:ratio_map}
\end{figure*}
\setcounter{figure}{1}
\begin{figure*}[t!]
 \begin{center}
  \includegraphics[width=16cm]{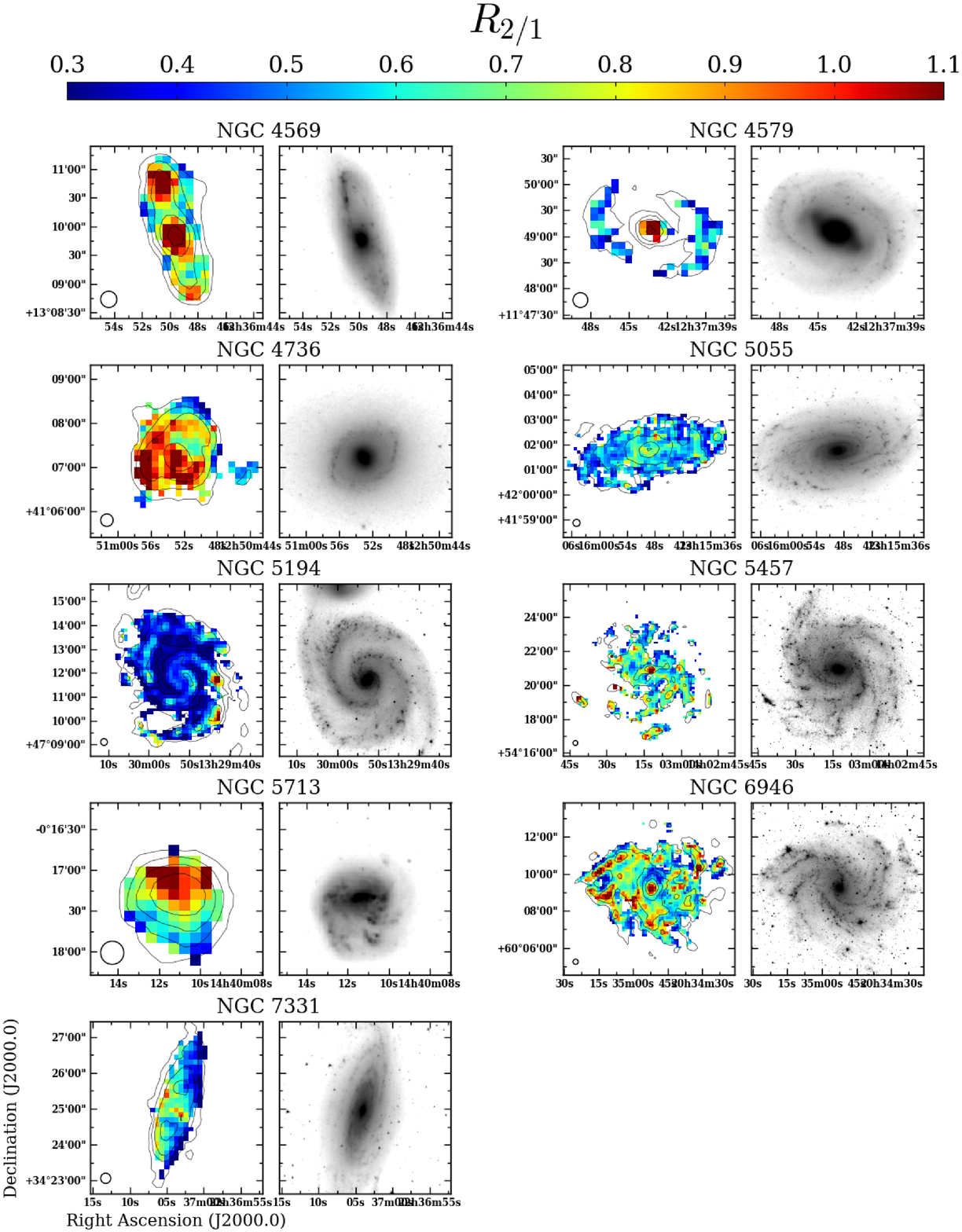}
 \end{center}
 \caption{(Continued.)
 }
 \label{fig:ratio_map}
\end{figure*}
\begin{figure}[t!]
 \begin{center}
  \includegraphics[width=8cm]{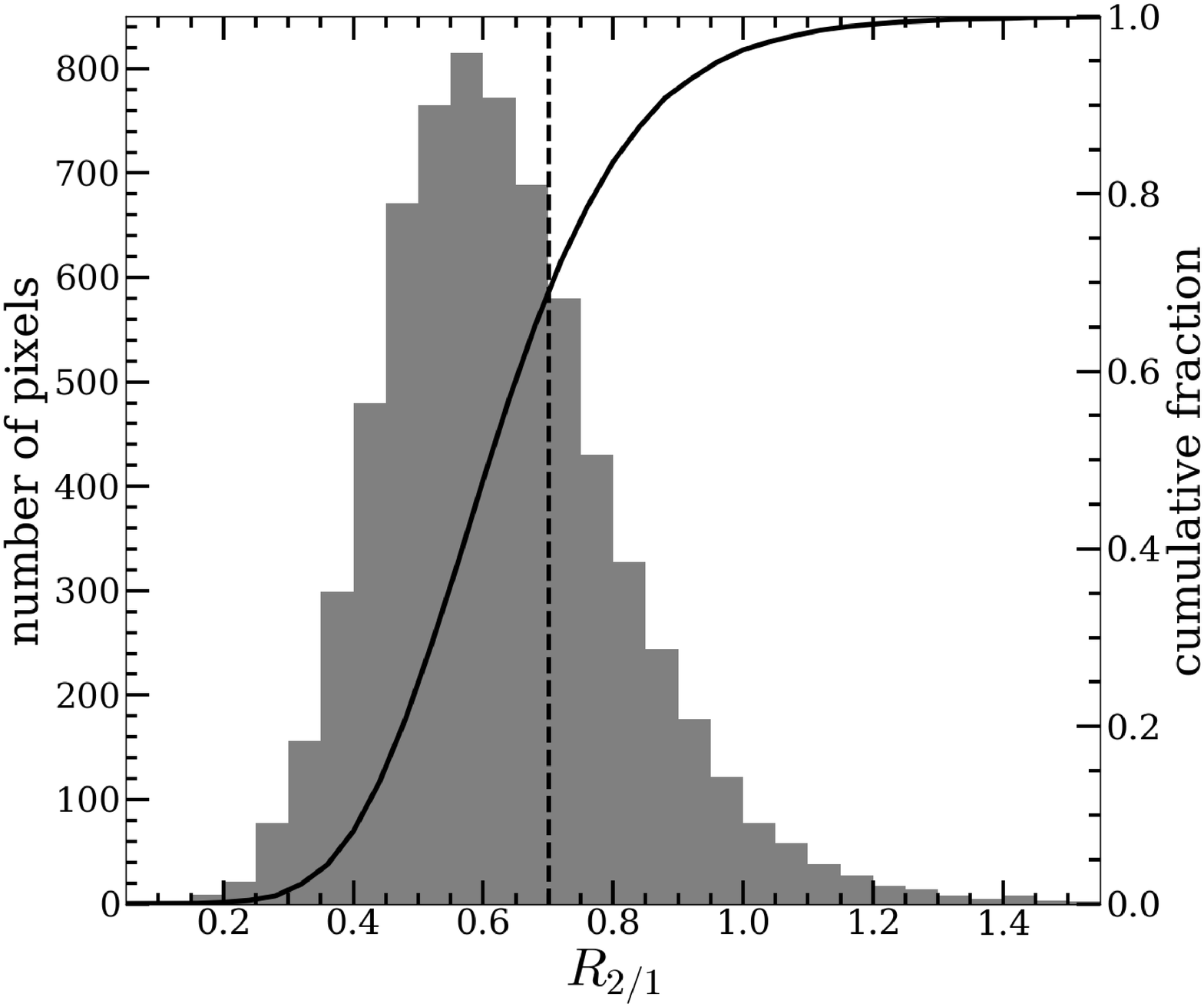}
 \end{center}
 \caption{Histogram of $R_{2/1}$ for all pixels of all galaxies.
 The solid line shows the cumulative distribution function and the dashed line indicates $R_{2/1} = 0.7$.
  }
 \label{fig:ratio_hist}
\end{figure}
\begin{figure}[t!]
 \begin{center}
  \includegraphics[width=8cm]{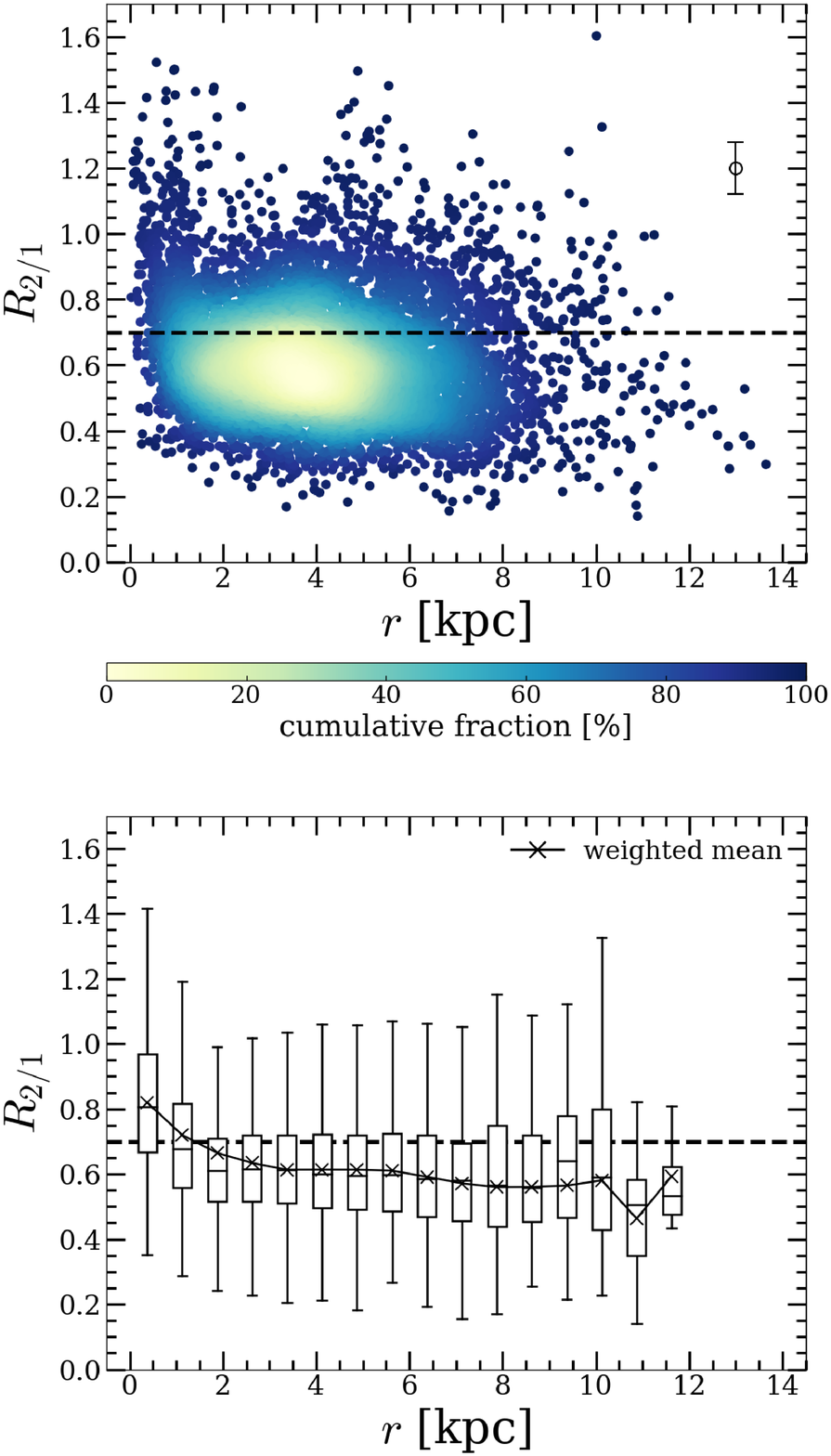}
 \end{center}
 \caption{Radial distribution of $R_{2/1}$ for all galaxies.
 The dashed line in the both panels indicates $R_{2/1} = 0.7$.
 (Top) Scatter plot of $R_{2/1}$ against the galactocentric radius.
 The colors indicate the cumulative fraction of data.
 The typical error of $R_{2/1}$ is shown on the top right with the open circle marker.
 (Bottom) Box plot for the top panel.
 Each bin is 0.75 kpc in width.
 Cross markers show mean $R_{2/1}$ in a bin weighted by $^{12}$CO($J$\,=\,1--0) integrated-intensity.
 The upper whisker extends up to the maximum value and the lower whisker extends down to the minimum value in each bin unless the maximum value is less than $Q_3+1.5{\rm IQR}$ and the minimum value is larger than $Q_1-1.5{\rm IQR}$, where $Q_1$ is the 25th percentile, $Q_3$ is 75th percentile, and IQR is the interquartile range defined as $Q_3-Q_1$.
 Otherwise, the upper whisker extends up to $Q_3+1.5{\rm IQR}$ and the lower whisker extends down to $Q_1-1.5{\rm IQR}$ without outliers for simplicity. 
  }
 \label{fig:ratio_radial_dist}
\end{figure}

\section{Results}
\subsection{Maps, histograms, statistics, and radial distribution of $R_{2/1}$}
Figure \ref{fig:ratio_map} shows the $R_{2/1}$ maps for our sample galaxies in the original resolution of COMING (\timeform{17''}).
Here, error of $R_{2/1}$ is derived from that of $I_{^{12}{\rm CO}(1-0)}$ and $I_{^{12}{\rm CO}(2-1)}$ based on propagation of their error.
Each galaxy shows various $R_{2/1}$.
For example, NGC\,2798 and NGC\,4736 show higher ($> 0.9$) $R_{2/1}$ in large area.
$R_{2/1}$ is low ($< 0.6$) in most positions of NGC\,2841 and NGC\,3627.
Some galaxies have significant variations within each galaxy.
NGC\,2146 and NGC\,5713 show an $R_{2/1}$ gradient from $\sim 1.2$ to 0.4.
In NGC\,4321, $R_{2/1}$ is clearly high ($\gtrsim 1.0$) in the center and bar ends, low ($\lesssim 0.6$) in inter-arms, and intermediate (0.7--0.8) in arms.
In NGC\,3351 and NGC\,4579, $R_{2/1}$ is low in the ring-like structure, whereas $R_{2/1}$ is high in the center.

Some galaxies show gradation of $R_{2/1}$ from a side to the other (e.g., NGC\,2146, NGC\,2798, NGC\,5713, NGC\,7331).
We tested these gradations are due to the systematic pointing offset between $^{12}$CO($J$\,=\,1--0) and $^{12}$CO($J$\,=\,2--1) observations with spectra at some doubtful positions and residual maps of the first moment maps derived from the lines.
According to the test, there is no indication of systematic pointing offset.
Gradations of $R_{2/1}$ may be caused by interactions (e.g., NGC\,2146, NGC\,2798, NGC\,5713) or appearance of disk due to three-dimensional warp (NGC\,7331).

Note that the result for NGC\,5194 is not consistent with \citet{Koda12}.
They reported that intensities of $^{12}$CO($J$\,=\,1--0) data obtained in K07 (the original map is from \cite{Nakai94}) for this galaxy is $\sim 2$ times higher than that of \citet{Koda11}.
They argued that a calibration error in the data of \citet{Nakai94} was caused by difficulties in the calibration method at that time.

Figure \ref{fig:ratio_hist} shows the histogram of $R_{2/1}$ and the cumulative distribution function for all pixels in all galaxies in the samples.
As shown in the figure, the constant value of 0.7 assumed in many cases so far is rather large.
The peak of the histogram is in the range of 0.55--0.60; the most frequently appearing value of $R_{2/1}$ is lower than 0.7.
Cumulative fraction also indicates $R_{2/1} = 0.7$ is quite higher in our samples (cumulative fraction at $R_{2/1}=0.7$ is $\sim 70$\%).
This is significant even considering $R_{2/1}$ error of 13\% due to calibration error of CO data.
The median, mean weighted by $^{12}$CO($J$\,=\,1--0) integrated-intensity, and standard deviation of $R_{2/1}$ combined for all galaxies are 0.61, 0.66, and 0.19, respectively.
 
The median we obtained ($R_{2/1} = 0.61$) is slightly lower than the value ($R_{2/1} = 0.67$) reported in \citet{Leroy13}, while this difference would not be significant considering $R_{2/1}$ error due to calibration error is 13\% (section 2.2).
If this difference is significant, this discrepancy may have originated from differences in observed areas.
Most $^{12}$CO($J$\,=\,1--0) data \citet{Leroy13} used are from \citet{Usero15}, which observed specific positions in a galaxy where $^{12}$CO($J$\,=\,2--1) is strong.
Their observed areas in each galaxy usually include the center of the galaxy where $R_{2/1}$ tends to be high as mentioned in the next paragraph.
These results and statistics of $R_{2/1}$ such as median, $Q_2(R_{2/1})$, and weighted mean, ${\overline{R_{2/1}}}$, for each galaxy are summarized in table \ref{table:ratio_stat} with some properties of galaxies related to star formation.
The relations between $R_{2/1}$ and these properties of galaxies are discussed in section 4.1.
The histograms of $R_{2/1}$ for each galaxy are shown in Appendix 1.

Figure \ref{fig:ratio_radial_dist} shows $R_{2/1}$ as a function of the galactocentric radius for all galaxies.
Both the median and weighted mean of $R_{2/1}$ in the disk region ($r >2$ kpc) are nearly constant at 0.60.
Although $R_{2/1}$ in the disk is lower than 0.7, it tends to exceed 0.7 in central regions.
Both the median and weighted mean is 0.83 in the inner 0.75 kpc (the innermost bin of the bottom panel of figure \ref{fig:ratio_radial_dist}).
The usually assumed constant value of 0.7 for $R_{2/1}$ is common only for the transition region of the galactic center and the disk (1 kpc $\lesssim r \lesssim$ 2 kpc).

\begin{table*}[bt]
  \begin{center}
  \tbl{Statistics of $R_{2/1}$ and properties of the samples.}{
  \begin{tabular}{lccccccc} \hline \\[-3.4mm]
galaxy & $Q_2 (R_{2/1})$ & ${\overline {R_{2/1}}}$ & $\sigma (R_{2/1})$ & ${\overline {\Sigma_{\rm{mol}}}}$ & ${\overline {\Sigma_{\rm{SFR}}}}$ & ${\overline{\tau_{\rm{dep}}}}$ & ${\overline {I_{70\micron}}/{\overline {I_{160\micron}}}}$ \\
 &  &  &  & [$M_{\odot}\> \mathrm{pc^{-2}}$] & [$10^{-2}\> M_{\odot}\> \mathrm{yr^{-1} \> kpc^{-2}}$] & [$\rm{Gyr}$] &  \\
 & (1) & (2) & (3) & (4) & (5) & (6) & (7) \\
 \hline
all & 0.61 & 0.66 & 0.19 & --- & --- & --- & --- \\
NGC\,337 & ---$^{*}$ & ---$^{*}$ & ---$^{*}$ & ---$^{*}$ & ---$^{*}$ & ---$^{*}$ & ---$^{*}$ \\
NGC\,628 & 0.54 & 0.54 & 0.14 & 17.5 & 0.962 & 1.82 & 0.363 \\
NGC\,2146 & 0.66 & 0.73 & 0.23 & 113 & 27.9 & 0.406 & 1.22 \\
NGC\,2798 & 0.99 & 1.0 & 0.33 & 44.0 & 17.8 & 0.248 & 1.45 \\
NGC\,2841 & 0.51 & 0.50 & 0.12 & 4.34 & 0.287 & 1.51 & 0.210 \\
NGC\,2903 & 0.59 & 0.62 & 0.11 & 21.8 & 1.93 & 1.13 & ---$^{\dagger}$ \\
NGC\,2976 & 0.67 & 0.67 & 0.20 & 3.13 & 0.527 & 0.594 & 0.483 \\
NGC\,3034 & 0.56 & 0.67 & 0.15 & 23.2 & 3.75 & 0.617 & 1.43 \\
NGC\,3077 & ---$^{*}$ & ---$^{*}$ & ---$^{*}$ & ---$^{*}$ & ---$^{*}$ & ---$^{*}$ & ---$^{*}$ \\
NGC\,3184 & 0.55 & 0.56 & 0.14 & 14.5 & 0.756 & 1.91 & 0.307 \\
NGC\,3198 & 0.46 & 0.47 & 0.10 & 6.37 & 0.915 & 0.697 & 0.493 \\
NGC\,3351 & 0.48 & 0.73 & 0.21 & 19.4 & 3.20 & 0.608 & 0.765 \\
NGC\,3521 & 0.61 & 0.63 & 0.12 & 31.2 & 2.17 & 1.43 & 0.423 \\
NGC\,3627 & 0.46 & 0.46 & 0.10 & 39.0 & 2.96 & 1.32 & 0.543 \\
NGC\,3938 & 0.56 & 0.56 & 0.11 & 20.5 & 1.43 & 1.44 & 0.409 \\
NGC\,4214 & ---$^{*}$ & ---$^{*}$ & ---$^{*}$ & ---$^{*}$ & ---$^{*}$ & ---$^{*}$ & ---$^{*}$ \\
NGC\,4254 & 0.70 & 0.72 & 0.17 & 39.8 & 3.27 & 1.21 & 0.483 \\
NGC\,4321 & 0.76 & 0.83 & 0.18 & 26.7 & 2.00 & 1.33 & 0.421 \\
NGC\,4536 & 0.79 & 0.84 & 0.24 & 54.7 & 11.5 & 0.474 & 1.18 \\
NGC\,4559 & ---$^{*}$ & ---$^{*}$ & ---$^{*}$ & ---$^{*}$ & ---$^{*}$ & ---$^{*}$ & ---$^{*}$ \\
NGC\,4569 & 0.76 & 0.89 & 0.25 & 27.5 & 1.63  & 1.69 & 0.420 \\
NGC\,4579 & 0.50 & 0.63 & 0.20 & 20.1 & 0.890  & 2.25 & 0.396 \\
NGC\,4736 & 0.84 & 0.88 & 0.24 & 20.8 & 2.90  & 0.718 & 0.870 \\
NGC\,5055 & 0.54 & 0.56 & 0.11 & 18.2 & 0.87  & 2.11 & 0.330 \\
NGC\,5194 & 0.40$^{\ddagger}$ & 0.41$^{\ddagger}$ & 0.16$^{\ddagger}$ & ---$^{\S}$ & 2.98  & ---$^{\S}$ & 0.436 \\
NGC\,5457 & 0.62 & 0.64 & 0.18 & 15.3 & 0.89  & 1.72 & 0.336 \\
NGC\,5713 & 0.72 & 0.80 & 0.24 & 61.3 & 9.35  & 0.656 & 0.804 \\
NGC\,6946 & 0.67 & 0.71 & 0.17 & 29.6 & 1.85  & 1.60 & 0.491 \\
NGC\,7331 & 0.53 & 0.55 & 0.17 & 18.3 & 1.11  & 1.65 & 0.426 \\			
  \hline
  \end{tabular}}
  \label{table:ratio_stat}
  \begin{tabnote}
    (1) Median of $R_{2/1}$.
    (2) Mean of $R_{2/1}$ weighted by integrated intensity of $^{12}$CO($J$\,=\,1--0) averaged over the pixels where $R_{2/1}$ is significantly measured (cf. figure \ref{fig:ratio_map}).
    (3) Standard deviation of $R_{2/1}$.
    (4)--(5) Mean surface density of molecular gas and SFR. The area used to derive these means are the same as column (2).
    (6) Mean depletion time derived as total molecular gas mass over total SFR within the area used in column (2).
    (7) Mean IR color derived as total luminosity of 70 $\micron$ over that of 160 $\micron$ within the area used in column (2).
    \footnotemark[$*$] There are few or no pixels to measure $R_{2/1}$.
    \footnotemark[$\dag$] Archival data is not available.
    \footnotemark[$\ddag$] There may be calibration error of CO data.
    \footnotemark[$\S$] Not derived due to the possibility of calibration error.
  \end{tabnote}
  \end{center}
\end{table*}

\subsection{Effects of $R_{2/1}$ on molecular gas mass derived from $^{12}$CO($J$\,=\,2--1)}
\begin{figure*}[t!]
 \begin{center}
  \includegraphics[width=16cm]{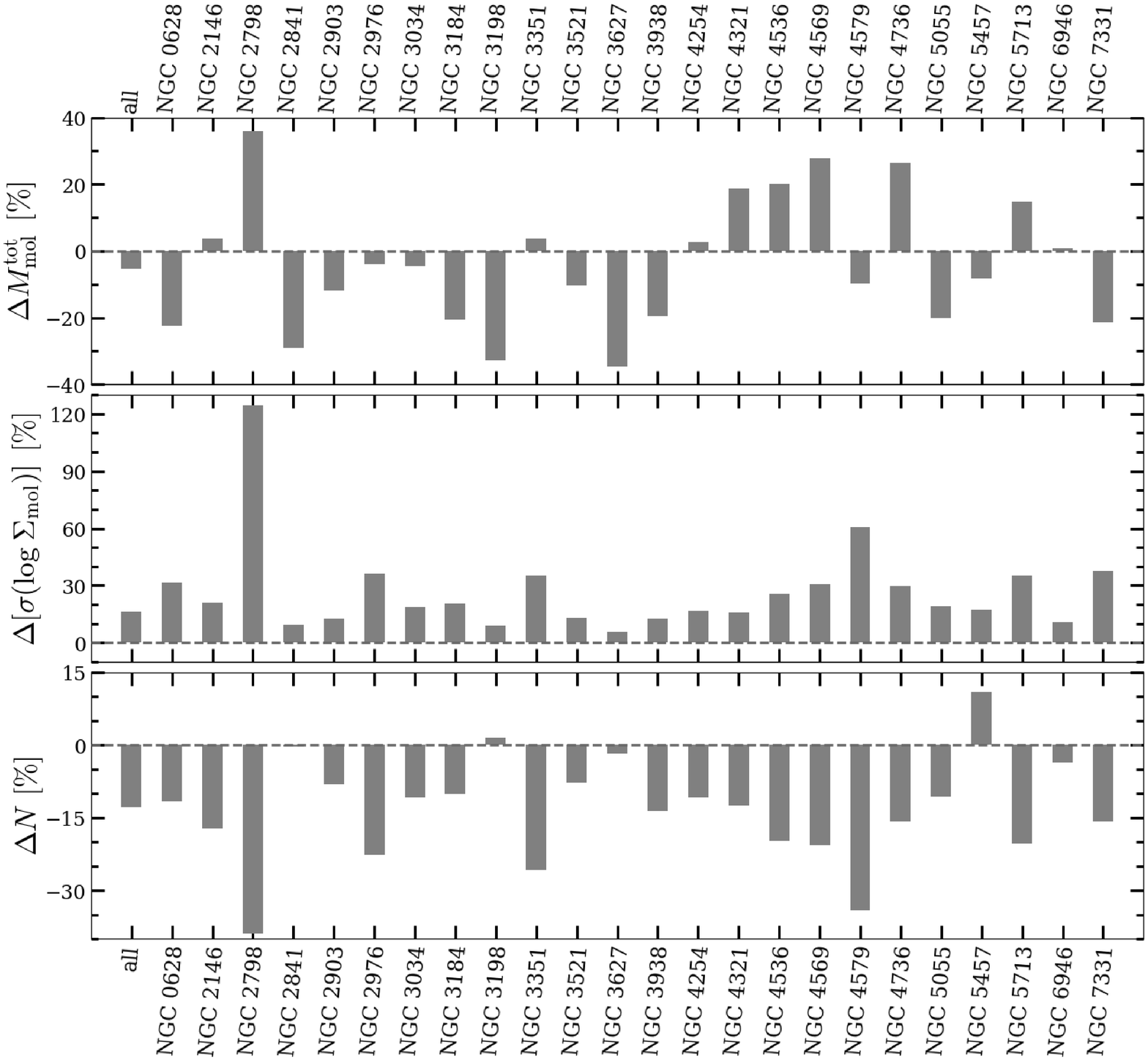}
 \end{center}
 \caption{Rate of change in the total molecular gas mass $\Delta M_{\mathrm{mol}}^{\mathrm{tot}}$ (top), standard deviation of molecular gas surface density $\Delta [\sigma (\log \Sigma_{\mathrm{mol}})]$ (middle), and index of the K--S relation $\Delta N$ (bottom) in each galaxy.
  }
 \label{fig:gas_mass_diff}
\end{figure*}

As effects originated from the assumption that $R_{2/1}$ is a constant, we report that how molecular gas mass is underestimated or overestimated and how the scatter of molecular gas surface density within a galaxy changes.
We adopted the standard CO-to-H$_2$ conversion factor $X_{\rm CO}$ of $2.0 \times 10^{20}\> {\rm cm^{-2}\> (K\>\> km\>\> s^{-1})^{-1}}$ \citep{Bolatto13} for the entire area of all galaxies based on the widely used method.

The top panel of figure \ref{fig:gas_mass_diff} shows how the total molecular gas mass within a galaxy changes (rate of change in total molecular gas mass; $\Delta M_{\mathrm{mol}}^{\mathrm{tot}}$) when the molecular gas mass is derived from $^{12}$CO($J$\,=\,2--1) with the $R_{2/1}$ of 0.7 {compared with the case wherein $M_{\mathrm{mol}}^{\mathrm{tot}}$ is derived from CO($J$\,=\,1--0)}.
For instance, $\Delta M_{\mathrm{mol}}^{\mathrm{tot}} = -20$\% indicates that the total molecular gas mass derived from $^{12}$CO($J$\,=\,2--1) is underestimated by 20\%.
NGC\,337, NGC\,3077, NGC\,4214, and NGC\,4559 are omitted because $R_{2/1}$ cannot be significantly measured in most positions in these galaxies (figure \ref{fig:ratio_map}) and NGC\,5194 is omitted due to the possibility of the calibration problem as mentioned in section 3.1.

As ${\overline{R_{2/1}}}$ deviates from 0.7 (table \ref{table:ratio_stat}), the total molecular gas mass within a galaxy is underestimated or overestimated when the molecular gas mass is derived from $^{12}$CO($J$\,=\,2--1) assuming the constant $R_{2/1}$ of 0.7.
For instance, the total molecular gas mass is underestimated by $\sim 30$\% for galaxies that show low ${\overline{R_{2/1}}}$ ($\lesssim 0.50$; e.g., NGC\,2841, NGC\,3198).
Further, it is overestimated by $\sim 30$\% when ${\overline{R_{2/1}}}$ is high ($\gtrsim 0.9$; e.g., NGC\,2798, NGC\,4569, NGC\,4736).
For the most deviated galaxy, the molecular gas mass is underestimated by $\sim 35$\% in NGC\,3627.

The middle panel of figure \ref{fig:gas_mass_diff} shows the change rate of standard deviation for $\log \Sigma_{\rm mol}$, $\Delta [\sigma (\log \Sigma_{\rm mol})]$, in each galaxy when $R_{2/1}$ is assumed to be the constant.
$\Delta [\sigma (\log \Sigma_{\rm mol})]$ is positive for all galaxies and exceeds 30\% in some galaxies (NGC\,628, {NGC\,2798}, NGC\,2976, NGC\,3351, NGC\,5713, and NGC\,7331), and the highest one reaches 120\% (NGC\,2798).
When the scatter of $R_{2/1}$, $\sigma (R_{2/1})$, is relatively large ($\gtrsim 0.2$) or when ${\overline{R_{2/1}}}$ deviates from 0.7, $\Delta [\sigma (\log \Sigma_{\rm mol})]$ tends to be large.
In contrast, when $\sigma (R_{2/1})$ is small, $\Delta [\sigma (\log \Sigma_{\rm mol})]$ is also small (e.g., NGC\,3627).
These results indicate that not only the total molecular gas mass but also molecular gas surface density and its relevant quantities in positions-to-position are misled by the assumption of the constant $R_{2/1}$.
This becomes an issue when spatially resolved data are used, which has been the standard recently.

\subsection{Effects of $R_{2/1}$ on the Kennicutt--Schmidt relation derived from $^{12}$CO($J$\,=\,2--1)}
\begin{figure}[t!]
 \begin{center}
  \includegraphics[width=8cm]{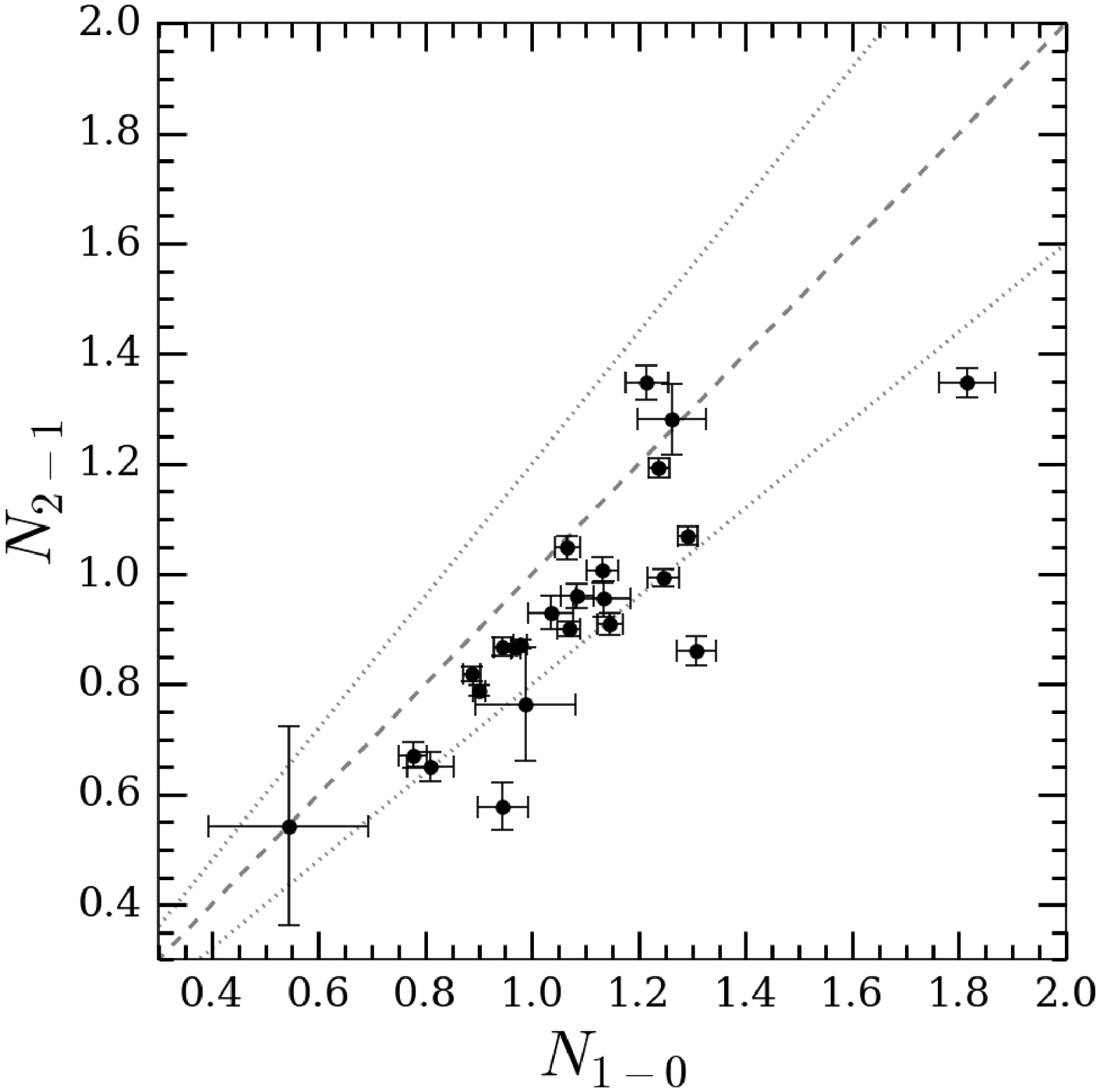}
 \end{center}
 \caption{Indices of the K--S relation derived from $^{12}$CO($J$\,=\,2--1) with the constant $R_{2/1}$ of 0.7 ($N_{2-1}$) against that from $^{12}$CO($J$\,=\,1--0) ($N_{1-0}$).
 The diagonal dashed line indicates $N_{2-1} = N_{1-0}$.
 Dotted lines indicate that $N_{2-1}$ is higher and lower by 20\% than $N_{1-0}$. 
  }
 \label{fig:index_change}
\end{figure}
\begin{figure*}[p!]
 \begin{center}
 \includegraphics[width=16cm]{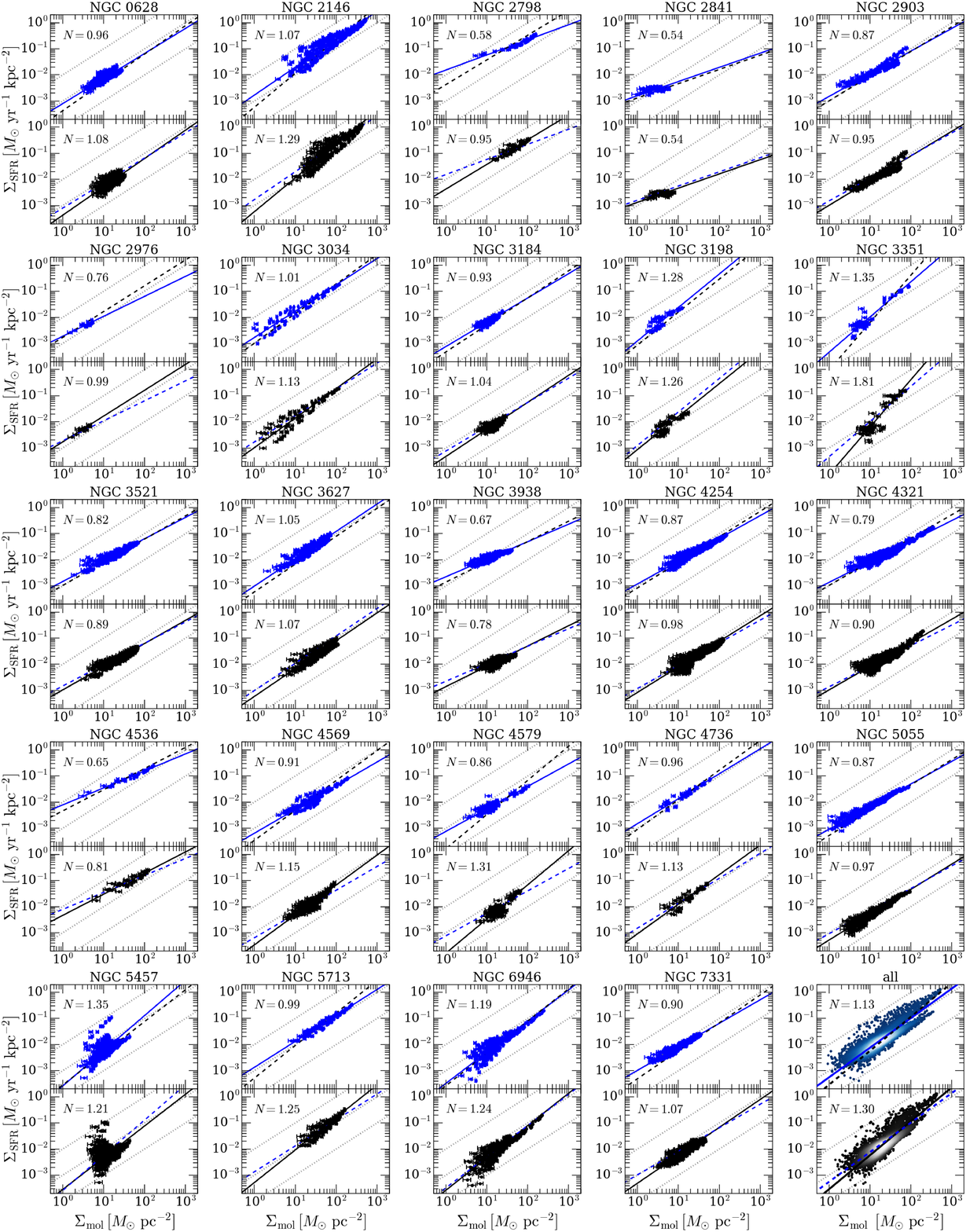}
 \end{center}
 \caption{The K--S relation of the sample galaxies derived from $^{12}$CO($J$\,=\,2--1) assuming $R_{2/1} = 0.7$ (top panels, blue plots) and from $^{12}$CO($J$\,=\,1--0) (bottom panels, black plots).
The index of each panel is provided on the top left.
Solid lines and dashed lines indicate regression lines of the panel and the other panel, respectively.
Dotted lines indicate that the depletion time of molecular gas is $10^{-1}$ Gyr, 10$^0$ Gyr, and $10^1$ Gyr from bottom to top.
The K--S plot of all galaxies (excluding NGC\,337, NGC\,3077, NGC\,4214, NGC\,4559, and NGC\,5194) is shown in the panel at the bottom right corner.
 }
 \label{fig:KSplot_total}
\end{figure*}
\begin{table*}[t!]
  \begin{center}
  \tbl{Fitted parameters of the K--S relation.}{
  \begin{tabular}{lcccccc} \hline
galaxy & $N_{1-0}$ & $N_{2-1}$ & $\Delta N$ & $A_{1-0}$ & $A_{2-1}$ & $\Delta A$ \\
 &  &  & [\%] &  &  & [\%] \\
 & (1) & (2) & (3) & (4) & (5) & (6) \\
 \hline
all & $1.299 \pm 0.005$ & $1.132 \pm 0.004$ & $-12.8$ & $-3.537 \pm 0.008$ & $-3.253 \pm 0.007$ & $-8.02$ \\
NGC\,337 & --- & --- & --- & --- & --- & --- \\
NGC\,628 & $1.08 \pm 0.03$ & $0.96 \pm 0.02$ & $-11.5$ & $-3.37 \pm 0.05$ & $-3.10 \pm 0.04$ & $-8.05$ \\
NGC\,2146$^*$ & $1.29 \pm 0.02$ & $1.07 \pm 0.02$ & $-17.2$ & $-3.24 \pm 0.03$ & $-2.77 \pm 0.03$ & $-14.3$ \\
NGC\,2798 & $0.95 \pm 0.05$ & $0.58 \pm 0.04$ & $-38.8$ & $-2.37 \pm 0.06$ & $-1.81 \pm 0.04$ & $-23.5$ \\
NGC\,2841 & $0.5 \pm 0.1$ & $0.5 \pm 0.2$ & $-0.24$ & $-2.9 \pm 0.4$ & $-2.8 \pm 0.4$ & $-2.87$ \\
NGC\,2903 & $0.95 \pm 0.02$ & $0.87 \pm 0.02$ & $-8.18$ & $-2.99 \pm 0.03$ & $-2.82 \pm 0.03$ & $-5.64$ \\
NGC\,2976 & $0.99 \pm 0.09$ & $0.8 \pm 0.1$ & $-22.7$ & $-2.8 \pm 0.2$ & $-2.7 \pm 0.2$ & $-1.54$ \\
NGC\,3034 & $1.13 \pm 0.03$ & $1.01 \pm 0.02$ & $-10.9$ & $-2.99 \pm 0.04$ & $-2.77 \pm 0.04$ & $-7.27$ \\
NGC\,3077 & --- & --- & --- & --- & --- & --- \\
NGC\,3184 & $1.04 \pm 0.04$ & $0.93 \pm 0.03$ & $-10.2$ & $-3.33 \pm 0.07$ & $-3.12 \pm 0.05$ & $-6.21$ \\
NGC\,3198 & $1.26 \pm 0.06$ & $1.28 \pm 0.06$ & $1.48$ & $-3.06 \pm 0.08$ & $-2.90 \pm 0.08$ & $-5.22$ \\
NGC\,3351 & $1.81 \pm 0.05$ & $1.35 \pm 0.03$ & $-25.7$ & $-4.02 \pm 0.07$ & $-3.34 \pm 0.04$ & $-16.9$ \\
NGC\,3521 & $0.89 \pm 0.02$ & $0.82 \pm 0.01$ & $-7.77$ & $-2.98 \pm 0.03$ & $-2.83 \pm 0.02$ & $-4.98$ \\
NGC\,3627 & $1.07 \pm 0.02$ & $1.05 \pm 0.02$ & $-1.73$ & $-3.23 \pm 0.04$ & $-3.01 \pm 0.03$ & $-6.89$ \\
NGC\,3938 & $0.78 \pm 0.03$ & $0.67 \pm 0.02$ & $-13.6$ & $-2.86 \pm 0.05$ & $-2.65 \pm 0.04$ & $-7.19$ \\
NGC\,4214 & --- & --- & --- & --- & --- & --- \\
NGC\,4254 & $0.98 \pm 0.01$ & $0.872 \pm 0.008$ & $-10.8$ & $-3.08 \pm 0.02$ & $-2.93 \pm 0.01$ & $-4.94$ \\
NGC\,4321 & $0.90 \pm 0.01$ & $0.79 \pm 0.01$ & $-12.5$ & $-2.98 \pm 0.02$ & $-2.87 \pm 0.02$ & $-3.92$ \\
NGC\,4536 & $0.81 \pm 0.04$ & $0.65 \pm 0.03$ & $-19.7$ & $-2.33 \pm 0.06$ & $-2.10 \pm 0.03$ & $-9.81$ \\
NGC\,4559 & --- & --- & --- & --- & --- & --- \\
NGC\,4569 & $1.15 \pm 0.02$ & $0.91 \pm 0.02$ & $-20.6$ & $-3.46 \pm 0.04$ & $-3.19 \pm 0.04$ & $-7.89$ \\
NGC\,4579 & $1.31 \pm 0.04$ & $0.86 \pm 0.03$ & $-34.2$ & $-3.78 \pm 0.06$ & $-3.12 \pm 0.05$ & $-17.5$ \\
NGC\,4736 & $1.13 \pm 0.05$ & $0.96 \pm 0.03$ & $-15.7$ & $-3.04 \pm 0.07$ & $-2.82 \pm 0.05$ & $-7.16$ \\
NGC\,5055 & $0.969 \pm 0.009$ & $0.866 \pm 0.007$ & $-10.6$ & $-3.28 \pm 0.02$ & $-3.04 \pm 0.01$ & $-7.36$ \\
NGC\,5457 & $1.21 \pm 0.04$ & $1.35 \pm 0.03$ & $10.9$ & $-3.55 \pm 0.06$ & $-3.58 \pm 0.04$ & $1.07$ \\
NGC\,5713$^*$ & $1.25 \pm 0.03$ & $0.99 \pm 0.01$ & $-20.2$ & $-3.28 \pm 0.05$ & $-2.87 \pm 0.02$ & $-12.5$ \\
NGC\,6946 & $1.24 \pm 0.02$ & $1.19 \pm 0.02$ & $-3.56$ & $-3.58 \pm 0.03$ & $-3.49 \pm 0.03$ & $-2.52$ \\
NGC\,7331 & $1.07 \pm 0.02$ & $0.90 \pm 0.01$ & $-15.8$ & $-3.31 \pm 0.03$ & $-2.99 \pm 0.02$ & $-9.7$ \\
  \hline
  \end{tabular}}
  \label{table:KS_results}
  \begin{tabnote}
    (1) Index of the K--S relation derived from $^{12}$CO($J$\,=\,1--0).
    (2) Index of the K--S relation derived from $^{12}$CO($J$\,=\,2--1) and the constant $R_{2/1}$ of 0.7.
    (3) Change rate of index.
    (4) Intercept of the K--S relation derived from $^{12}$CO($J$\,=\,1--0).
    (5) Intercept of the K--S relation derived from $^{12}$CO($J$\,=\,2--1) and the constant $R_{2/1}$ of 0.7.
    (6) Change rate of intercept.
    \footnotemark[$*$] The spatial resolution is different from 1.5 kpc (cf. table \ref{table:sample}).
  \end{tabnote}
  \end{center}
\end{table*}

Next, we investigate how the Kennicutt--Schmidt (K--S) relation (\cite{Schmidt59}; \cite{Kennicutt89}) changes when the molecular gas surface density is derived from $^{12}$CO($J$\,=\,2--1) and the fixed $R_{2/1}$ compared with that derived from $^{12}$CO($J$\,=\,1--0).
The molecular K--S relation is described as
\begin{equation}
\log \left( \frac{\Sigma_{\mathrm{SFR}}}{M_{\odot}\> \mathrm{yr^{-1}\> kpc^{-2}}} \right) =
 N \log \left( \frac{\Sigma_{\mathrm{mol}}}{M_{\odot}\> \mathrm{pc^{-2}}} \right) + A,
\end{equation}
where $N$ is the index and $A$ is the intercept on the double-logarithmic plot.
Several studies suggested that this relation, in particular $N$, reflects the processes of star formation in galaxies (e.g., \cite{Elmegreen02}; \cite{Krumholz05}; \cite{Komugi06}; \cite{Tan10}; Takeuchi, T., T. et al. 2020 in preparation).
Therefore, we focused on how $N$ changes in this study.

When the spatial resolution is changed, the result of the K--S relation also changes \citep{Onodera10}.
Thus, we smoothed $^{12}$CO cubes and $\Sigma_{\mathrm{SFR}}$ maps so that the spatial resolution is the same value of 1.5 kpc for all galaxies except for NGC\,337, NGC\,2146, and NGC\,5713.
Since the original spatial resolution of these three galaxies is larger than 1.5 kpc (table \ref{table:sample}), we did not smooth them.
We also regridded $^{12}$CO cubes and $\Sigma_{\mathrm{SFR}}$ maps to fix spatial sampling (i.e., pixel size) for all galaxies.
The pixel size is set to 650 pc which is slightly smaller than the Nyquist sampling.
For convolved and regridded data, pixels whose $S/N$ of integrated-intensities does not reach 4.5$\,\sigma$ were masked.
We made K--S plots from $^{12}$CO($J$\,=\,1--0) by adopting the standard $X_{\rm{CO}}$ of $2.0 \times 10^{20}\>\> {\rm cm^{-2}\>\> (\mathrm{K\>\> km\>\> s^{-1}})^{-1}}$ for the entire area of all galaxies according to the widely used method.
The K--S plots from $^{12}$CO($J$\,=\,2--1) are obtained by converting $^{12}$CO($J$\,=\,2--1) intensity into that of $^{12}$CO($J$\,=\,1--0) with the constant $R_{2/1}$ of 0.7, which is the same method employed in previous studies about the K--S relation with $^{12}$CO($J$\,=\,2--1) (e.g., \cite{Bigiel08}).

We fitted the K--S relation with the ordinary least-squares (OLS) bisector method \citep{Isobe90} and derived $N$ and $A$ of the relation made with $^{12}$CO($J$\,=\,1--0) and $^{12}$CO($J$\,=\,2--1), respectively.
Indices derived from $^{12}$CO($J$\,=\,1--0) and $^{12}$CO($J$\,=\,2--1) (hereafter, $N_{1-0}$ and $N_{2-1}$, respectively), intercepts from the two $^{12}$CO lines (similarly, $A_{1-0}$ and $A_{2-1}$, respectively), and their rates of change ($\Delta N$ and $\Delta A$) are listed in table \ref{table:KS_results}.
The bottom panel of figure \ref{fig:gas_mass_diff} indicates the change rate of indices; figure \ref{fig:index_change} shows the correlation plot of $N_{2-1}$ against $N_{1-0}$.
The K--S plots of each galaxy are shown in figure \ref{fig:KSplot_total}.
We could not fit the K--S relation for NGC\,337, NGC\,3077, NGC\,4214, and NGC\,4559 because the number of pixels at which $R_{2/1}$ is significantly measured is not enough.
Therefore, these four galaxies were excluded from the K--S relation of all compiled galaxies in the samples (the last panel in figure \ref{fig:KSplot_total}).

We find that indices decrease typically by 10--20\%, up to {39\%}, when the relation is derived from $^{12}$CO($J$\,=\,2--1) with the constant $R_{2/1} = 0.7$ in 17 galaxies.
The K--S relation using molecular gas surface density derived by this method produces a lower index than that derived by $^{12}$CO($J$\,=\,1--0) for most galaxies.
This tendency is the same as \citet{Momose13} that discussed the discrepancy between their super-linear slope of the K--S relation derived by $^{12}$CO($J$\,=\,1--0) and the linear slope derived by $^{12}$CO($J$\,=\,2--1) reported in \citet{Bigiel08}.
The significant change ($\Delta N < -20\%$) of the index is shown in NGC\,2798, NGC\,2976, and NGC\,3351.
In these galaxies, $\sigma (R_{2/1})$ within a galaxy tends to be relatively larger than others (table \ref{table:ratio_stat}).
When the variation of $R_{2/1}$ in a galaxy is small (e.g., NGC\,3198, NGC\,3627), $\Delta N$ is small ($|\Delta N | \lesssim 2\%$).
The differences of the K--S relation for all sample galaxies also shows a lower index ($\Delta N = -13\%$).

We interpret these underestimated indices of the K--S relation derived from $^{12}$CO($J$\,=\,2--1) as follows.
$R_{2/1}$ is often higher than 0.7 when $\Sigma_{\rm SFR}$ and $\Sigma_{\rm mol}$ are high (the top right on the K--S plot), while $R_{2/1}$ is prone to be lower than 0.7 when $\Sigma_{\rm SFR}$ and $\Sigma_{\rm mol}$ are low (the bottom left on the K--S plot).
Here, we mean that ``$\Sigma_{\rm{mol}}$'' is derived from $^{12}$CO($J$\,=\,1--0).
As a result, molecular gas surface density derived from $^{12}$CO($J$\,=\,2--1) with the fixed $R_{2/1}$ of 0.7 is overestimated (data points move to right in the plot) where $\Sigma_{\rm{SFR}}$ and $\Sigma_{\rm{mol}}$ are high, and it is underestimated (data points move to left in the plot) where $\Sigma_{\rm{SFR}}$ and $\Sigma_{\rm{mol}}$ are low (we examine correlations of $R_{2/1}$, $\Sigma_{\rm SFR}$, and $\Sigma_{\rm mol}$ in section 4.1; figures \ref{fig:R21_vs_SFR} and \ref{fig:R21_vs_gasmass}).
This becomes prominent in galaxies whose $\sigma(R_{2/1})$ is large, as seen in the plot of NGC\,2798, NGC\,3351, and NGC\,4579 of figure \ref{fig:KSplot_total}.
We note that indices are underestimated even if $R_{2/1}$ is assumed as the mean or median of $R_{2/1}$ in a galaxy.

When the K--S relation is derived from higher-$J$ $^{12}$CO lines such as $^{12}$CO($J$\,=\,3--2), the same tendency of lower indices is seen (\cite{Morokuma17}).
Dense gas tracers also produce nearly linear K--S relations, whereas $^{12}$CO($J$\,=\,1--0) produces super-linear relations (e.g., \cite{Gao07}; \cite{Kennicutt89}).
According to these results, it seems that indices of the K--S relation become low when the relation is derived from the molecular gas, that is more related to star-formation activity.
We conclude that the K--S relation, the basic relation of molecular gas with star formation, will be misinterpreted, in addition to total molecular gas within a galaxy and surface density of molecular gas when $^{12}$CO($J$\,=\,2--1) is used as a molecular gas tracer with a fixed $R_{2/1}$.
Previous studies of the K--S relation using molecular gas surface density derived from this method would underestimate their indices.

\section{Discussion}
\subsection{Correlations of $R_{2/1}$ with basic properties of galaxies}
To investigate how $R_{2/1}$ changes depending on basic properties of galaxies such as $\Sigma_{\rm SFR}$, correlations of $R_{2/1}$ with representative quantities are examined in this section.
The spatial resolution and grid spacing are fixed to be 1.5 kpc and 650 pc except for NGC\,337, NGC\,2146, and NGC\,5713, similar to the K--S relation case (these three galaxies were analyzed with their original resolution as shown in table \ref{table:sample}).

\begin{figure}[t!]
 \begin{center}
  \includegraphics[width=8cm]{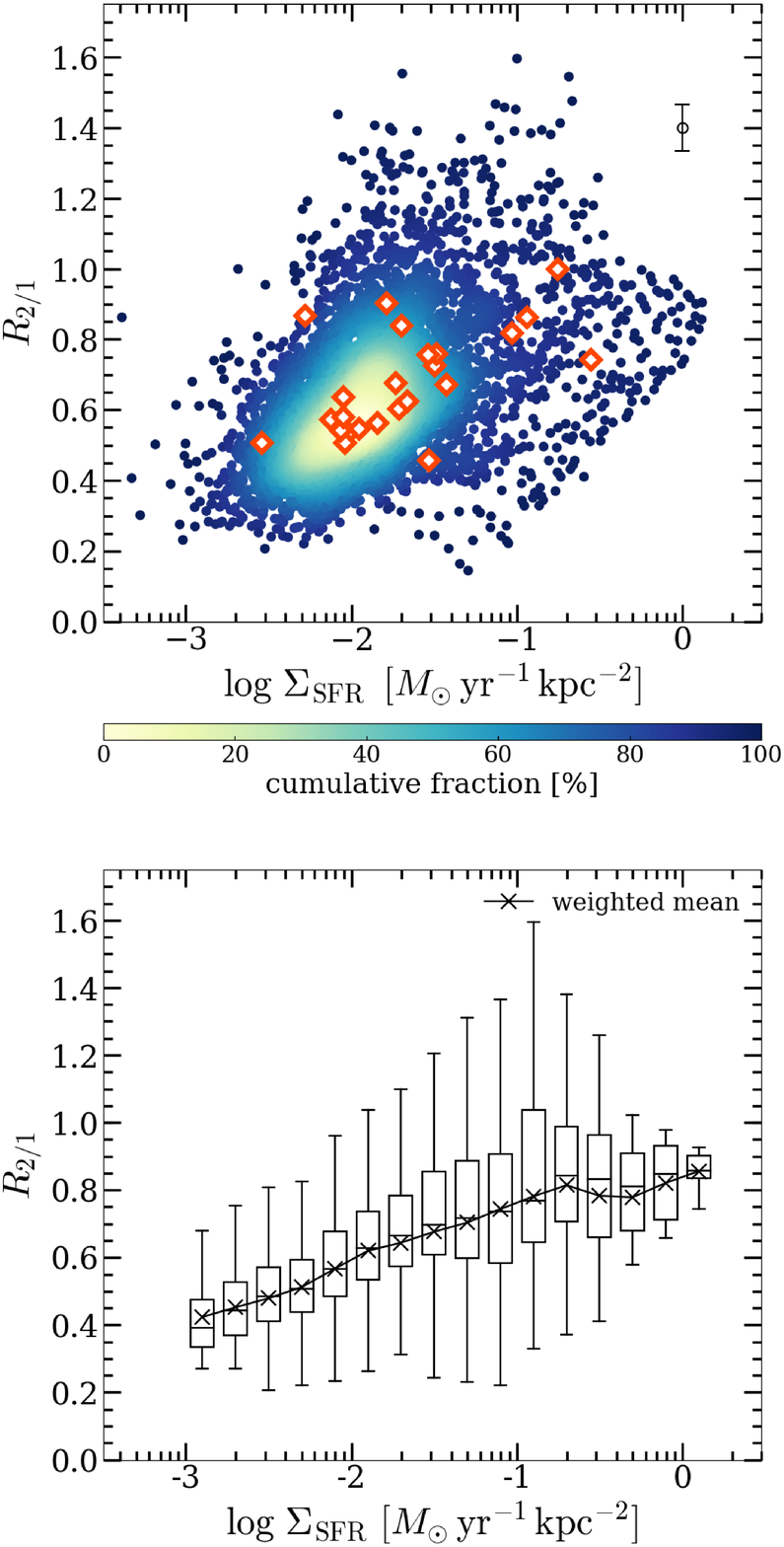}
 \end{center}
 \caption{(Top) Correlation of spatially resolved data (circles) and integrated data over the whole galaxy (open orange diamonds) of $R_{2/1}$ and $\Sigma_{\rm{SFR}}$.
 The colors indicate the cumulative fraction of data points.
 The typical error of $R_{2/1}$ is shown on the top right in the panel.
 Errors of $\Sigma_{\rm SFR}$ is negligible.
  (Bottom) Box plot and ${\overline{R_{2/1}}}$ in each bin (cross markers) for the top panel.
 The bin width is 0.2 dex.
 Settings of the box plot are the same as figure \ref{fig:ratio_radial_dist}.
  }
 \label{fig:R21_vs_SFR}
\end{figure}

First, we describe the correlation of $R_{2/1}$ with $\Sigma_{\rm SFR}$.
The top panel of figure \ref{fig:R21_vs_SFR} shows the correlation plot of $R_{2/1}$ against $\Sigma_{\rm SFR}$ for spatially resolved data and integrated data over the pixels where $R_{2/1}$ is significantly measured (table \ref{table:ratio_stat}, figure \ref{fig:ratio_map}).
The bottom panel of the figure shows the box plot for the top panel binned with 0.2 dex and the mean $R_{2/1}$ weighted by $^{12}$CO($J$\,=\,1--0) integrated-intensity in each bin.
The Spearman's rank correlation coefficient ($\rho_s$) for the spatially resolved case and the integrated one are 0.47 and 0.51, respectively.
Although the scatter is rather large, $Q_2(R_{2/1})$ and ${\overline{R_{2/1}}}$ in each bin clearly increase from $\sim 0.4$ to $\sim 0.9$ as $\Sigma_{\rm SFR}$ increases.
This result is consistent with \citet{Koda12} and \citet{Koda20}, whereas the spatial resolution of the former is approximately twice higher than ours.
From this result, possibilities are as follows: one main physical factor to change $R_{2/1}$ can be star-formation feedback (i.e., high $R_{2/1}$ is a result and high $\Sigma_{\rm SFR}$ is a cause) or $R_{2/1}$ becomes high because the molecular gas is dense and as a result, the star-formation rate is high (i.e., high-density condition is a cause and high $R_{2/1}$ and high $\Sigma_{\rm SFR}$ are results).
Otherwise, this correlation might be just mere coincidence.

\begin{figure}[t!]
 \begin{center}
  \includegraphics[width=8cm]{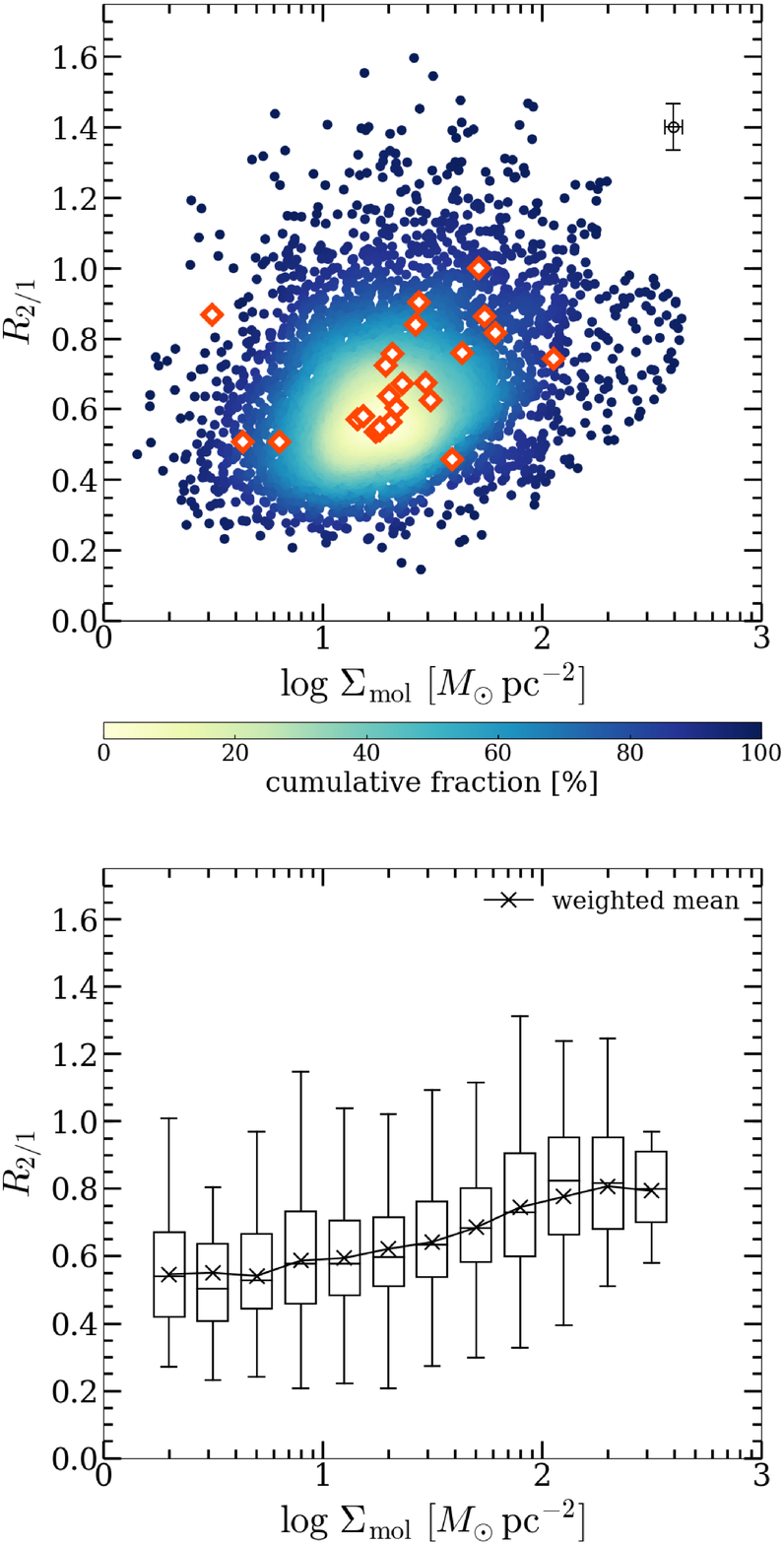}
 \end{center}
 \caption{The same as figure \ref{fig:R21_vs_SFR} but for $\Sigma_{\rm mol}$.
  }
 \label{fig:R21_vs_gasmass}
\end{figure}

Second, we investigate the dependence of $R_{2/1}$ on $\Sigma_{\rm mol}$ as shown in figure \ref{fig:R21_vs_gasmass}.
Mean $\Sigma_{\rm mol}$ is derived by the same method of that for mean $\Sigma_{\rm SFR}$ case.
$R_{2/1}$ loosely increases from $\sim 0.55$ to $\sim 0.8$ as $\Sigma_{\rm mol}$ increases according to the bottom panel, while there seems to be no clear tendency according to the scatter plot in the top panel.
The $\rho_s$ for the spatially resolved and global result is 0.28 and 0.51, respectively.
Since the tendency of $R_{2/1}$ against $\Sigma_{\rm SFR}$ is clearer than that against $\Sigma_{\rm mol}$, the lower indices of the K--S relation using molecular gas mass derived from $^{12}$CO($J$\,=\,2--1) (discussed in section 3.3) are mainly due to fluctuated $R_{2/1}$ by $\Sigma_{\rm SFR}$ rather than $\Sigma_{\rm mol}$.

\begin{figure}[t!]
 \begin{center}
  \includegraphics[width=8cm]{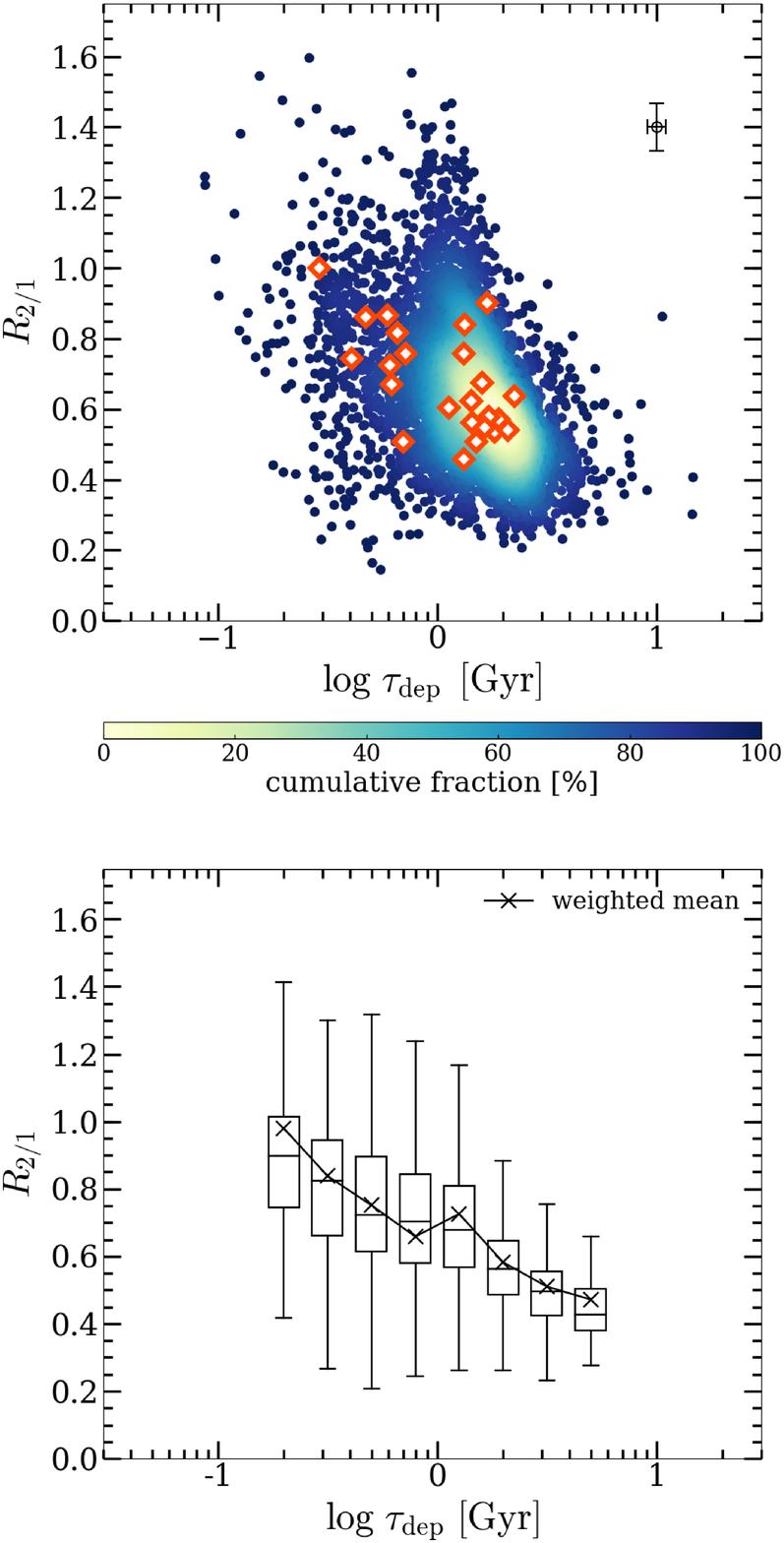}
 \end{center}
 \caption{The same as figure \ref{fig:R21_vs_SFR} but for $\tau_{\rm dep}$.
  }
 \label{fig:R21_vs_gasdep}
\end{figure}

Third, the correlation between $R_{2/1}$ and the depletion time of molecular gas ($\tau_{\rm dep}$) is examined.
The $\tau_{\rm dep}$ is derived as the following equation:
\begin{equation}
\left( \frac{\tau_{\rm dep}}{\rm yr} \right) = \left( \frac{\Sigma_{\rm mol}}{M_{\odot}\> {\rm pc^{-2}}} \right)
\bigg / \left( \frac{\Sigma_{\rm SFR}}{M_{\odot}\> {\rm yr^{-1}\> pc^{-2}}} \right).
\end{equation}
Figure \ref{fig:R21_vs_gasdep} shows the relation between $R_{2/1}$ and $\tau_{\rm dep}$ in the same way as figure \ref{fig:R21_vs_SFR}.
Mean $\tau_{\rm dep}$ is derived by dividing total molecular gas mass by total SFR within the pixels where $R_{2/1}$ is significantly measured.
Both the top and bottom panels show significantly decreased $R_{2/1}$ as $\tau_{\rm dep}$ increases.
This relation is also consistent with \citet{Koda12} (note that their horizontal axis is proportional to SFE that is equivalent to the reciprocal number of $\tau_{\rm dep}$).
The $\rho_s$ for the spatially resolved and integrated case is $-0.47$ and $-0.50$, respectively.
$Q_2(R_{2/1})$ and ${\overline{R_{2/1}}}$ decrease from $\sim 1.2$ to $\sim 0.45$ when $\tau_{\rm dep}$ increases from $\sim 0.1$ Gyr to $\sim 5$ Gyr.
Therefore, $R_{2/1}$ may be related to an efficient conversion from molecular gas into stars because of the dense condition (e.g., \cite{Muraoka16}; \cite{Yajima19}).
As another possibility, molecular gas is easily warmed by active star formation (heat source) with poor molecular gas (low heat capacity) when $\tau_{\rm dep}$ is short.
These possibilities are relevant to the dense or warm conditions of molecular gas likely mentioned in the $\Sigma_{\rm SFR}$ case.

\begin{figure}[t!]
 \begin{center}
  \includegraphics[width=8cm]{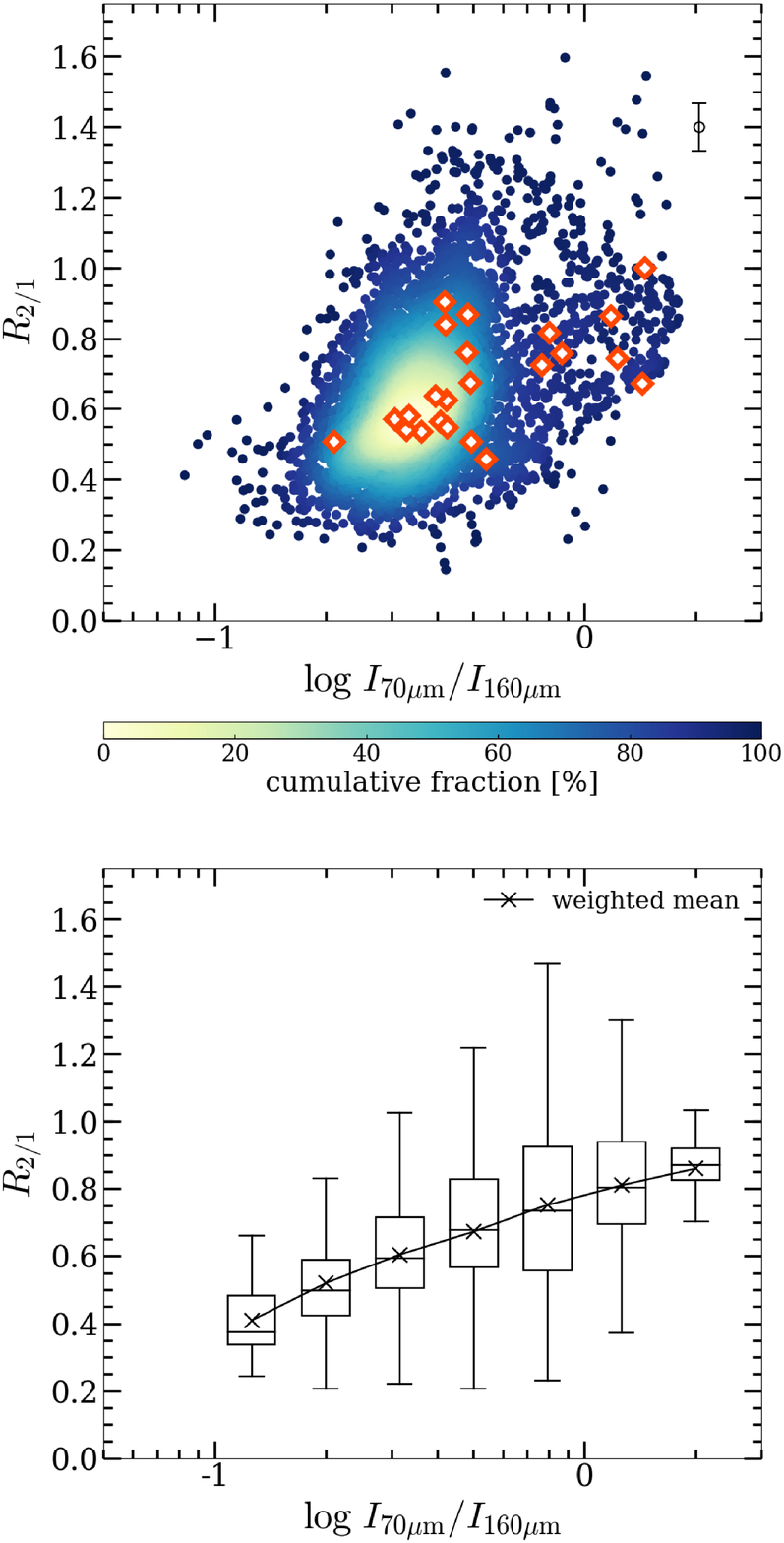}
 \end{center}
 \caption{The same as figure \ref{fig:R21_vs_SFR} but for IR color ($I_{70 \micron} / I_{160 \micron}$).
 Errors of IR color is negligible.
  }
 \label{fig:R21_vs_IRcolor}
\end{figure}

Finally, the correlation of $R_{2/1}$ with IR color is examined.
The IR color is derived from Herschel/PACS 70-$\micron$ and 160-$\micron$ intensity ratios.
Usually, the peak of spectral energy distribution (SED) for cool dust, which is dominant in ISM, is around the 160-$\micron$ band.
Therefore, this band selection well reflects dust temperature, especially in the present case of lacking data of the long-wavelength side for most of the sample galaxies.
Cool dust temperature may also be a good probe of ISM conditions.
Since there is no archival data of PACS for NGC\,2903, we do not derive IR color for this galaxy.
Figure \ref{fig:R21_vs_IRcolor} shows the correlation between $R_{2/1}$ and IR color and the box plot in the same manner as figure \ref{fig:R21_vs_SFR}.
The mean IR color is derived from the ratio of total intensity of $I_{70\micron}$ to that of $I_{160\micron}$ within the pixels where $R_{2/1}$ is significantly measured.
The $\rho_s$ of the spatially resolved and integrated case is 0.39 and 0.49.
Both $Q_2(R_{2/1})$ and ${\overline{R_{2/1}}}$ increase from $\sim 0.4$ to $\sim 0.85$ as IR color increases (corresponding dust temperature is from $\sim 15$ K to $\sim 35$ K) similarly to the $\Sigma_{\rm SFR}$ case.
The tendency is consistent with \citet{Koda20}.
However, the correlation coefficient between $R_{2/1}$ and IR color is higher than that for $R_{2/1}$ and $\Sigma_{\rm SFR}$ in their paper, whereas it is not higher (0.47 and 0.39) in our results.
This difference may be originated from far IR band selection and sample selection because their resolution is comparable to ours (their resolution is 1.2 kpc and ours is 1.5 kpc).

\subsection{$R_{2/1}$ and physical properties of molecular gas}
\subsubsection{$R_{2/1}$ and $^{12}$CO($J$\,=\,1--0)/$^{13}$CO($J$\,=\,1--0) ratio}
\begin{figure}[tbh]
 \begin{center}
  \includegraphics[width=8cm]{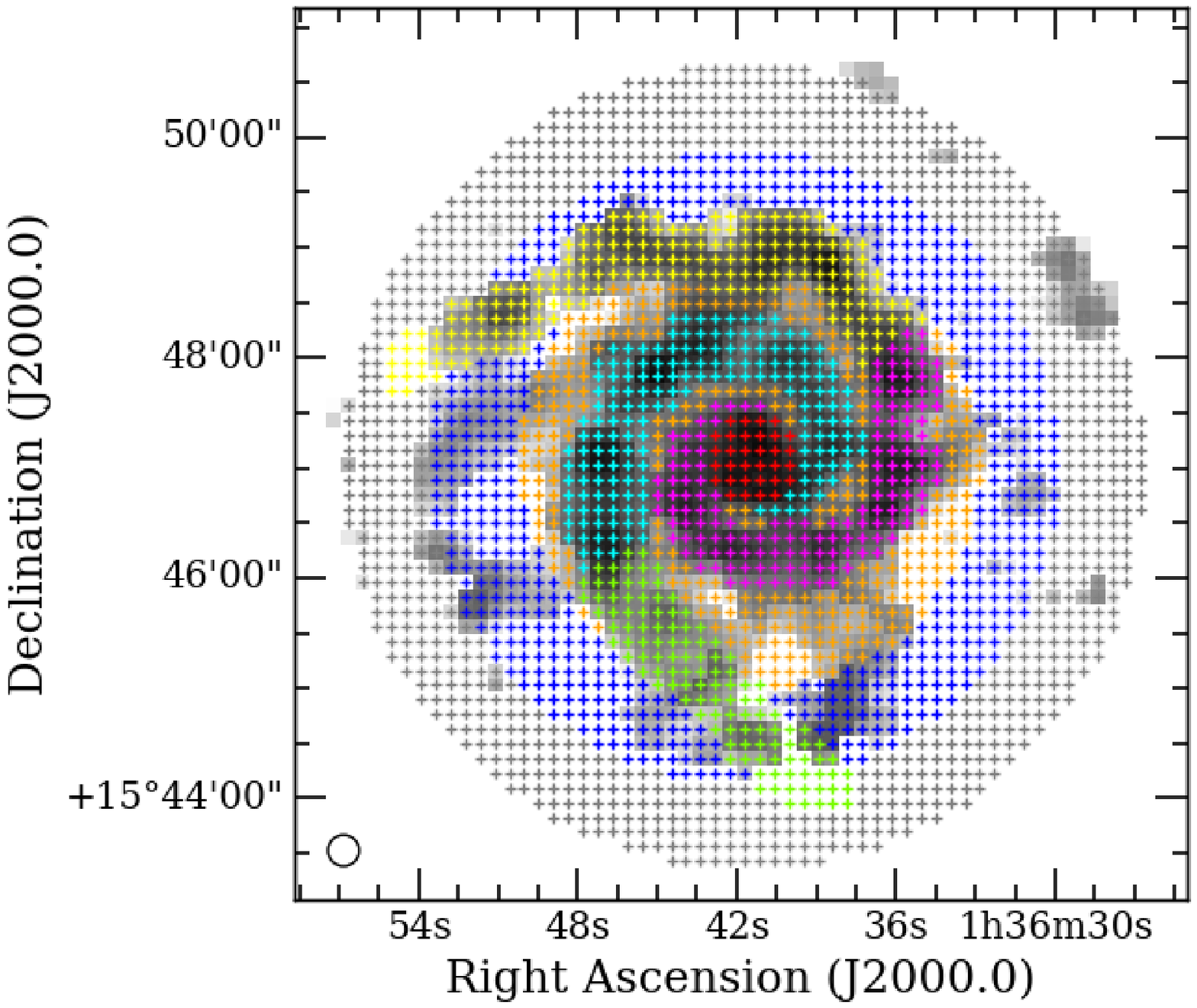}
 \end{center}
 \caption{Regions for spectra stacking in NGC\,628: center (red), inner arm 1 (magenta), inner arm 2 (cyan), outer arm 1 (yellow), outer arm 2 (yellow green), inter-arm 1 (orange), inter-arm 2 (blue), and outer disk (gray).
 Gray scale is the integrated-intensity map of $^{12}$CO($J$\,=\,2--1) obtained by L09.
  }
 \label{fig:n0628_region}
\end{figure}
\begin{figure}[tbh]
 \begin{center}
  \includegraphics[width=8cm]{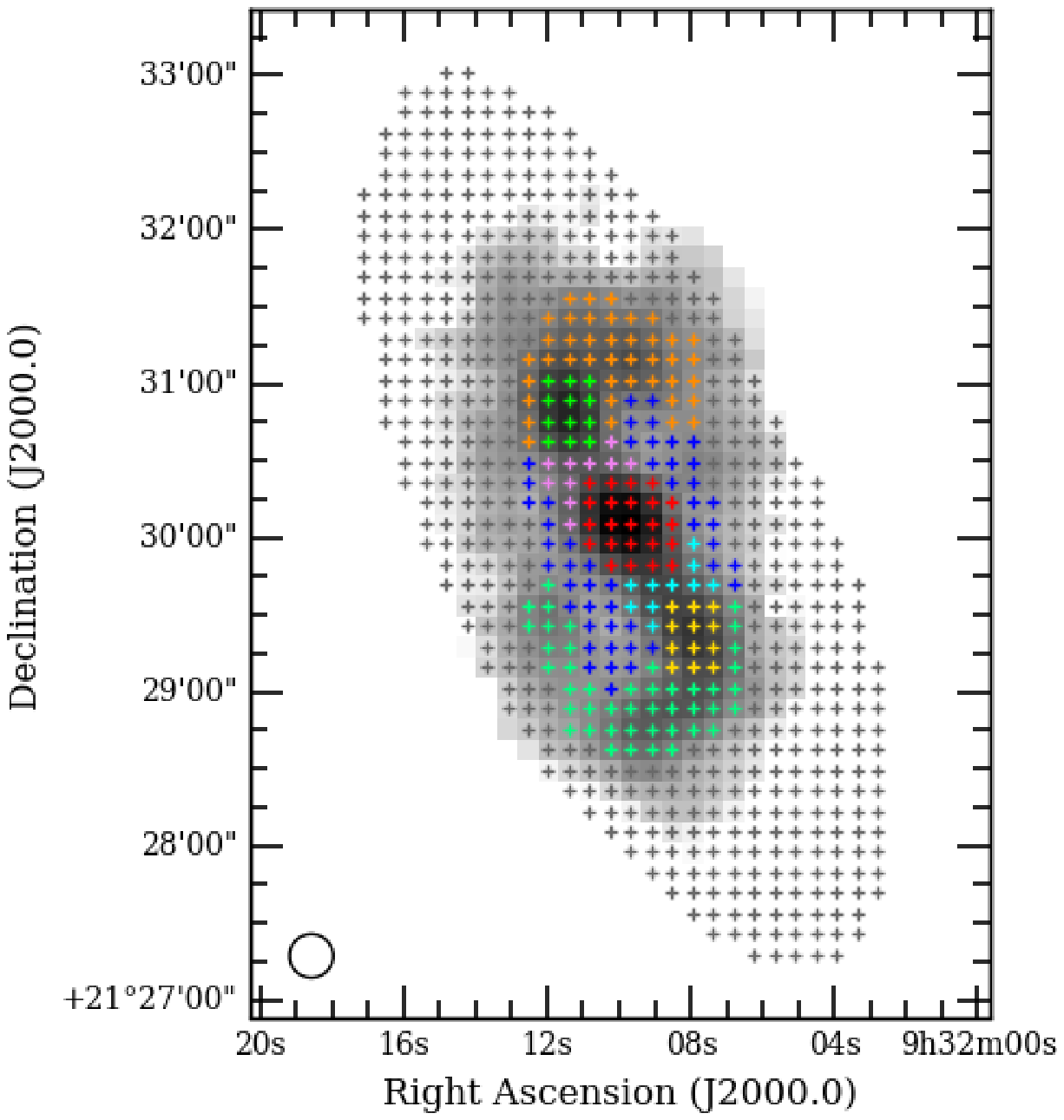}
 \end{center}
 \caption{The same as figure \ref{fig:n0628_region} but for NGC\,2903: center (red), northern bar (purple), southern bar (cyan), northern bar end (yellow green), southern bar end (yellow), northern arm (orange), southern arm (mint), inter-arm (blue), and outer disk (gray).
 The reference of $^{12}$CO($J$\,=\,2--1) data is the same as figure \ref{fig:n0628_region}.
  }
 \label{fig:n2903_region}
\end{figure}
\begin{figure}[tbh]
 \begin{center}
  \includegraphics[width=8cm]{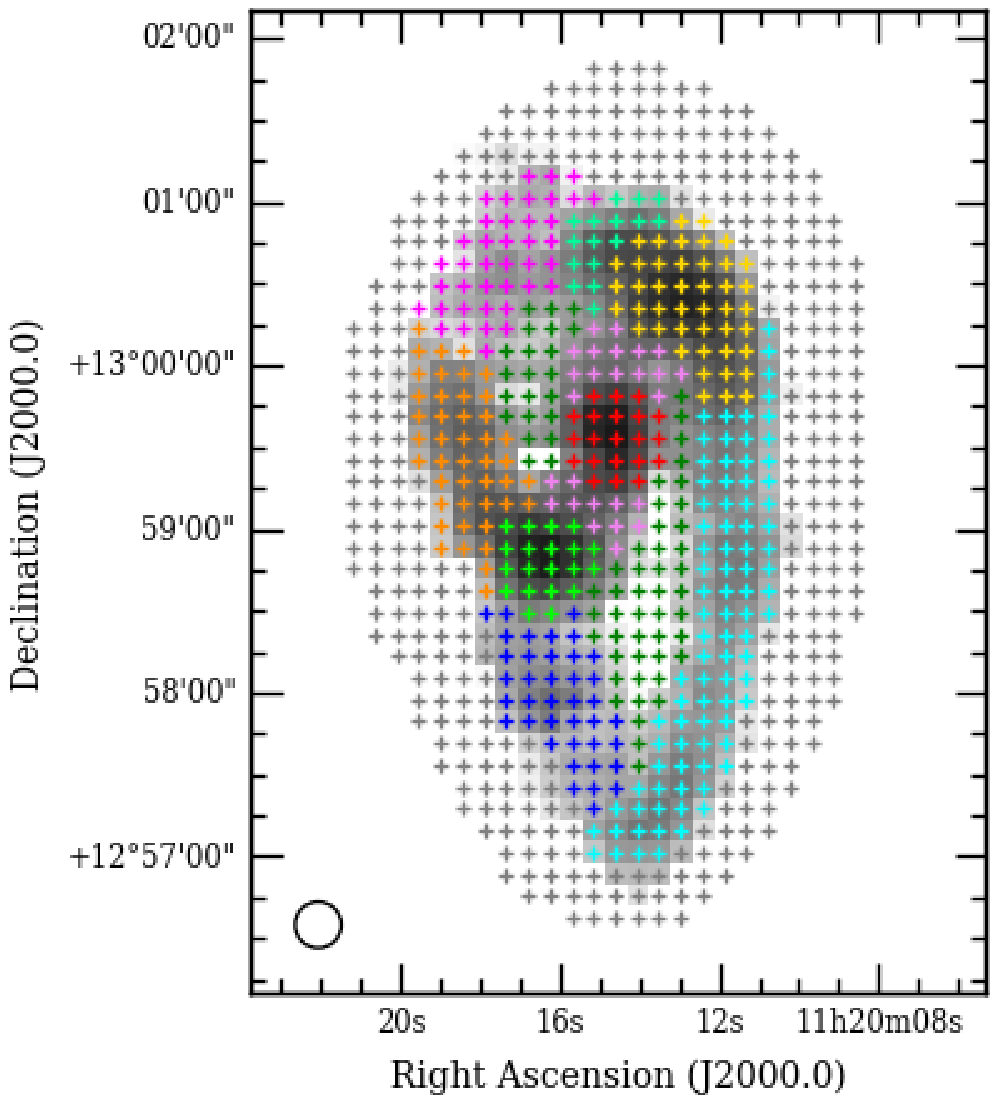}
 \end{center}
 \caption{The same as figure \ref{fig:n0628_region} but for NGC\,3627: center (red), bar (purple), northern bar end (yellow), southern bar end (yellow green), western arm (cyan), eastern arm (orange), southern stream (blue), northern offset stream (mint), arm-bar end interacting region (magenta), inter-arm (green), and outer disk (gray).
 The reference of $^{12}$CO($J$\,=\,2--1) data is the same as figure \ref{fig:n0628_region}.
  }
 \label{fig:n3627_region}
\end{figure}
\begin{figure}[tbh]
 \begin{center}
  \includegraphics[width=8cm]{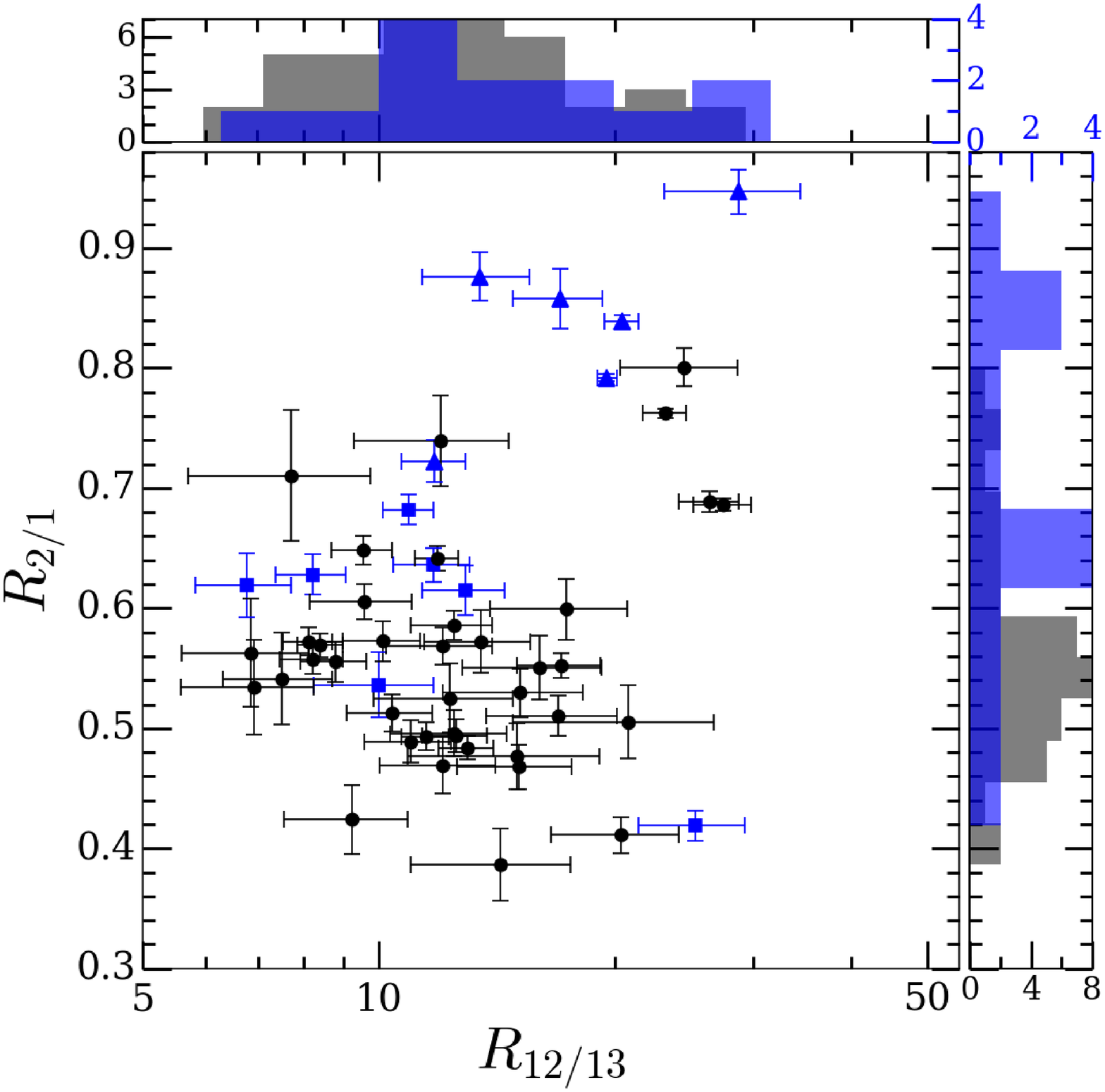}
 \end{center}
 \caption{Correlation of $R_{2/1}$ with $R_{12/13}$ measured from stacked spectra.
 Black circle markers represent results in disks, blue triangles indicate galactic centers with active star formation (mean $\Sigma_{\rm SFR}$ in inner $r_{25}/8$ or the region defined as ``center'' is $> 0.1$ $M_{\odot}$ yr$^{-1}$ kpc$^{-2}$; NGC\,2146, NGC\,2903, NGC\,3034, NGC\,3351, NGC\,4536, and NGC\,5713), and blue squares indicate galactic centers with quiescent star formation (mean $\Sigma_{\rm SFR}$ in the region is lower than the value).
 Black histograms are results of disk regions (black plots) and blue histograms are results of galactic centers (blue plots).
 Colors of tick labels correspond to those of histograms.
  }
 \label{fig:R21_vs_R1213}
\end{figure}

In this section, we investigate how properties of molecular gas themselves (such as density and temperature) change $R_{2/1}$.
We obtained $^{13}$CO($J$\,=\,1--0) maps of the COMING galaxies in our samples.
Therefore, the number density of molecular gas, $n({\rm H_2})$, and the kinetic temperature of molecular gas, $T_{\rm kin}$, can be derived with a non-local thermodynamic equilibrium (non-LTE) analysis with integrated-intensity ratios of the three lines (\cite{Scoville74}; \cite{Goldreich74}).

Since $^{13}$CO($J$\,=\,1--0) emission is weak compared with $^{12}$CO lines, we measured the integrated intensity with the velocity-alignment stacking analysis (\cite{Schruba11}; \cite{Morokuma15}) within concentric annuli or galactic structures such as arms and the bar.
We determined galactic structures for NGC\,628, NGC\,2903, and NGC\,3627 because the structures of these three galaxies can be clearly seen in the integrated-intensity maps of the $^{12}$CO lines.
The determined structures are shown in figures \ref{fig:n0628_region}--\ref{fig:n3627_region}.
For other galaxies, the regions for spectra stacking are determined as concentric annuli whose width is $r_{25}/8$.
To align spectra along the velocity axis, we used the first-moment maps of H\emissiontype{I} obtained by The H\emissiontype{I} Nearby Galaxy Survey (THINGS; \cite{Walter08}) if a galaxy was observed in the survey.
The H\emissiontype{I} first-moment maps enable us to stack spectra even in outer disks. 
When H\emissiontype{I} data is not available, the first-moment maps of $^{12}$CO($J$\,=\,2--1) are used.

Integrated intensities, integrated-intensity ratios, and full width at half maximum (FWHM) of stacked spectra are summarized in Appendix 2.
The 3\,$\sigma$ lower limit of $^{12}$CO($J$\,=\,1--0)/$^{13}$CO($J$\,=\,1--0) integrated-intensity ratio (hereafter, denoted as $R_{12/13}$) is adopted when $S/N$ of $^{13}$CO($J$\,=\,1--0) integrated-intensity is lower than 3\,$\sigma$.
Prior to the non-LTE analysis, we discuss the relation of $R_{2/1}$ and $R_{12/13}$ as a bare observed quantity.
Figure \ref{fig:R21_vs_R1213} shows the correlation plot of $R_{2/1}$ against $R_{12/13}$.

Regions where $S/N$ of $^{13}$CO($J$\,=\,1--0) is poor (derived as only the upper limit of integrated intensity) even with stacking are mainly inter-arms and outer disks.
Since the area where $^{13}$CO($J$\,=\,1--0) is emitted is much smaller than that of $^{12}$CO lines in such regions, a different beam-filling factor may be effective, i.e., the line ratio is no longer a probe of molecular gas properties.

In figure \ref{fig:R21_vs_R1213}, it appears that there are two components (or groups): one is that both $R_{2/1}$ and $R_{12/13}$ are high and the other is not so high $R_{2/1}$ ($\lesssim 0.7$) with various $R_{12/13}$ ($\sim 7$--20).
The former largely includes central regions of galaxies with active star formation (mean $\Sigma_{\rm SFR}$ within inner $r_{25}/8$ or within the region defined as ``center'' is higher than 0.1 $M_{\odot}$ yr$^{-1}$ kpc$^{-2}$; NGC\,2146, NGC\,2903, NGC\,3034, NGC\,3351, NGC\,4536, and NGC\,5713).
The latter corresponds to galactic disks and centers with quiescent star formation (mean $\Sigma_{\rm SFR}$ in the region is lower than the value above).
The same tendency of decreasing $R_{2/1}$ with increasing $R_{12/13}$ can be seen in results of another survey \citep{Cormier18}.
It is likely that systematic differences of molecular gas properties between the central region and the disk influencing $R_{2/1}$ and $R_{12/13}$ are seen in the figure.
Among galactic structures in disks (arms, inter-arms, and bars etc.), $R_{2/1}$ tends to be relatively low in bars and inter-arms, and high in arms and bar ends, although the contrast of $R_{2/1}$ is low (see results of stacked spectra for NGC\,628, NGC\,2903, and NGC\,3627 in Appendix 2).
In addition, the tendency of $R_{2/1}$ and $R_{12/13}$ among these structures roughly follow the disk-phase feature (i.e., $R_{2/1}$ decreases as $R_{12/13}$ increases).
Previous studies about $R_{12/13}$ suggested that high $R_{12/13}$ indicates low $n({\rm H_2})$ (\cite{Meier04}), high $T_{\rm kin}$ (\cite{Paglione01}), or low abundance of $^{13}$CO due to selective photodissociation (e.g., \cite{Davis14}).
Since $R_{2/1}$ is also high in galactic centers with active star formation, both high $n({\rm H_2})$ and high $T_{\rm kin}$ conditions are possible.

\subsubsection{Non-LTE analysis using $R_{2/1}$}
\begin{figure*}[t!]
 \begin{center}
  \includegraphics[width=16cm]{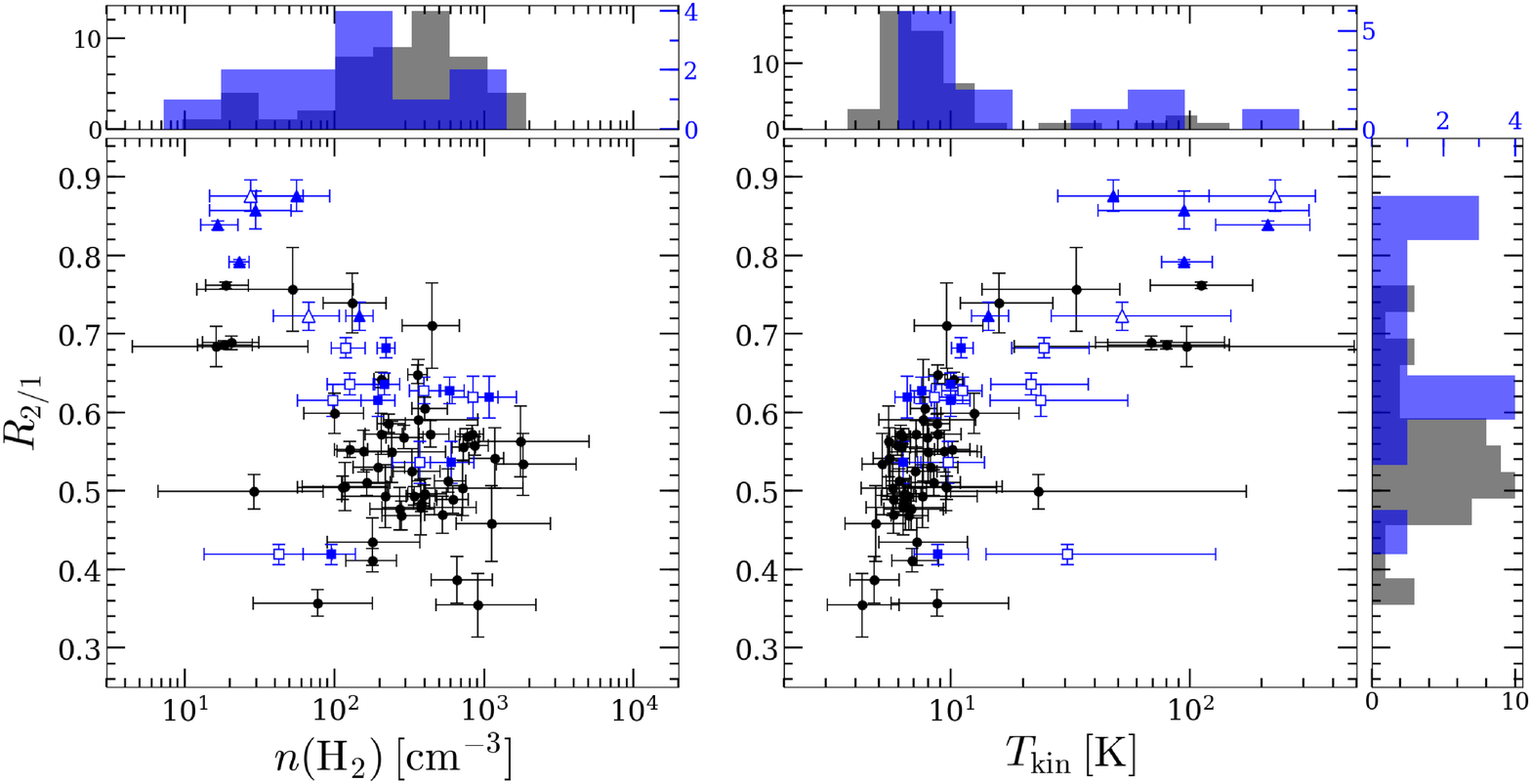}
 \end{center}
 \caption{Correlation of $R_{2/1}$ with $n(\rm{H_2})$ (left) and $T_{\rm{kin}}$ (right) and their histograms.
 Black circle markers represent results in disks and those of blue filled triangle and square are galactic centers with active and quiescent star formation as with figure \ref{fig:R21_vs_R1213}.
 Blue open markers correspond to central regions in which different [$^{12}$CO]/[$^{13}$CO] and $X_{\rm CO}$ are adopted.
 As with the plots, black histograms are results of disk regions and blue histograms are results of galactic centers.
 Colors of tick labels correspond to those of histograms.
 Blue histograms exclude the results for which different [$^{12}$CO]/[$^{13}$CO] and $X_{\rm CO}$ are adopted (open blue markers).
 }
 \label{fig:R21_vs_PP}
\end{figure*}

\begin{figure}[t!]
 \begin{center}
  \includegraphics[width=8cm]{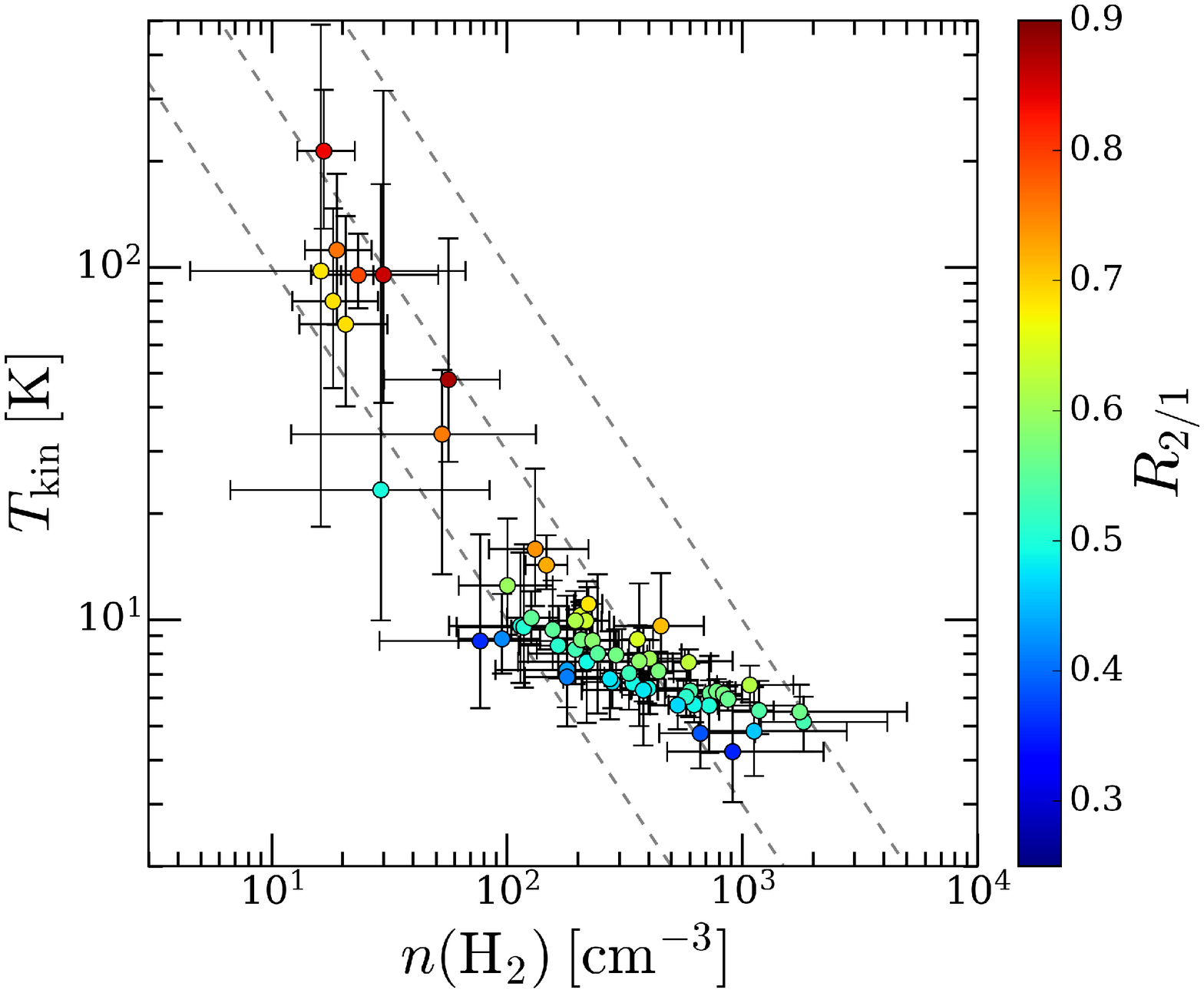}
 \end{center}
 \caption{$R_{2/1}$ distribution on the $T_{\rm kin}$--$n({\rm H_2})$ plot.
 Dashed lines indicate that pressure divided by $k_{\rm B}$ is $10^3$ K cm$^{-3}$, $3 \times 10^3$ K cm$^{-3}$, and $10^4$ K cm$^{-3}$ from bottom to top.
 }
 \label{fig:R21_vs_PP_3D}
\end{figure}

To compare $R_{2/1}$ with intrinsic molecular gas properties, $n({\rm H_2})$ and $T_{\rm kin}$ are derived with the non-LTE analysis.
In this study, we made use of the 1D non-LTE radiative transfer code RADEX \citep{vanderTak07}.
Settings for RADEX are based on those in \citet{Yajima19}.

Input parameters for calculations are $R_{2/1}$, $R_{12/13}$, column density of $^{12}$CO of all energy levels ($\mathcal{N_{\rm ^{12}CO}}$), that of $^{13}$CO ($\mathcal{N_{\rm ^{13}CO}}$), and FWHM of a GMC's spectrum ($dv$).
To derive $\mathcal{N_{\rm ^{12}CO}}$ and $\mathcal{N_{\rm ^{13}CO}}$ from column density of H$_2$ with observed $I_{^{12}{\rm CO(1-0)}}$, abundance ratios of [$^{12}$CO]/[H$_2$], [$^{12}$CO]/[$^{13}$CO], and $X_{\rm CO}$ were basically adopted as $8.5 \times 10^{-5}$ (\cite{Pineda08}), 70 (the solar neighborhood value; \cite{Milam05}), and $2.0 \times 10^{20}$ cm$^{-2}$ (K km s$^{-1}$)$^{-1}$ (\cite{Bolatto13}), respectively.
For the central region of galaxies, [$^{12}$CO]/[$^{13}$CO] and $X_{\rm CO}$ were also adopted as 40 and $1.0 \times 10^{20}$ cm$^{-2}$ (K km s$^{-1}$)$^{-1}$, respectively, to reflect the environment of galactic centers (e.g., \cite{Meier01}; \cite{Oka01}).
The escape probability of photon was calculated based on the Sobolev approximation \citep{Sobolev60}.

The FWHM of GMCs $dv$ was estimated by assuming that $dv$ in a region is proportional to the FWHM of the stacked spectrum in the region as described below.
This is because molecular gas is a continuous medium and GMCs do not have a rigid boundary.
Thus, internal kinematics of GMCs is likely to be influenced by dynamics of surrounding gas (i.e., large-scale dynamics), especially in bar ends and bars where large velocity dispersion can be seen.
In addition, since our velocity resolution is 20 km s$^{-1}$ and our spatial resolution is a kpc order, which are much wider than the typical velocity width of GMCs and much larger than the typical size of them, FWHM of a stacked spectrum, $\Delta V$, reflects velocity dispersion among molecular clouds within the beam.
This is a good indicator of large-scale dynamics of molecular gas.

At first, $\Delta V$ of a stacked spectrum within the entire disk in each galaxy was measured (hereafter, this $\Delta V$ is denoted as $\Delta V_{\rm disk}$) and the $dv$ of 5.0 km s$^{-1}$ was adopted corresponding to $\Delta V_{\rm disk}$.
This velocity of 5.0 km s$^{-1}$ as the standard of $dv$ in the entire disk was determined based on the typical FWHM of GMCs in the disk of the Milky Way (e.g., \cite{Heyer09}).
Next, the $dv$ of 20 km s$^{-1}$ was adopted corresponding to $\Delta V$ of the galaxy center (hereafter, denoted as $\Delta V_{\rm center}$).
This 20 km s$^{-1}$ is based on FWHM of GMCs found in the central region of galaxies and the Milky Way (e.g., \cite{Oka01}; \cite{Leroy15}).
Then, the $dv$ for each region (i.e., each annulus or galactic structure) was calculated by linearly interpolating or extrapolating with its $\Delta V$, $\Delta V_{\rm disk}$, and $\Delta V_{\rm center}$.
That is, the $dv$ for region $i$ ($dv_i$) is determined by the following equation:
\begin{equation}
\left( \frac{dv_i}{\rm km\> s^{-1}} \right) =
a \left( \frac{\Delta V_i}{\rm km\> s^{-1}} \right) + b,
\end{equation}
where
\begin{eqnarray}
a & \equiv & \left( 20\>{\rm km\> s^{-1}} - 5 \>{\rm km\> s^{-1}} \right) \hspace{-1.5mm} \bigg/ \hspace{-1mm}
\left[ \hspace{-0.5mm} \left( \frac{\Delta V_{\rm center}}{\rm km\> s^{-1}} \right) \hspace{-0.5mm} - \hspace{-0.5mm}
\left( \frac{\Delta V_{\rm disk}}{\rm km\> s^{-1}} \right) \hspace{-0.5mm} \right], \\
b & \equiv & 20\> {\rm km\> s^{-1}} -
a \left( \frac{\Delta V_{\rm center}}{\rm km\> s^{-1}} \right),
\end{eqnarray}
and $\Delta V_i$ is the FWHM of the stacked spectrum in the region.
The reason why the $dv$ for the innermost region is fixed to be 20 km s$^{-1}$ is $\Delta V_{\rm center}$ mainly reflects not velocity dispersion in a large scale but the velocity gradient due to the rigid rotation of the galactic disk.
High-resolution observations support increasing $dv$ as close to the galactic center in inner disks (e.g., \cite{Sun20}).
Note that $dv$ is not monotonically increases towards the center in outer disks.
Since $\Delta V$ in outer disks is almost constant, $dv$ is also constant (see the result of NGC\,5055 in Appendix 2).

Derived $n({\rm H_2})$, $T_{\rm kin}$, and ancillary results of excitation temperature for $^{12}$CO($J$\,=\,1--0), $^{12}$CO($J$\,=\,2--1), and $^{13}$CO($J$\,=\,1--0) [$T_{\rm ex}^{12(1-0)}$, $T_{\rm ex}^{12(2-1)}$, $T_{\rm ex}^{13(1-0)}$], and optical depth of these lines [$\tau_{12(1-0)}$, $\tau_{12(2-1)}$, $\tau_{13(1-0)}$] are listed in table \ref{table:radex_results}.
Errors of $n({\rm H_2})$ and $T_{\rm kin}$ are derived from errors of $R_{2/1}$ and $R_{12/13}$.
Namely, four pairs of [$n({\rm H_2})$, $T_{\rm kin}$] are obtained from $R_{2/1} \pm dR_{2/1}$ and $R_{12/13} \pm dR_{12/13}$ (where $dR_{2/1}$ and $dR_{12/13}$ are errors of $R_{2/1}$ and $R_{12/13}$, respectively).
Errors of $n({\rm H_2})$, $T_{\rm kin}$ toward the positive direction are adopted as $\max[X_i, i=1,2,3,4]- X_0$, where $X_0$ is $n({\rm H_2})$ and $T_{\rm kin}$ derived from $R_{2/1}$ and $R_{12/13}$, and $X_i$ is $n({\rm H_2})$ and $T_{\rm kin}$ derived from $R_{2/1} \pm dR_{2/1}$ and $R_{12/13} \pm dR_{12/13}$.
Errors of $n({\rm H_2})$, $T_{\rm kin}$ toward the negative direction are adopted as $X_0 - \min[X_i, i=1,2,3,4]$.
Errors of other ancillary quantities such as intrinsic intensities and excitation temperature are nearly the same factor of errors for $n({\rm H_2})$ and $T_{\rm kin}$.
There are no solutions of RADEX calculations in several regions.
These are caused by the low $S/N$ of $^{13}$CO($J$\,=\,1--0) even with stacking.
It is also possible that the one-zone model adopted for RADEX is not valid because the area seen in $^{13}$CO($J$\,=\,1--0) is much smaller than those seen in $^{12}$CO($J$\,=\,1--0) and $^{12}$CO($J$\,=\,2--1).
Based on these RADEX results, we examined the $R_{2/1}$ dependence on $n({\rm H_2})$ and $T_{\rm kin}$ as shown in figure \ref{fig:R21_vs_PP}.

There is a clear tendency that $R_{2/1}$ increases with $T_{\rm kin}$, while $R_{2/1}$ seems to decrease with increasing $n({\rm H_2})$.
The Spearman's rank correlation for $R_{2/1}$ and $T_{\rm kin}$ is 0.67 with a $p$-value of $\mathcal{O}(10^{-9})$ and that for $R_{2/1}$ and $n({\rm H_2})$ is $-0.36$; however, it is not statistically significant with the significance level of 5\% ($p$-value is 0.06).
In addition, the tendency of increased $R_{2/1}$ with $T_{\rm kin}$ is not changed even when different [$^{12}$CO]/[$^{13}$CO] and $X_{\rm CO}$ for galactic centers \{[$^{12}$CO]/[$^{13}$CO] = 40 and $X_{\rm CO} = 1.0 \times 10^{20}$ cm$^{-2}$ (K km s$^{-1}$)$^{-1}$\} are adopted.
The result indicates that the variations of $R_{2/1}$ in galaxies seen in kpc-scale resolutions reflect the temperature of molecular gas.
As discussed in section 4.1, the positive correlation between $R_{2/1}$ and $\Sigma_{\rm SFR}$ indicates that $R_{2/1}$ varies depending on $T_{\rm kin}$ or $n({\rm H_2})$ [high $\Sigma_{\rm SFR}$ enhances $T_{\rm kin}$ and $R_{2/1}$ or high $n({\rm H_2}$) induces high $\Sigma_{\rm SFR}$ and $R_{2/1}$].
Based on the results of the non-LTE analysis, the correlation implies the former case.
Radiation feedback from stars warms dust and molecular gas directly and/or dust warms molecular gas; therefore, $R_{2/1}$ becomes high, as suggested by \citet{Koda20}.

Moreover, it seems that there are two phases of molecular gas.
One is relatively dense (10$^2$--10$^3$ cm$^{-3}$) and cold (5--10 K) with $R_{2/1} \lesssim 0.7$.
Its correlation of $R_{2/1}$ with $T_{\rm kin}$ is tight and steep but no clear correlation with $n({\rm H_2})$.
The other is the diffuse (20--300 cm$^{-3}$), warm (30--200 K) with high $R_{2/1}$, and it is loosely correlated with $T_{\rm kin}$ and $n({\rm H_2})$.
The former corresponds to results for disk regions and galactic centers with quiescent star formation (low $R_{2/1}$ and low $R_{12/13}$ indicated by black markers and blue squares in figure \ref{fig:R21_vs_R1213}).
The latter is the central regions of galaxies with active star formation (high $R_{2/1}$ and high $R_{12/13}$) indicated by blue triangles in the figure.
Differences of molecular gas properties in disks and galactic centers are tested with the Kolmogorov--Smirnov test (i.e., for black and blue histograms in figure \ref{fig:R21_vs_PP}).
It is confirmed that the hypothesis that $T_{\rm kin}$ and $R_{2/1}$ are the same between the disks and galactic centers is rejected at the significance level of 5\%; however, that for $n({\rm H_2})$ cannot be rejected at the same level ($p$-value is 0.15).
In short, differences of molecular gas phases in disks and centers are clearly seen in $R_{2/1}$ and physical properties planes, especially for $T_{\rm kin}$, at the kpc-scale resolutions.

The relation between $R_{2/1}$ and $n({\rm H_2})$ seems inconsistent with other studies for GMCs (\cite{Sakamoto94}; \cite{Nishimura15}; \cite{Penaloza18}), although the negative correlation of $R_{2/1}$ with $n({\rm H_2})$ is not statistically significant.
For instance, $R_{2/1}$ is high in not only warm environments but also dense regions such as ridges and filaments in molecular clouds in these studies.
The discrepancy is because spatial scale is quite different between ours (kpc scale) and theirs (order of pc scale), and dynamic range of characteristic $n({\rm H_2})$ is much wider for cloud-scale studies than galactic-scales.
If observations with such high spatial resolution for galaxies is conducted, high $R_{2/1}$ with high $n({\rm H_2})$ would appear.
Furthermore, observations for molecular clouds in the Milky Way do not contain high $R_{2/1}$ and high $R_{12/13}$ environments (common in galactic centers with active star formation) that show warm and diffuse conditions via non-LTE calculations.
Thus, the discrepancy possibly originated from observed environments (represented on the $R_{2/1}$--$R_{12/13}$ plane).
Related to it, this negative correlation would be produced by data distribution of the bimodality of molecular gas properties seen in figure \ref{fig:R21_vs_R1213}.

Moreover, studies of molecular gas in other galactic centers support our results of high $R_{2/1}$ with warm diffuse molecular gas seen in central regions with active star formation.
For example, \citet{Meier00} suggested that there are two phases of molecular gas: warm diffuse layers and cold dense clumps in the starburst nucleus of IC\,342.
In addition, low-$J$ $^{12}$CO lines, in particular $^{12}$CO($J$\,=\,2--1), are dominantly emitted in such warm diffuse molecular gas.
\citet{Israel03} modeled properties of molecular gas in the center of IC\,342 and Maffei\,2 (this galaxy also has the modest starburst nucleus) with transitions of $^{12}$CO from $J$\,=\,1 to $J$\,=\,4; those of $^{13}$CO from $J$\,=\,1 to $J$\,=\,3 and [C \emissiontype{I}].
Further, they argued the existence of the warm diffuse molecular gas of $T_{\rm kin} \sim 100$--150 K and $n({\rm H_2}) \sim 10^{2}$ cm$^{-3}$ associated with the photon-dominated region (PDR), and the cold dense clumps of $T_{\rm kin} \sim 10$--20 K and $n({\rm H_2}) \sim 10^3$ cm$^{-3}$.
It was also suggested that this warm diffuse molecular gas accounts for approximately 2/3 of the total molecular gas.
Our results of physical properties with high $R_{2/1}$ reflect such hot diffuse molecular gas related to PDR.

Figure \ref{fig:R21_vs_PP_3D} shows $R_{2/1}$ distribution on the $T_{\rm kin}$--$n({\rm H_2})$ plot.
Most of molecular gas is nearly under the pressure equilibrium of a few $10^3$ K cm$^{-3}$.
This figure indicates that both $T_{\rm kin}$ and pressure of molecular gas are important for variations of $R_{2/1}$ in galaxies.
That is, even when pressure is so high ($\gtrsim 3 \times 10^3$ K cm$^{-3}$), $R_{2/1}$ is not always high ($\sim 0.5$--0.6) as long as $T_{\rm kin}$ is low ($\lesssim 10$ K).
Similarly, $R_{2/1}$ is not always very high ($< 0.8$) even when $T_{\rm kin}$ is high ($\gtrsim 30$ K) as long as pressure is not relatively high ($\sim 10^3$ K cm$^{-3}$).

Our results of $R_{2/1}$ and molecular gas properties suggest an issue of molecular gas mass derived from $^{12}$CO($J$\,=\,2--1) according to fluctuated $R_{2/1}$ and $X_{\rm CO}$ as follows.
Some studies reported that $X_{\rm CO}$ is lower in galactic centers, (ultra-) luminous infrared galaxies, and mergers than that in disks (i.e., hosting active star formation; e.g., \cite{Oka01}; {\cite{Narayanan11}; \cite{Papadopoulos12}}; \cite{Sandstrom13}\footnote{\citet{Sandstrom13} assumed the constant $R_{2/1}$ to derive molecular gas mass from $^{12}$CO($J$\,=\,2--1), however, the tendency of lower CO-to-H$_2$ conversion factors in galactic centers can also be seen in $^{12}$CO($J$\,=\,1--0)-based studies.}; Yasuda et al. 2020 in prep).
In these systems, $R_{2/1}$ should be higher than in normal disks considering our results (e.g., figure \ref{fig:ratio_radial_dist}, \ref{fig:R21_vs_SFR}, \ref{fig:R21_vs_IRcolor}, \ref{fig:R21_vs_PP}, and \ref{fig:R21_vs_PP_3D}).
In addition, the recent synthetic observation of $^{12}$CO($J$\,=\,1--0) and $^{12}$CO($J$\,=\,2--1) in a 3D magneto-hydrodynamics simulation \citep{Gong20} shows that $X_{\rm CO}$ is low in the environments where $R_{2/1}$ is high.
Hence, the molecular gas mass derived from $^{12}$CO($J$\,=\,2--1) is more overestimated in active star-formation environments when the constant $R_{2/1}$ and $X_{\rm CO}$ are adopted because $M_{\rm mol} \propto I_{\rm ^{12}CO(2-1)} \times X_{\rm CO}/ R_{2/1}$.
When $R_{2/1}$ increases and $X_{\rm CO}$ decreases, the fluctuation of $X_{\rm CO}/R_{2/1}$ is amplified.
Based on the reported variations of $X_{\rm CO}/ R_{2/1}$ so far, active star-formation environments lead to more overestimated molecular gas if $^{12}$CO($J$\,=\,2--1), constant $X_{\rm CO}$, and fixed $R_{2/1}$ are used compared with the $^{12}$CO($J$\,=\,1--0) case.

\section{Conclusions}
We present the variations of $^{12}$CO($J$\,=\,2--1)/$^{12}$CO($J$\,=\,1--0) line ratio ($R_{2/1}$) within and among galaxies, the effects of the assumption that $R_{2/1}$ is a constant on the derivation of molecular gas mass, and the properties of molecular gas reflecting variations of $R_{2/1}$ in {24} nearby galaxies on kpc scales.
The main conclusions of this paper are as follows:

\begin{enumerate}
\renewcommand{\labelenumi}{(\arabic{enumi})}
\item The median and mean $R_{2/1}$ weighted by $^{12}$CO($J$\,=\,1--0) integrated-intensity for spatially resolved data of galaxies in our samples are 0.61 and 0.66, respectively, with the standard deviation of 0.19.
$R_{2/1}$ varies from 0.4 to 0.9 among and within galaxies.
While $R_{2/1}$ in the galactic centers (inner $\sim 1$ kpc) is typically higher ($\sim 0.8$), the median of $R_{2/1}$ in disks (exterior to $\sim 2$ kpc) is nearly constant at 0.60.

\item The total molecular gas mass within a galaxy derived from $^{12}$CO($J$\,=\,2--1) is underestimated or overestimated by $\sim 20$\% for most galaxies, and at most by 35\%, when $R_{2/1}$ is assumed to be the constant of 0.7.
In addition, the scatter of molecular gas surface density ($\Sigma_{\rm mol}$) by $^{12}$CO($J$\,=\,2--1) within a galaxy increases in all galaxies.
Its change rate exceeds $\sim 30$\% in some galaxies, and it is 120\% in the highest case when the constant $R_{2/1}$ is adopted.
This increased scatter of $\Sigma_{\rm mol}$ is serious for mapping (spatially resolved) observations.

\item The indices of the Kennicutt--Schmidt relation using molecular gas surface density derived from $^{12}$CO($J$\,=\,2--1) and $R_{2/1} = 0.7$ become lower by 10--20\%, up to 39\% compared with that using $^{12}$CO($J$\,=\,1--0) for 17 galaxies out of 24.
This is because $R_{2/1}$ tends to be higher than 0.7 when $\Sigma_{\rm SFR}$ and $\Sigma_{\rm mol}$ are high while $R_{2/1}$ is often lower than 0.7 when $\Sigma_{\rm SFR}$ and $\Sigma_{\rm mol}$ are low.
Underestimated indices are prominent when the scatter of $R_{2/1}$ within a galaxy is large.

\item $R_{2/1}$ has positive correlations with $\Sigma_{\rm SFR}$ and IR color and the negative correlation with depletion time of molecular gas.
These suggest that $R_{2/1}$ becomes high by warmed molecular gas from stars (stars influence on molecular gas properties and $R_{2/1}$).
Otherwise, $R_{2/1}$ is high because molecular gas is dense; as a result, star formation is active (molecular gas properties influence on star formation activity and $R_{2/1}$).
There is no clear tendency between $R_{2/1}$ and $\Sigma_{\rm mol}$.

\item Comparing $R_{2/1}$ with $^{12}$CO($J$\,=\,1--0)/$^{13}$CO($J$\,=\,1--0) line ratio ($R_{12/13}$) measured within each galactic structure or concentric annulus, there seems to be two components; one is both $R_{2/1}$ and $R_{12/13}$ are high, and the other is relatively low $R_{2/1}$ with various $R_{12/13}$.
The former corresponds to molecular gas in galactic centers with active star formation (mean $\Sigma_{\rm SFR}$ within inner $r_{25}/8$ or the region defined as ``center'' is higher than 0.1 $M_{\odot}$ yr$^{-1}$ kpc$^{-2}$) and the latter corresponds to that in disk regions and centers of galaxies with quiescent star formation (mean $\Sigma_{\rm SFR}$ in the region is lower than the above value).

\item According to the non-LTE analysis, there is a clear tendency that $R_{2/1}$ increases with $T_{\rm kin}$; however, $R_{2/1}$ and $n({\rm H_2})$ show a rather negative correlation but it is not statistically significant.
This suggests that variations of $R_{2/1}$ on kpc scales imply the temperature of molecular gas.
The dependence of $R_{2/1}$ on $n({\rm H_2})$ would appear when the spatial resolution is higher.
Stellar radiation feedback influences molecular gas properties and $R_{2/1}$.
The bimodality caused by differences of properties between galactic centers and disks seen on the $R_{2/1}$--$R_{12/13}$ plane can also be seen on the $R_{2/1}$--$T_{\rm kin}$ and $R_{2/1}$--$n({\rm H_2})$ plots.
Molecular gas in the centers is warm and diffuse, and its $R_{2/1}$ has relatively loose correlations between $T_{\rm kin}$ and $n({\rm H_2})$.
The disk phase molecular gas is cold, relatively dense, and its $R_{2/1}$ is tightly correlated with $T_{\rm kin}$; however, it has no relation with $n({\rm H_2})$.

\item Not only $T_{\rm kin}$ but also pressure of molecular gas is important to understand $R_{2/1}$ variations in galaxies.
Namely, even when pressure is so high ($\gtrsim 3 \times 10^3$ K cm$^{-3}$), $R_{2/1}$ is not always high ($\sim 0.5$--0.6) as long as $T_{\rm kin}$ is low ($\lesssim 10$ K).
Likewise, even when $T_{\rm kin}$ is high ($\gtrsim 30$ K), $R_{2/1}$ is not always very high ($< 0.8$) as long as pressure is not relatively high ($\sim 10^3$ K cm$^{-3}$).

\item Considering variations of the CO-to-H$_2$ conversion factor $X_{\rm CO}$, molecular gas mass in active star-formation environments such as galactic centers and (U)LIRGs is more overestimated when it is derived from $^{12}$CO($J$\,=\,2--1) and the constant $R_{2/1}$ of 0.7.
This is because molecular gas mass is $\propto I_{^{12}{\rm CO}(2-1)} \times X_{\rm CO} / R_{2/1}$, $X_{\rm CO}$ decreases, and $R_{2/1}$ increases [where $I_{^{12}{\rm CO}(2-1)}$ is integrated intensity of $^{12}$CO($J$\,=\,2--1)].
As a result, fluctuations of $X_{\rm CO} / R_{2/1}$ are amplified.
\end{enumerate}

\section*{Acknowledgments}
We are grateful to the anonymous referee for the constructive, meaningful comments and suggestions that significantly improve the manuscript.
We also thank the staff of Nobeyama Radio Observatory for their help in our observations with the 45-m radio telescope and the continuous efforts to improve the performance of the instruments.
This work has been partially supported by JSPS Grants-in-Aid for Scientific Research (17H01110, 18K13593, 19H05076).
This work has also been supported in part by the Sumitomo Foundation Fiscal 2018 Grant for Basic Science Research Projects (180923), and the Collaboration Funding of the Institute of Statistical Mathematics ``New Development of the Studies on Galaxy Evolution with a Method of Data Science''.
The Nobeyama 45-m radio telescope is operated by Nobeyama Radio Observatory, a branch of National Astronomical Observatory of Japan.
This work made use of HERACLES, `The HERA CO-Line Extragalactic Survey' (Leroy et al. 2009).
This research has made use of the NASA/IPAC Extragalactic Database, which is operated by the Jet Propulsion Laboratory, California Institute of Technology, under contract with the National Aeronautics and Space Administration.
This research has made use of the NASA/ IPAC Infrared Science Archive, which is operated by the Jet Propulsion Laboratory, California Institute of Technology, under contract with the National Aeronautics and Space Administration.
This research also made use of APLpy, an open-source plotting package for Python \citep{Robitaille12}.
We would like to thank Editage (www.editage.com) for English language editing.

\renewcommand\arraystretch{1.15}
\begin{table*}[p!]
  \begin{center}
  \tbl{Derived physical properties of molecular gas with the non-LTE analysis.}{
  \begin{tabular}{llcccccccc} \hline
galaxy & region$^{*}$ & $n({\rm H_2})$ & $T_{\rm kin}$ & $T_{\rm ex}^{\rm 12(1-0)}$ & $T_{\rm ex}^{\rm 12(2-1)}$ & $T_{\rm ex}^{\rm 13(1-0)}$ & $\tau_{12(1-0)}$ & $\tau_{12(2-1)}$ & $\tau_{13(1-0)}$ \\
  &   & $[{\rm cm^{-3}}]$ & $[{\rm K}]$ & $[{\rm K}]$ & $[{\rm K}]$ & $[{\rm K}]$ &   &   &   \\
  &   & (1) & (2) & (3) & (4) & (5) & (6) & (7) & (8) \\
  \hline
NGC\,628 & center & $\left(1.1^{+0.6}_{-0.3} \right) \times 10^{3}$ & $6.5^{+0.9}_{-0.7}$ & $6.3$ & $6.1$ & $4.4$ & $18$ & $22$ & $0.41$ \\
  & inner arm1 & $\left(1.9^{+1.0}_{-0.6} \right) \times 10^{2}$ & $8.2^{+2.5}_{-1.7}$ & $6.9$ & $6.1$ & $3.4$ & $19$ & $26$ & $0.64$ \\
  & inner arm2 & $\left(2.1^{+0.8}_{-0.5} \right) \times 10^{2}$ & $8.8^{+2.2}_{-1.6}$ & $7.4$ & $6.6$ & $3.5$ & $20$ & $29$ & $0.7$ \\
  & outer arm2 & $\left(6.6^{+4.7}_{-2.2} \right) \times 10^{2}$ & $4.8^{+1.3}_{-1.0}$ & $4.5$ & $4.0$ & $3.3$ & $13$ & $10$ & $0.26$ \\
  & inter-arm1 & $\left(1.8^{+1.9}_{-0.9} \right) \times 10^{2}$ & $7.2^{+4.5}_{-2.2}$ & $5.9$ & $5.0$ & $3.2$ & $17$ & $20$ & $0.49$ \\
  & center$^{\dagger}$ & $\left(8.5^{+3.9}_{-2.3} \right) \times 10^{2}$ & $8.5^{+1.9}_{-1.5}$ & $7.8$ & $7.2$ & $4.9$ & $9.8$ & $15$ & $0.45$ \\
NGC\,2146 & center & $\left(1.7^{+0.6}_{-0.4} \right) \times 10^{1}$ & $210^{+110}_{-80}$ & $14$ & $14$ & $3.5$ & $18$ & $40$ & $1.6$ \\
  & ring1 & $\left(2.1^{+1.1}_{-0.7} \right) \times 10^{1}$ & $69^{+71}_{-29}$ & $10$ & $9.4$ & $3.2$ & $20$ & $38$ & $1.2$ \\
  & ring2 & $\left(2.9^{+5.5}_{-2.2} \right) \times 10^{1}$ & $23^{+149}_{-13}$ & $7.0$ & $6.0$ & $3.0$ & $19$ & $27$ & $0.74$ \\
  & center$^{\dagger}$ & --- & --- & --- & --- & --- & --- & --- & --- \\
NGC\,2841 & ring2 & $\left(1.2^{+0.2}_{-0.4} \right) \times 10^{3}$ & $5.5^{+1.2}_{-0.8}$ & $5.4$ & $5.1$ & $4.0$ & $17$ & $17$ & $0.36$ \\
  & ring3 & $\left(1.8^{+2.3}_{-0.6} \right) \times 10^{3}$ & $5.1^{+0.9}_{-0.9}$ & $5.0$ & $4.8$ & $4.0$ & $15$ & $14$ & $0.3$ \\
  & ring4 & $\left(1.1^{+1.7}_{-0.5} \right) \times 10^{3}$ & $4.8^{+1.6}_{-1.2}$ & $4.7$ & $4.3$ & $3.6$ & $14$ & $12$ & $0.27$ \\
NGC\,2903 & center & $\left(1.5^{+0.3}_{-0.3} \right) \times 10^{2}$ & $14^{+3.0}_{-2.0}$ & $11$ & $9.9$ & $3.9$ & $20$ & $38$ & $1.1$ \\
  & northern bar & $\left(1.0^{+0.6}_{-0.4} \right) \times 10^{2}$ & $13^{+7.0}_{-3.0}$ & $8.5$ & $7.5$ & $3.4$ & $20$ & $33$ & $0.9$ \\
  & southern bar & $\left(1.1^{+1.2}_{-0.6} \right) \times 10^{2}$ & $9.6^{+5.9}_{-3.0}$ & $7.0$ & $6.0$ & $3.2$ & $19$ & $27$ & $0.68$ \\
  & northern bar end & $\left(3.6^{+0.6}_{-0.5} \right) \times 10^{2}$ & $8.8^{+0.9}_{-0.7}$ & $7.9$ & $7.4$ & $4.0$ & $19$ & $30$ & $0.67$ \\
  & southern bar end & $\left(2.3^{+0.7}_{-0.4} \right) \times 10^{2}$ & $8.7^{+1.2}_{-1.1}$ & $7.5$ & $6.7$ & $3.6$ & $20$ & $29$ & $0.69$ \\
  & northern arm & $\left(4.0^{+1.7}_{-0.8} \right) \times 10^{2}$ & $7.8^{+1.0}_{-1.0}$ & $7.1$ & $6.6$ & $3.8$ & $19$ & $26$ & $0.59$ \\
  & southern arm & $\left(7.3^{+1.5}_{-1.1} \right) \times 10^{2}$ & $6.2^{+0.5}_{-0.6}$ & $5.9$ & $5.5$ & $3.9$ & $18$ & $20$ & $0.42$ \\
  & inter-arm & --- & --- & --- & --- & --- & --- & --- & --- \\
  & outer disk & $\left(1.2^{+1.1}_{-0.6} \right) \times 10^{2}$ & $9.5^{+6.8}_{-3.1}$ & $7.0$ & $6.0$ & $3.2$ & $19$ & $26$ & $0.63$ \\
  & center$^{\dagger}$ & $\left(6.8^{+4.1}_{-2.9} \right) \times 10^{1}$ & $52^{+97}_{-26}$ & $12$ & $11$ & $4.1$ & $10$ & $22$ & $1.1$ \\
NGC\,2976 & center & --- & --- & --- & --- & --- & --- & --- & --- \\
  & ring1 & $\left(5.3^{+8.0}_{-4.1} \right) \times 10^{1}$ & $34^{+17}_{-20}$ & $12$ & $11$ & $3.6$ & $18$ & $38$ & $1.2$ \\
  & ring2 & $\left(1.3^{+0.9}_{-0.5} \right) \times 10^{2}$ & $16^{+11}_{-5.0}$ & $11$ & $10$ & $3.9$ & $19$ & $39$ & $1.1$ \\
  & ring3 & $\left(4.5^{+2.3}_{-1.7} \right) \times 10^{2}$ & $9.6^{+4.0}_{-2.6}$ & $8.8$ & $8.4$ & $4.5$ & $19$ & $32$ & $0.68$ \\
NGC\,3034 & center & $\left(2.3^{+0.4}_{-0.4} \right) \times 10^{1}$ & $95^{+30}_{-19}$ & $13$ & $12$ & $3.5$ & $18$ & $40$ & $1.6$ \\
  & ring1 & $\left(1.9^{+0.8}_{-0.5} \right) \times 10^{1}$ & $110^{+70}_{-40}$ & $12$ & $11$ & $3.4$ & $19$ & $40$ & $1.4$ \\
  & ring2 & $\left(1.8^{+1.0}_{-0.6} \right) \times 10^{1}$ & $80^{+66}_{-35}$ & $10$ & $9.3$ & $3.2$ & $20$ & $38$ & $1.2$ \\
  & center$^{\dagger}$ & --- & --- & --- & --- & --- & --- & --- & --- \\
NGC\,3198 & center & $\left(6.0^{+2.6}_{-1.6} \right) \times 10^{2}$ & $6.3^{+1.2}_{-1.0}$ & $5.9$ & $5.5$ & $3.7$ & $17$ & $19$ & $0.42$ \\
  & ring1 & $\left(2.2^{+2.6}_{-1.1} \right) \times 10^{2}$ & $7.6^{+5.3}_{-2.5}$ & $6.4$ & $5.6$ & $3.3$ & $18$ & $22$ & $0.53$ \\
  & center$^{\dagger}$ & $\left(3.7^{+2.0}_{-1.3} \right) \times 10^{2}$ & $9.7^{+4.1}_{-2.5}$ & $7.7$ & $6.7$ & $4.1$ & $9.3$ & $14$ & $0.51$ \\
NGC\,3351 & center & $\left(5.6^{+3.7}_{-2.6} \right) \times 10^{1}$ & $48^{+73}_{-20}$ & $16$ & $16$ & $4.1$ & $18$ & $43$ & $1.8$ \\
  & ring1 & $\left(3.6^{+5.5}_{-1.5} \right) \times 10^{2}$ & $7.6^{+5.1}_{-2.6}$ & $7.0$ & $6.4$ & $3.7$ & $20$ & $27$ & $0.62$ \\
  & ring2 & $\left(7.2^{+11.0}_{-3.2} \right) \times 10^{2}$ & $5.7^{+2.2}_{-1.5}$ & $5.5$ & $5.0$ & $3.7$ & $16$ & $17$ & $0.36$ \\
  & center$^{\dagger}$ & $\left(2.8^{+3.5}_{-1.3} \right) \times 10^{1}$ & $230^{+110}_{-180}$ & $16$ & $16$ & $4.2$ & $12$ & $30$ & $1.6$ \\
NGC\,3521 & center & $\left(2.2^{+0.6}_{-0.4} \right) \times 10^{2}$ & $9.9^{+1.4}_{-1.2}$ & $8.3$ & $7.6$ & $3.7$ & $20$ & $32$ & $0.79$ \\
  & ring1 & $\left(2.1^{+0.2}_{-0.2} \right) \times 10^{2}$ & $10^{+1.0}_{-1.0}$ & $8.5$ & $7.8$ & $3.7$ & $20$ & $33$ & $0.81$ \\
  & ring2 & $\left(1.3^{+0.4}_{-0.3} \right) \times 10^{2}$ & $10^{+2.0}_{-2.0}$ & $7.6$ & $6.6$ & $3.3$ & $20$ & $30$ & $0.76$ \\
  & ring3 & $\left(1.6^{+0.7}_{-0.5} \right) \times 10^{2}$ & $9.4^{+3.5}_{-2.2}$ & $7.4$ & $6.5$ & $3.4$ & $19$ & $28$ & $0.71$ \\
  & center$^{\dagger}$ & $\left(1.3^{+0.6}_{-0.4} \right) \times 10^{2}$ & $22^{+16}_{-7.0}$ & $10$ & $8.9$ & $4.0$ & $9.8$ & $19$ & $0.83$ \\
  \hline
  \end{tabular}}
  \label{table:radex_results}
  \end{center}
\end{table*}

\renewcommand\arraystretch{1.15}
\setcounter{table}{3}
\begin{table*}[t!]
  \begin{center}
  \tbl{(Continued.)}{
  \begin{tabular}{llcccccccc} \hline
galaxy & region$^{*}$ & $n({\rm H_2})$ & $T_{\rm kin}$ & $T_{\rm ex}^{\rm 12(1-0)}$ & $T_{\rm ex}^{\rm 12(2-1)}$ & $T_{\rm ex}^{\rm 13(1-0)}$ & $\tau_{12(1-0)}$ & $\tau_{12(2-1)}$ & $\tau_{13(1-0)}$ \\
  &   & $[{\rm cm^{-3}}]$ & $[{\rm K}]$ & $[{\rm K}]$ & $[{\rm K}]$ & $[{\rm K}]$ &   &   &   \\
  &   & (1) & (2) & (3) & (4) & (5) & (6) & (7) & (8) \\
  \hline
NGC\,3627 & center & $\left(9.5^{+4.3}_{-3.3} \right) \times 10^{1}$ & $8.8^{+3.0}_{-1.8}$ & $6.0$ & $5.0$ & $3.1$ & $18$ & $21$ & $0.54$ \\
  & bar & $\left(1.8^{+0.8}_{-0.6} \right) \times 10^{2}$ & $6.9^{+2.0}_{-1.2}$ & $5.6$ & $4.8$ & $3.2$ & $17$ & $18$ & $0.46$ \\
  & northern bar end & $\left(3.4^{+0.4}_{-0.4} \right) \times 10^{2}$ & $6.6^{+0.6}_{-0.5}$ & $6.0$ & $5.3$ & $3.5$ & $18$ & $20$ & $0.55$ \\
  & southern bar end & $\left(3.9^{+0.8}_{-0.6} \right) \times 10^{2}$ & $6.4^{+0.6}_{-0.6}$ & $5.8$ & $5.2$ & $3.5$ & $17$ & $19$ & $0.45$ \\
  & western arm & $\left(2.8^{+1.0}_{-0.7} \right) \times 10^{2}$ & $6.7^{+1.4}_{-1.0}$ & $5.9$ & $5.1$ & $3.3$ & $18$ & $20$ & $0.48$ \\
  & eastern arm & $\left(6.2^{+1.6}_{-1.4} \right) \times 10^{2}$ & $5.7^{+0.8}_{-0.6}$ & $5.4$ & $5.0$ & $3.6$ & $16$ & $17$ & $0.37$ \\
  & offset stream & $\left(1.7^{+0.7}_{-0.5} \right) \times 10^{2}$ & $8.4^{+2.5}_{-1.6}$ & $6.8$ & $5.9$ & $3.3$ & $19$ & $26$ & $0.64$ \\
  & southern arm & $\left(2.7^{+1.9}_{-1.0} \right) \times 10^{2}$ & $6.8^{+2.5}_{-1.6}$ & $6.0$ & $5.2$ & $3.4$ & $18$ & $20$ & $0.49$ \\
  & arm-bar end inter. region & $\left(7.7^{+10.3}_{-4.8} \right) \times 10^{1}$ & $8.7^{+8.8}_{-3.1}$ & $5.4$ & $4.5$ & $3.0$ & $17$ & $17$ & $0.45$ \\
  & center$^{\dagger}$ & $\left(4.3^{+5.4}_{-2.9} \right) \times 10^{1}$ & $31^{+98}_{-17}$ & $6.1$ & $5.1$ & $3.2$ & $9.1$ & $11$ & $0.42$ \\
NGC\,3938 & center & $\left(5.9^{+1.4}_{-0.9} \right) \times 10^{2}$ & $7.6^{+0.7}_{-0.8}$ & $7.1$ & $6.7$ & $4.1$ & $18$ & $25$ & $0.51$ \\
  & ring1 & $\left(4.4^{+1.4}_{-0.7} \right) \times 10^{2}$ & $7.1^{+1.0}_{-0.8}$ & $6.6$ & $6.1$ & $3.8$ & $19$ & $24$ & $0.52$ \\
  & ring2 & $\left(2.9^{+1.0}_{-0.7} \right) \times 10^{2}$ & $8.0^{+1.4}_{-1.2}$ & $7.0$ & $6.4$ & $3.6$ & $19$ & $26$ & $0.6$ \\
  & ring3 & $\left(1.8^{+3.2}_{-0.8} \right) \times 10^{3}$ & $5.5^{+1.1}_{-0.1}$ & $5.4$ & $5.2$ & $4.2$ & $16$ & $16$ & $0.3$ \\
  & center$^{\dagger}$ & $\left(4.0^{+1.1}_{-0.8} \right) \times 10^{2}$ & $11^{+2.0}_{-2.0}$ & $9.0$ & $8.1$ & $4.5$ & $9.8$ & $17$ & $0.6$ \\
NGC\,4536 & center & $\left(3.0^{+2.1}_{-1.5} \right) \times 10^{1}$ & $95^{+222}_{-54}$ & $16$ & $15$ & $3.8$ & $18$ & $42$ & $1.8$ \\
  & center$^{\dagger}$ & --- & --- & --- & --- & --- & --- & --- & --- \\
NGC\,4579 & center & --- & --- & --- & --- & --- & --- & --- & --- \\
  & ring1 & $\left(2.4^{+3.1}_{-1.2} \right) \times 10^{2}$ & $8.0^{+5.4}_{-2.6}$ & $6.9$ & $6.2$ & $3.5$ & $19$ & $26$ & $0.62$ \\
  & ring2 & $\left(3.3^{+1.8}_{-0.8} \right) \times 10^{2}$ & $7.1^{+1.7}_{-1.5}$ & $6.3$ & $5.7$ & $3.5$ & $18$ & $22$ & $0.52$ \\
  & center$^{\dagger}$ & --- & --- & --- & --- & --- & --- & --- & --- \\
NGC\,5055 & center & $\left(2.2^{+0.3}_{-0.3} \right) \times 10^{2}$ & $11^{+1.0}_{-1.0}$ & $9.2$ & $8.6$ & $3.9$ & $20$ & $35$ & $0.87$ \\
  & ring1 & $\left(7.8^{+1.6}_{-1.4} \right) \times 10^{2}$ & $6.3^{+0.5}_{-0.4}$ & $6.0$ & $5.6$ & $4.0$ & $18$ & $20$ & $0.42$ \\
  & ring2 & $\left(3.8^{+0.5}_{-0.4} \right) \times 10^{2}$ & $6.3^{+0.5}_{-0.4}$ & $5.8$ & $5.1$ & $3.4$ & $17$ & $19$ & $0.45$ \\
  & ring3 & $\left(5.8^{+1.8}_{-1.1} \right) \times 10^{2}$ & $6.1^{+0.7}_{-0.7}$ & $5.7$ & $5.2$ & $3.6$ & $17$ & $18$ & $0.41$ \\
  & ring4 & $\left(4.0^{+1.6}_{-0.9} \right) \times 10^{2}$ & $6.4^{+1.1}_{-1.0}$ & $5.9$ & $5.2$ & $3.5$ & $17$ & $19$ & $0.45$ \\
  & ring5 & $\left(3.8^{+5.0}_{-1.7} \right) \times 10^{2}$ & $6.3^{+3.2}_{-1.9}$ & $5.7$ & $5.1$ & $3.4$ & $17$ & $19$ & $0.43$ \\
  & ring6 & $\left(9.1^{+13.0}_{-4.3} \right) \times 10^{2}$ & $4.2^{+1.8}_{-1.2}$ & $4.0$ & $3.7$ & $3.2$ & $11$ & $7.1$ & $0.19$ \\
  & center$^{\dagger}$ & $\left(1.2^{+0.4}_{-0.2} \right) \times 10^{2}$ & $24^{+13}_{-6.0}$ & $11$ & $9.8$ & $4.1$ & $10$ & $21$ & $0.95$ \\
NGC\,5713 & center & --- & --- & --- & --- & --- & --- & --- & --- \\
  & ring1 & --- & --- & --- & --- & --- & --- & --- & --- \\
  & ring2 & $\left(1.6^{+5.1}_{-1.2} \right) \times 10^{1}$ & $97^{+390}_{-79}$ & $9.9$ & $9.2$ & $3.2$ & $19$ & $37$ & $1.1$ \\
  & center$^{\dagger}$ & --- & --- & --- & --- & --- & --- & --- & --- \\
NGC\,7331 & center & $\left(2.0^{+0.6}_{-0.4} \right) \times 10^{2}$ & $9.9^{+2.1}_{-1.6}$ & $8.2$ & $7.4$ & $3.6$ & $19$ & $31$ & $0.77$ \\
  & ring1 & $\left(8.3^{+1.6}_{-1.4} \right) \times 10^{2}$ & $6.2^{+0.5}_{-0.3}$ & $6.0$ & $5.6$ & $4.0$ & $18$ & $21$ & $0.42$ \\
  & ring2 & $\left(8.7^{+2.1}_{-1.8} \right) \times 10^{2}$ & $6.0^{+0.5}_{-0.4}$ & $5.8$ & $5.4$ & $3.9$ & $18$ & $20$ & $0.41$ \\
  & ring3 & $\left(5.3^{+1.8}_{-1.3} \right) \times 10^{2}$ & $5.7^{+1.1}_{-0.8}$ & $5.4$ & $4.8$ & $3.5$ & $16$ & $16$ & $0.38$ \\
  & center$^{\dagger}$ & $\left(9.8^{+5.3}_{-4.1} \right) \times 10^{1}$ & $24^{+32}_{-9.0}$ & $9.5$ & $8.4$ & $3.8$ & $10$ & $19$ & $0.86$ \\
 \hline
  \end{tabular}}
  \label{table:radex_results}
  \begin{tabnote}
    (1) Number density of molecular gas.
    (2) Kinetic temperature of molecular gas.
    (3) Excitation temperature of $^{12}$CO($J$\,=\,1--0).
    (4) The same as (3) but for $^{12}$CO($J$\,=\,2--1).
    (5) The same as (3) but for $^{13}$CO($J$\,=\,1--0).
    (6) Optical depth of $^{12}$CO($J$\,=\,1--0).
    (7) The same as (6) but for $^{12}$CO($J$\,=\,2--1).
    (8) The same as (6) but for $^{13}$CO($J$\,=\,1--0).
    Results of (1)--(8) are obtained from the one-zone model of RADEX.
    $^*$ In annuli cases, regions of each annulus are named `ring1', `ring2'... in order from the galactic center.
    $^{\dagger}$ Adopted $X_{\rm CO}$ and [$^{12}$CO]/[$^{13}$CO] are different to reflect environments in galactic centers.
  \end{tabnote}
  \end{center}
\end{table*}

\appendix
\section{Histograms of $R_{2/1}$ in each galaxy}
Figure \ref{fig:R21hist_allgals} shows histograms of $R_{2/1}$ in each galaxy for data of the original resolution of 17$''$.

\begin{figure*}[t!]
 \begin{center}
  \includegraphics[width=16cm]{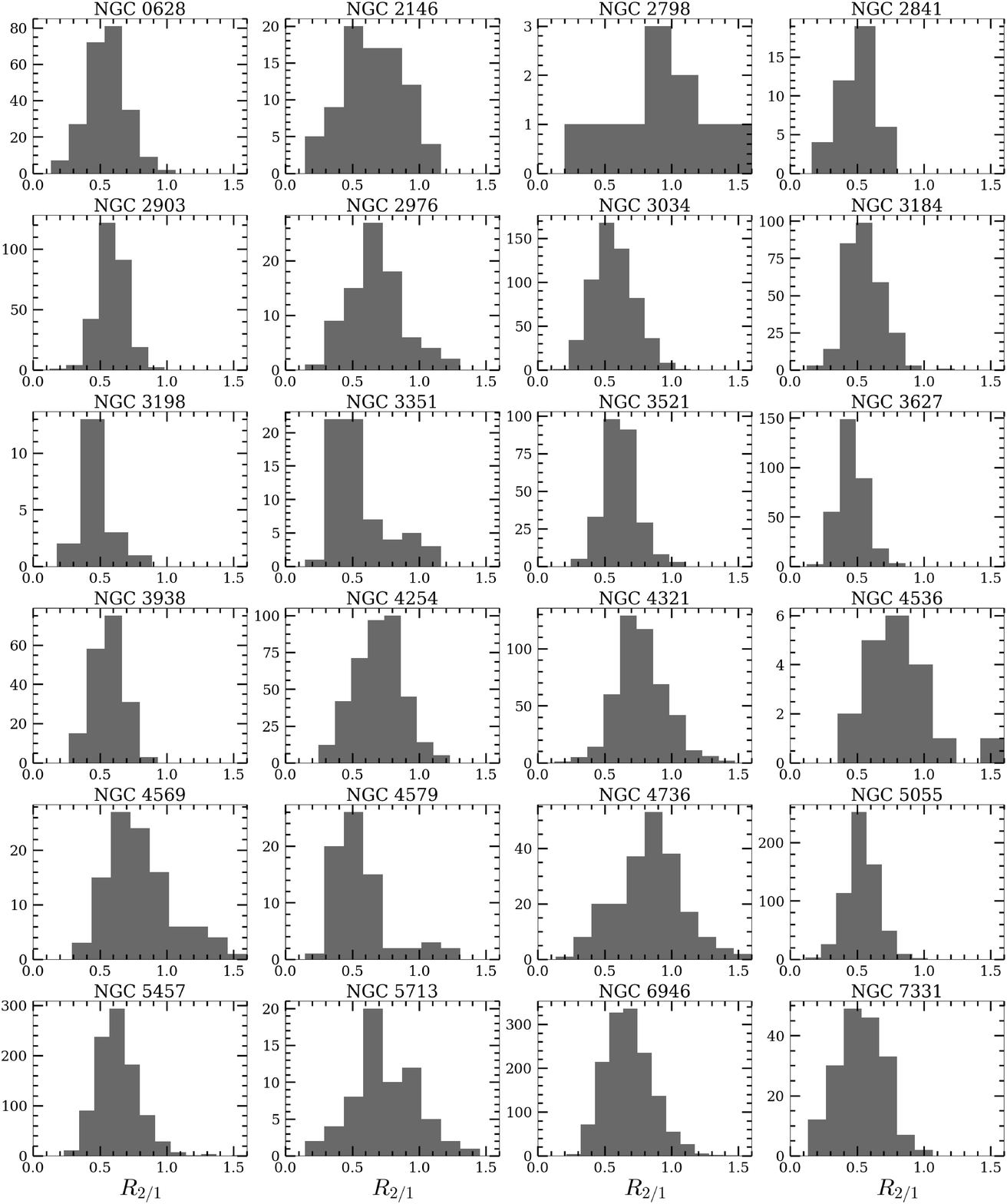}
 \end{center}
 \caption{Histograms of $R_{2/1}$ in each galaxy for original resolution data.
  }
 \label{fig:R21hist_allgals}
\end{figure*}

\section{Results of stacked spectra}
Table \ref{table:stacking_results} summarizes parameters of stacked spectra in each region of galaxies.

\renewcommand\arraystretch{1.0}
\begin{table*}[t!]
  \begin{center}
  \tbl{Results of the stacking analysis.}{
  \begin{tabular}{llcccccc} \hline
galaxy & region & $I_{^{12}{\rm CO}(1-0)}$ & $I_{^{12}{\rm CO}(2-1)}$ & $I_{^{13}{\rm CO}(1-0)}$ & $R_{2/1}$ & $R_{12/13}$ & $\Delta V$ \\
 &  & [K km s$^{-1}$] & [K km s$^{-1}$] & [K km s$^{-1}$] &  &  & [km s$^{-1}$] \\
 &  & (1) & (2) & (3) & (4) & (5) & (6) \\ \hline
NGC\,628 & center & $6.8 \pm 0.3$ & $4.23 \pm 0.09$ & $1.0 \pm 0.1$ & $0.62 \pm 0.03$ & $6.8 \pm 0.9$ & $30 \pm 2$ \\
 & inner arm1 & $5.3 \pm 0.2$ & $2.80 \pm 0.05$ & $0.35 \pm 0.07$ & $0.53 \pm 0.02$ & $15.1 \pm 3.1$ & $28 \pm 2$ \\
 & inner arm2 & $5.1 \pm 0.2$ & $2.93 \pm 0.05$ & $0.38 \pm 0.06$ & $0.57 \pm 0.03$ & $13.5 \pm 2.1$ & $37 \pm 2$ \\
 & outer arm1 & $3.0 \pm 0.2$ & $1.92 \pm 0.06$ & $ <0.07$ & $0.65 \pm 0.06$ & $ >41.1$ & $28 \pm 4$ \\
 & outer arm2 & $3.8 \pm 0.3$ & $1.47 \pm 0.06$ & $0.27 \pm 0.06$ & $0.39 \pm 0.03$ & $14.3 \pm 3.3$ & $27 \pm 3$ \\
 & inter-arm1 & $3.1 \pm 0.2$ & $1.37 \pm 0.03$ & $ <0.07$ & $0.43 \pm 0.03$ & $ >42.9$ & $32 \pm 4$ \\
 & inter-arm2 & $1.6 \pm 0.1$ & $0.69 \pm 0.04$ & $ <0.04$ & $0.44 \pm 0.04$ & $ >37.3$ & $30 \pm 5$ \\
NGC\,2146 & center & $148 \pm 1$ & $123.9 \pm 0.3$ & $7.2 \pm 0.4$ & $0.839 \pm 0.006$ & $20.4 \pm 1.0$ & $202 \pm 3$ \\
 & ring1 & $52.3 \pm 0.6$ & $36.0 \pm 0.2$ & $2.0 \pm 0.2$ & $0.689 \pm 0.009$ & $26.4 \pm 2.3$ & $150 \pm 4$ \\
 & ring2 & $16.4 \pm 0.6$ & $8.2 \pm 0.1$ & $ <0.18$ & $0.50 \pm 0.02$ & $ >92.8$ & $160 \pm 21$ \\
 & ring3 & $6.5 \pm 0.5$ & $3.7 \pm 0.2$ & $ <0.2$ & $0.57 \pm 0.05$ & $ >31.8$ & $71 \pm 12$ \\
NGC\,2841 & ring1 & $5.1 \pm 0.2$ & $2.2 \pm 0.1$ & $0.55 \pm 0.1$ & $0.42 \pm 0.03$ & $9.2 \pm 1.7$ & $88 \pm 8$ \\
 & ring2 & $3.7 \pm 0.2$ & $2.01 \pm 0.09$ & $0.49 \pm 0.07$ & $0.54 \pm 0.04$ & $7.5 \pm 1.2$ & $58 \pm 6$ \\
 & ring3 & $2.4 \pm 0.1$ & $1.29 \pm 0.06$ & $0.35 \pm 0.06$ & $0.53 \pm 0.04$ & $6.9 \pm 1.3$ & $59 \pm 5$ \\
 & ring4 & $1.6 \pm 0.1$ & $0.74 \pm 0.06$ & $ <0.05$ & $0.46 \pm 0.05$ & $ >30.8$ & $67 \pm 10$ \\
NGC\,2903 & center & $46 \pm 1$ & $33.5 \pm 0.2$ & $3.9 \pm 0.4$ & $0.72 \pm 0.02$ & $11.8 \pm 1.1$ & $157 \pm 13$ \\
 & northern bar & $21.7 \pm 0.8$ & $13.0 \pm 0.3$ & $1.3 \pm 0.2$ & $0.60 \pm 0.03$ & $17.3 \pm 3.4$ & $94 \pm 7$ \\
 & southern bar & $28.0 \pm 0.7$ & $14.1 \pm 0.2$ & $ <0.35$ & $0.50 \pm 0.02$ & $ >79.2$ & $151 \pm 4$ \\
 & northern bar end & $41.7 \pm 0.7$ & $27.0 \pm 0.2$ & $4.4 \pm 0.4$ & $0.65 \pm 0.01$ & $9.5 \pm 0.8$ & $101 \pm 3$ \\
 & southern bar end & $36.5 \pm 0.7$ & $21.4 \pm 0.2$ & $2.9 \pm 0.3$ & $0.59 \pm 0.01$ & $12.5 \pm 1.5$ & $119 \pm 4$ \\
 & northern arm & $19.7 \pm 0.4$ & $11.9 \pm 0.1$ & $2.1 \pm 0.3$ & $0.61 \pm 0.01$ & $9.6 \pm 1.4$ & $65 \pm 2$ \\
 & southern arm & $18.8 \pm 0.5$ & $10.4 \pm 0.1$ & $2.1 \pm 0.2$ & $0.56 \pm 0.02$ & $8.8 \pm 0.8$ & $65 \pm 4$ \\
 & inter-arm & $13.8 \pm 0.4$ & $7.0 \pm 0.1$ & $ <0.16$ & $0.51 \pm 0.02$ & $ >85.5$ & $82 \pm 5$ \\
 & outer disk & $3.9 \pm 0.2$ & $1.98 \pm 0.06$ & $0.19 \pm 0.05$ & $0.51 \pm 0.03$ & $20.8 \pm 6.0$ & $53 \pm 5$ \\
NGC\,2976 & center & $3.1 \pm 0.3$ & $2.61 \pm 0.09$ & $ <0.12$ & $0.84 \pm 0.08$ & $ >24.9$ & $33 \pm 5$ \\
 & ring1 & $2.6 \pm 0.2$ & $1.98 \pm 0.08$ & $ <0.05$ & $0.76 \pm 0.05$ & $ >49.0$ & $31 \pm 4$ \\
 & ring2 & $3.3 \pm 0.1$ & $2.41 \pm 0.05$ & $0.27 \pm 0.06$ & $0.74 \pm 0.04$ & $12.0 \pm 2.7$ & $30 \pm 2$ \\
 & ring3 & $2.0 \pm 0.1$ & $1.44 \pm 0.07$ & $0.26 \pm 0.07$ & $0.71 \pm 0.05$ & $7.7 \pm 2.0$ & $31 \pm 4$ \\
 & ring4 & $0.6 \pm 0.1$ & $0.41 \pm 0.04$ & $ <0.05$ & $0.7 \pm 0.2$ & $ >11.0$ & $28 \pm 9$ \\
NGC\,3034 & center & $462 \pm 2$ & $365.5 \pm 0.4$ & $23.6 \pm 0.7$ & $0.792 \pm 0.003$ & $19.5 \pm 0.6$ & $147 \pm 1$ \\
 & ring1 & $204 \pm 1$ & $155.4 \pm 0.2$ & $8.8 \pm 0.6$ & $0.762 \pm 0.004$ & $23.2 \pm 1.5$ & $137 \pm 2$ \\
 & ring2 & $115.0 \pm 0.8$ & $78.9 \pm 0.2$ & $4.2 \pm 0.4$ & $0.686 \pm 0.005$ & $27.4 \pm 2.3$ & $126 \pm 2$ \\
 & ring3 & $60.8 \pm 0.8$ & $38.5 \pm 0.1$ & $1.5 \pm 0.2$ & $0.633 \pm 0.008$ & $41.7 \pm 7.0$ & $112 \pm 3$ \\
 & ring4 & $33 \pm 1$ & $24.56 \pm 0.1$ & $1.1 \pm 0.2$ & $0.74 \pm 0.03$ & $29.5 \pm 6.3$ & $106 \pm 13$ \\
 & ring5 & $33.9 \pm 0.9$ & $19.33 \pm 0.07$ & $ <0.26$ & $0.57 \pm 0.01$ & $ >128$ & $105 \pm 6$ \\
 & ring6 & $27.9 \pm 0.8$ & $15.6 \pm 0.1$ & $ <0.26$ & $0.56 \pm 0.02$ & $ >108$ & $102 \pm 7$ \\
 & ring7 & $23.1 \pm 0.9$ & $12.83 \pm 0.07$ & $ <0.21$ & $0.56 \pm 0.02$ & $ >107$ & $91 \pm 9$ \\
 & ring8 & $17.0 \pm 0.7$ & $9.9 \pm 0.1$ & $ <0.19$ & $0.58 \pm 0.03$ & $ >88.5$ & $82 \pm 8$ \\
 & ring9 & $13.5 \pm 0.7$ & $8.23 \pm 0.08$ & $ <0.22$ & $0.61 \pm 0.03$ & $ >60.9$ & $82 \pm 12$ \\
 & ring10 & $11.6 \pm 0.5$ & $7.15 \pm 0.09$ & $ <0.21$ & $0.62 \pm 0.03$ & $ >56.0$ & $87 \pm 15$ \\
 & ring11 & $11.2 \pm 0.6$ & $7.25 \pm 0.09$ & $ <0.25$ & $0.65 \pm 0.04$ & $ >44.1$ & $91 \pm 18$ \\
NGC\,3198 & center & $8.0 \pm 0.3$ & $4.3 \pm 0.1$ & $0.8 \pm 0.1$ & $0.54 \pm 0.03$ & $10.0 \pm 1.7$ & $92 \pm 7$ \\
 & ring1 & $4.0 \pm 0.3$ & $1.98 \pm 0.06$ & $ <0.1$ & $0.49 \pm 0.04$ & $ >41.9$ & $63 \pm 8$ \\
 & ring2 & $1.9 \pm 0.2$ & $1.26 \pm 0.04$ & $ <0.05$ & $0.68 \pm 0.07$ & $ >34.5$ & $42 \pm 9$ \\
NGC\,3351 & center & $17.5 \pm 0.4$ & $15.4 \pm 0.1$ & $1.3 \pm 0.2$ & $0.88 \pm 0.02$ & $13.5 \pm 2.1$ & $194 \pm 8$ \\
 & ring1 & $3.1 \pm 0.4$ & $1.8 \pm 0.1$ & $ <0.13$ & $0.59 \pm 0.08$ & $ >23.2$ & $96 \pm 29$ \\
 & ring2 & $3.6 \pm 0.2$ & $1.79 \pm 0.06$ & $ <0.13$ & $0.50 \pm 0.03$ & $ >28.1$ & $41 \pm 5$ \\
 & ring3 & $2.7 \pm 0.3$ & $1.08 \pm 0.05$ & $ <0.09$ & $0.41 \pm 0.05$ & $ >29.8$ & $40 \pm 9$ \\
 & ring4 & $1.5 \pm 0.2$ & $0.64 \pm 0.04$ & $ <0.07$ & $0.42 \pm 0.07$ & $ >21.5$ & $33 \pm 13$ \\
NGC\,3521 & center & $32.3 \pm 0.6$ & $20.5 \pm 0.2$ & $2.8 \pm 0.3$ & $0.64 \pm 0.01$ & $11.7 \pm 1.3$ & $165 \pm 13$ \\
 & ring1 & $27.7 \pm 0.4$ & $17.8 \pm 0.1$ & $2.3 \pm 0.1$ & $0.64 \pm 0.01$ & $11.9 \pm 0.8$ & $118 \pm 3$ \\
 & ring2 & $16.8 \pm 0.3$ & $9.28 \pm 0.08$ & $1.0 \pm 0.1$ & $0.55 \pm 0.01$ & $17.1 \pm 2.1$ & $92 \pm 3$ \\
 & ring3 & $5.7 \pm 0.3$ & $3.15 \pm 0.06$ & $0.36 \pm 0.07$ & $0.55 \pm 0.03$ & $16.0 \pm 3.2$ & $88 \pm 9$ \\
 & ring4 & $2.1 \pm 0.2$ & $0.96 \pm 0.05$ & $ <0.09$ & $0.46 \pm 0.05$ & $ >24.1$ & $49 \pm 9$ \\
 & ring5 & $1.2 \pm 0.1$ & $0.59 \pm 0.04$ & $ <0.06$ & $0.51 \pm 0.07$ & $ >19.3$ & $56 \pm 13$ \\
 \hline
 \end{tabular}}
  \label{table:stacking_results}
  \begin{tabnote}
  \end{tabnote}
  \end{center}
\end{table*}

\setcounter{table}{4}
\begin{table*}[t!]
  \begin{center}
  \tbl{(Continued.)}{
  \begin{tabular}{llcccccc} \hline
galaxy & region & $I_{^{12}{\rm CO}(1-0)}$ & $I_{^{12}{\rm CO}(2-1)}$ & $I_{^{13}{\rm CO}(1-0)}$ & $R_{2/1}$ & $R_{12/13}$ & $\Delta V$ \\
 &  & [K km s$^{-1}$] & [K km s$^{-1}$] & [K km s$^{-1}$] &  &  & [km s$^{-1}$] \\
 &  & (1) & (2) & (3) & (4) & (5) & (6) \\ \hline
NGC\,3627 & center & $50.4 \pm 0.7$ & $21.1 \pm 0.6$ & $2.0 \pm 0.3$ & $0.42 \pm 0.01$ & $25.3 \pm 3.9$ & $164 \pm 9$ \\
 & bar & $24.7 \pm 0.8$ & $10.2 \pm 0.2$ & $1.2 \pm 0.2$ & $0.41 \pm 0.01$ & $20.3 \pm 3.8$ & $108 \pm 11$ \\
 & northern bar end & $31.9 \pm 0.6$ & $15.7 \pm 0.2$ & $2.8 \pm 0.2$ & $0.49 \pm 0.01$ & $11.5 \pm 0.8$ & $68 \pm 3$ \\
 & southern bar end & $41 \pm 1$ & $20.3 \pm 0.2$ & $3.3 \pm 0.3$ & $0.49 \pm 0.01$ & $12.5 \pm 1.2$ & $96 \pm 11$ \\
 & western arm & $13.7 \pm 0.4$ & $6.4 \pm 0.1$ & $0.9 \pm 0.1$ & $0.47 \pm 0.02$ & $15.1 \pm 2.5$ & $48 \pm 2$ \\
 & eastern arm & $18.4 \pm 0.4$ & $9.0 \pm 0.2$ & $1.7 \pm 0.2$ & $0.49 \pm 0.02$ & $11.0 \pm 1.4$ & $68 \pm 3$ \\
 & offset stream & $19.1 \pm 0.5$ & $9.7 \pm 0.2$ & $1.1 \pm 0.2$ & $0.51 \pm 0.02$ & $16.9 \pm 3.2$ & $60 \pm 3$ \\
 & southern arm & $11.1 \pm 0.6$ & $5.3 \pm 0.1$ & $0.7 \pm 0.2$ & $0.48 \pm 0.03$ & $15.0 \pm 4.1$ & $45 \pm 3$ \\
 & arm-bar end inter. region & $12.3 \pm 0.4$ & $4.4 \pm 0.1$ & $ <0.19$ & $0.36 \pm 0.02$ & $ >65.4$ & $51 \pm 3$ \\
 & inter-arm & $10.0 \pm 0.5$ & $4.7 \pm 0.1$ & $ <0.14$ & $0.47 \pm 0.03$ & $ >70.5$ & $86 \pm 9$ \\
 & outer disk & $2.6 \pm 0.2$ & $1.03 \pm 0.07$ & $ <0.09$ & $0.39 \pm 0.05$ & $ >30.4$ & $58 \pm 12$ \\
NGC\,3938 & center & $9.5 \pm 0.2$ & $5.98 \pm 0.09$ & $1.2 \pm 0.1$ & $0.63 \pm 0.02$ & $8.2 \pm 0.8$ & $43 \pm 2$ \\
 & ring1 & $6.4 \pm 0.2$ & $3.68 \pm 0.05$ & $0.63 \pm 0.07$ & $0.57 \pm 0.02$ & $10.1 \pm 1.2$ & $29 \pm 1$ \\
 & ring2 & $4.3 \pm 0.1$ & $2.46 \pm 0.04$ & $0.36 \pm 0.05$ & $0.57 \pm 0.02$ & $12.1 \pm 1.9$ & $28 \pm 1$ \\
 & ring3 & $2.4 \pm 0.2$ & $1.36 \pm 0.04$ & $0.35 \pm 0.06$ & $0.56 \pm 0.05$ & $6.9 \pm 1.3$ & $27 \pm 3$ \\
 & ring4 & $1.02 \pm 0.09$ & $0.55 \pm 0.03$ & $ <0.03$ & $0.54 \pm 0.06$ & $ >29.7$ & $24 \pm 3$ \\
NGC\,4536 & center & $34 \pm 1$ & $29.7 \pm 0.2$ & $2.0 \pm 0.3$ & $0.86 \pm 0.02$ & $17.0 \pm 2.2$ & $161 \pm 17$ \\
 & ring1 & $6.5 \pm 0.4$ & $4.9 \pm 0.1$ & $ <0.14$ & $0.74 \pm 0.05$ & $ >45.1$ & $136 \pm 23$ \\
 & ring2 & $2.2 \pm 0.2$ & $1.34 \pm 0.07$ & $ <0.1$ & $0.61 \pm 0.07$ & $ >22.8$ & $80 \pm 16$ \\
NGC\,4579 & center & $13.0 \pm 0.7$ & $10.1 \pm 0.2$ & $ <0.35$ & $0.78 \pm 0.05$ & $ >37.0$ & $154 \pm 36$ \\
 & ring1 & $5.5 \pm 0.4$ & $3.05 \pm 0.09$ & $ <0.2$ & $0.55 \pm 0.04$ & $ >28.0$ & $50 \pm 6$ \\
 & ring2 & $5.1 \pm 0.2$ & $2.68 \pm 0.07$ & $0.41 \pm 0.08$ & $0.52 \pm 0.03$ & $12.3 \pm 2.5$ & $50 \pm 4$ \\
 & ring3 & $2.8 \pm 0.3$ & $1.30 \pm 0.08$ & $ <0.13$ & $0.47 \pm 0.06$ & $ >21.0$ & $46 \pm 11$ \\
NGC\,5055 & center & $40.0 \pm 0.6$ & $27.3 \pm 0.3$ & $3.7 \pm 0.3$ & $0.68 \pm 0.01$ & $10.9 \pm 0.8$ & $120 \pm 4$ \\
 & ring1 & $20.1 \pm 0.3$ & $11.4 \pm 0.1$ & $2.4 \pm 0.2$ & $0.569 \pm 0.01$ & $8.4 \pm 0.6$ & $71 \pm 2$ \\
 & ring2 & $14.0 \pm 0.2$ & $6.79 \pm 0.08$ & $1.08 \pm 0.09$ & $0.484 \pm 0.01$ & $12.9 \pm 1.0$ & $53 \pm 1$ \\
 & ring3 & $8.5 \pm 0.2$ & $4.38 \pm 0.07$ & $0.82 \pm 0.1$ & $0.51 \pm 0.02$ & $10.4 \pm 1.3$ & $47 \pm 2$ \\
 & ring4 & $5.5 \pm 0.2$ & $2.74 \pm 0.07$ & $0.44 \pm 0.07$ & $0.5 \pm 0.02$ & $12.4 \pm 2.1$ & $44 \pm 2$ \\
 & ring5 & $2.7 \pm 0.2$ & $1.28 \pm 0.05$ & $ <0.06$ & $0.48 \pm 0.04$ & $ >42.4$ & $43 \pm 4$ \\
 & ring6 & $1.8 \pm 0.2$ & $0.65 \pm 0.04$ & $ <0.06$ & $0.35 \pm 0.04$ & $ >30.5$ & $49 \pm 8$ \\
 & ring7 & $1.1 \pm 0.2$ & $0.54 \pm 0.06$ & $ <0.06$ & $0.51 \pm 0.09$ & $ >18.1$ & $49 \pm 12$ \\
NGC\,5713 & center & $36.3 \pm 0.7$ & $34.4 \pm 0.2$ & $1.3 \pm 0.2$ & $0.95 \pm 0.02$ & $28.8 \pm 5.7$ & $112 \pm 5$ \\
 & ring1 & $20.7 \pm 0.4$ & $16.6 \pm 0.1$ & $0.8 \pm 0.1$ & $0.80 \pm 0.02$ & $24.5 \pm 4.2$ & $80 \pm 4$ \\
 & ring2 & $10.4 \pm 0.4$ & $7.1 \pm 0.1$ & $ <0.12$ & $0.68 \pm 0.03$ & $ >86.2$ & $61 \pm 6$ \\
 & ring3 & $5.7 \pm 0.3$ & $3.4 \pm 0.1$ & $ <0.11$ & $0.59 \pm 0.03$ & $ >50.2$ & $54 \pm 5$ \\
NGC\,7331 & center & $28.6 \pm 0.9$ & $17.6 \pm 0.2$ & $2.2 \pm 0.3$ & $0.61 \pm 0.02$ & $12.9 \pm 1.6$ & $143 \pm 21$ \\
 & ring1 & $35.2 \pm 0.7$ & $20.1 \pm 0.1$ & $4.3 \pm 0.3$ & $0.57 \pm 0.01$ & $8.1 \pm 0.6$ & $119 \pm 9$ \\
 & ring2 & $23.0 \pm 0.4$ & $12.8 \pm 0.1$ & $2.8 \pm 0.3$ & $0.56 \pm 0.01$ & $8.2 \pm 0.8$ & $114 \pm 4$ \\
 & ring3 & $11.6 \pm 0.6$ & $5.46 \pm 0.1$ & $1.0 \pm 0.2$ & $0.47 \pm 0.02$ & $12.0 \pm 2.0$ & $86 \pm 10$ \\
 & ring4 & $5.8 \pm 0.4$ & $2.60 \pm 0.07$ & $ <0.14$ & $0.45 \pm 0.04$ & $ >40.6$ & $93 \pm 23$ \\
 & ring5 & $3.1 \pm 0.3$ & $1.18 \pm 0.07$ & $ <0.11$ & $0.38 \pm 0.04$ & $ >29.1$ & $74 \pm 20$ \\
 & ring6 & $1.9 \pm 0.2$ & $0.66 \pm 0.05$ & $ <0.08$ & $0.34 \pm 0.05$ & $ >23.5$ & $67 \pm 14$ \\
 \hline
  \end{tabular}}
  \label{table:stacking_results}
  \begin{tabnote}
    (1) Integrated intensity of stacked $^{12}$CO($J$\,=\,1--0) spectra.
    (2) The same as (1) but for $^{12}$CO($J$\,=\,2--1).
    (3) The same as (1) but for $^{13}$CO($J$\,=\,1--0).
    (4) Integrated-intensity ratio of stacked $^{12}$CO($J$\,=\,2--1) to that for $^{12}$CO($J$\,=\,1--0).
    (5) The same as (4) but for stacked $^{12}$CO($J$\,=\,1--0)/$^{13}$CO($J$\,=\,1--0).
    (6) FWHM of stacked $^{12}$CO($J$\,=\,1--0) spectra.
  \end{tabnote}
  \end{center}
\end{table*}


\begin{thebibliography}{99}
\bibitem[Bigiel et al.(2008)]{Bigiel08} Bigiel, F., Leroy, A., Walter, F., et al.\ 2008, \aj, 136, 2846
\bibitem[Bolatto et al.(2013)]{Bolatto13} Bolatto, A.~D., Wolfire, M., \& Leroy, A.~K.\ 2013, \araa, 51, 207
\bibitem[Braine \& Combes(1992)]{Braine92} Braine, J., \& Combes, F.\ 1992, \aap, 264, 433
\bibitem[Casasola et al.(2017)]{Casasola17} Casasola, V., Cassar{\`a}, L.~P., Bianchi, S., et al.\ 2017, \aap, 605, A18
\bibitem[Cormier et al.(2018)]{Cormier18} Cormier, D., Bigiel, F., Jim{\'e}nez-Donaire, M.~J., et al.\ 2018, \mnras, 475, 3909
\bibitem[Davis(2014)]{Davis14} Davis, T.~A.\ 2014, \mnras, 445, 2378
\bibitem[Druard et al.(2014)]{Druard14} Druard, C., Braine, J., Schuster, K.~F., et al.\ 2014, \aap, 567, A118
\bibitem[Elmegreen(2002)]{Elmegreen02} Elmegreen, B.~G.\ 2002, \apj, 577, 206
\bibitem[Gao et al.(2007)]{Gao07} Gao, Y., Carilli, C.~L., Solomon, P.~M., et al.\ 2007, \apjl, 660, L93
\bibitem[Gil de Paz et al.(2007)]{GildePaz07} Gil de Paz, A., Boissier, S., Madore, B.~F., et al.\ 2007, ApJS, 173, 185
\bibitem[Garcia-Burillo et al.(1993)]{Garcia-Burillo93} Garcia-Burillo, S., Guelin, M., \& Cernicharo, J.\ 1993, \aap, 274, 123
\bibitem[Goldreich \& Kwan(1974)]{Goldreich74} Goldreich, P., \& Kwan, J.\ 1974, \apj, 189, 441
\bibitem[Gong et al.(2020)]{Gong20} Gong, M., Ostriker, E.~C., Kim, C.-G., et al.\ 2020, \apj, 903, 142
\bibitem[Heyer et al.(2009)]{Heyer09} Heyer, M., Krawczyk, C., Duval, J., et al.\ 2009, \apj, 699, 1092
\bibitem[Isobe et al.(1990)]{Isobe90} Isobe, T., Feigelson, E.~D., Akritas, M.~G., et al.\ 1990, \apj, 364, 104
\bibitem[Israel \& Baas(2003)]{Israel03} Israel, F.~P., \& Baas, F.\ 2003, \aap, 404, 495
\bibitem[Kennicutt(1989)]{Kennicutt89} Kennicutt, R.~C.\ 1989, \apj, 344, 685
\bibitem[Koda et al.(2011)]{Koda11} Koda, J., Sawada, T., Wright, M.~C.~H., et al.\ 2011, \apjs, 193, 19
\bibitem[Koda et al.(2012)]{Koda12} Koda, J., Scoville, N., Hasegawa, T., et al.\ 2012, \apj, 761, 41
\bibitem[Koda et al.(2020)]{Koda20} Koda, J., Sawada, T., Sakamoto, K., et al.\ 2020, \apjl, 890, L10
\bibitem[Komugi et al.(2006)]{Komugi06} Komugi, S., Sofue, Y., \& Egusa, F.\ 2006, \pasj, 58, 793
\bibitem[Kroupa(2001)]{Kroupa01} Kroupa, P.\ 2001, \mnras, 322, 231
\bibitem[Krumholz, \& McKee(2005)]{Krumholz05} Krumholz, M.~R., \& McKee, C.~F.\ 2005, \apj, 630, 250
\bibitem[Kuno et al.(2007)]{Kuno07} Kuno, N., Sato, N., Nakanishi, H., et al.\ 2007, \pasj, 59, 117 (K07)
\bibitem[Leroy et al.(2008)]{Leroy08} Leroy, A.~K., Walter, F., Brinks, E., et al.\ 2008, \aj, 136, 2782
\bibitem[Leroy et al.(2009)]{Leroy09} Leroy, A.~K., Walter, F., Bigiel, F., et al.\ 2009, \aj, 137, 4670 (L09)
\bibitem[Leroy et al.(2013)]{Leroy13} Leroy, A.~K., Walter, F., Sandstrom, K., et al.\ 2013, \aj, 146, 19
\bibitem[Leroy et al.(2015)]{Leroy15} Leroy, A.~K., Bolatto, A.~D., Ostriker, E.~C., et al.\ 2015, \apj, 801, 25
\bibitem[Meier et al.(2000)]{Meier00} Meier, D.~S., Turner, J.~L., \& Hurt, R.~L.\ 2000, \apj, 531, 200
\bibitem[Meier \& Turner(2001)]{Meier01} Meier, D.~S., \& Turner, J.~L.\ 2001, \apj, 551, 687
\bibitem[Meier \& Turner(2004)]{Meier04} Meier, D.~S., \& Turner, J.~L.\ 2004, \aj, 127, 2069
\bibitem[Milam et al.(2005)]{Milam05} Milam, S.~N., Savage, C., Brewster, M.~A., et al.\ 2005, \apj, 634, 1126
\bibitem[Minamidani et al.(2016)]{Minamidani16} Minamidani, T., Nishimura, A., Miyamoto, Y., et al.\ 2016, Proc. SIPE, 9914, 99141Z
\bibitem[Momose et al.(2013)]{Momose13} Momose, R., Koda, J., Kennicutt, R.~C., et al.\ 2013, \apj, 772, L13
\bibitem[Morokuma-Matsui et al.(2015)]{Morokuma15} Morokuma-Matsui, K., Sorai, K., Watanabe, Y., et al.\ 2015, \pasj, 67, 2
\bibitem[Morokuma-Matsui \& Muraoka(2017)]{Morokuma17} Morokuma-Matsui, K., \& Muraoka, K.\ 2017, \apj, 837, 137
\bibitem[Muraoka et al.(2016)]{Muraoka16} Muraoka, K., Sorai, K., Kuno, N., et al.\ 2016, \pasj, 68, 89
\bibitem[Muraoka et al.(2019)]{Muraoka19} Muraoka, K., Sorai, K., Miyamoto, Y., et al.\ 2019, \pasj, 71, S15
\bibitem[Nakai et al.(1994)]{Nakai94} Nakai, N., Kuno, N., Handa, T., et al.\ 1994, \pasj, 46, 527
\bibitem[Narayanan et al.(2011)]{Narayanan11} Narayanan, D., Krumholz, M., Ostriker, E.~C., et al.\ 2011, \mnras, 418, 664
\bibitem[Nishimura et al.(2015)]{Nishimura15} Nishimura, A., Tokuda, K., Kimura, K., et al.\ 2015, \apjs, 216, 18
\bibitem[Oka et al.(1998)]{Oka98} Oka, T., Hasegawa, T., Hayashi, M., et al.\ 1998, \apj, 493, 730
\bibitem[Oka et al.(2001)]{Oka01} Oka, T., Hasegawa, T., Sato, F., et al.\ 2001, \apj, 562, 348
\bibitem[Onodera et al.(2010)]{Onodera10} Onodera, S., Kuno, N., Tosaki, T., et al.\ 2010, \apjl, 722, L127
\bibitem[Paglione et al.(2001)]{Paglione01} Paglione, T.~A.~D., Wall, W.~F., Young, J.~S., et al.\ 2001, \apjs, 135, 183
\bibitem[Papadopoulos et al.(2012)]{Papadopoulos12} Papadopoulos, P.~P., van der Werf, P., Xilouris, E., et al.\ 2012, \apj, 751, 10
\bibitem[Pe{\~n}aloza et al.(2017)]{Penaloza17} Pe{\~n}aloza, C.~H., Clark, P.~C., Glover, S.~C.~O., et al.\ 2017, \mnras, 465, 2277
\bibitem[Pe{\~n}aloza et al.(2018)]{Penaloza18} Pe{\~n}aloza, C.~H., Clark, P.~C., Glover, S.~C.~O., et al.\ 2018, \mnras, 475, 1508
\bibitem[Pineda et al.(2008)]{Pineda08} Pineda, J.~E., Caselli, P., \& Goodman, A.~A.\ 2008, \apj, 679, 481
\bibitem[Robitaille, \& Bressert(2012)]{Robitaille12} Robitaille, T., \& Bressert, E.\ 2012, APLpy: Astronomical Plotting Library in Python, ascl:1208.017
\bibitem[Sakamoto et al.(1994)]{Sakamoto94} Sakamoto, S., Hayashi, M., Hasegawa, T., et al.\ 1994, \apj, 425, 641
\bibitem[Sakamoto et al.(1995)]{Sakamoto95} Sakamoto, S., Hasegawa, T., Hayashi, M., et al.\ 1995, \apjs, 100, 125
\bibitem[Sakamoto et al.(1997)]{Sakamoto97} Sakamoto, S., Hasegawa, T., Handa, T., et al.\ 1997, \apj, 486, 276
\bibitem[Sandstrom et al.(2013)]{Sandstrom13} Sandstrom, K.~M., Leroy, A.~K., Walter, F., et al.\ 2013, \apj, 777, 5
\bibitem[Sawada et al.(2001)]{Sawada01} Sawada, T., Hasegawa, T., Handa, T., et al.\ 2001, \apjs, 136, 189
\bibitem[Schmidt(1959)]{Schmidt59} Schmidt, M.\ 1959, \apj, 129, 243
\bibitem[Schruba et al.(2011)]{Schruba11} Schruba, A., Leroy, A.~K., Walter, F., et al.\ 2011, \aj, 142, 37
\bibitem[Schruba et al.(2019)]{Schruba19} Schruba, A., Kruijssen, J.~M.~D., \& Leroy, A.~K.\ 2019, \apj, 883, 2
\bibitem[Scoville \& Solomon(1974)]{Scoville74} Scoville, N.~Z., \& Solomon, P.~M.\ 1974, \apjl, 187, L67
\bibitem[Sheth et al.(2010)]{Sheth10} Sheth, K., Regan, M., Hinz, J.~L., et al.\ 2010, \pasp, 122, 1397
\bibitem[Solomon et al.(1987)]{Solomon87} Solomon, P.~M., Rivolo, A.~R., Barrett, J., et al.\ 1987, \apj, 319, 730
\bibitem[Sorai et al.(2001)]{Sorai01} Sorai, K., Hasegawa, T., Booth, R.~S., et al.\ 2001, \apj, 551, 794
\bibitem[Sorai et al.(2019)]{Sorai19} Sorai, K., Kuno, N., Muraoka, K., et al. 2019, \pasj, 71, S14 (S19)
\bibitem[Sobolev(1960)]{Sobolev60} Sobolev, V.~V.\ 1960, Cambridge: Harvard University Press
\bibitem[Sun et al.(2018)]{Sun18} Sun, J., Leroy, A.~K., Schruba, A., et al.\ 2018, \apj, 860, 172
\bibitem[Sun et al.(2020)]{Sun20} Sun, J., Leroy, A.~K., Schinnerer, E., et al.\ 2020, \apjl, 901, L8
\bibitem[]{} Takeuchi, T. T., et al. 2020, \pasj, in preparation
\bibitem[Tan(2010)]{Tan10} Tan, J.~C.\ 2010, \apjl, 710, L88
\bibitem[Usero et al.(2015)]{Usero15} Usero, A., Leroy, A.~K., Walter, F., et al.\ 2015, \aj, 150, 115
\bibitem[van der Tak et al.(2007)]{vanderTak07} van der Tak, F.~F.~S., Black, J.~H., Sch{\"o}ier, F.~L., et al.\ 2007, \aap, 468, 627
\bibitem[Walter et al.(2008)]{Walter08} Walter, F., Brinks, E., de Blok, W.~J.~G., et al.\ 2008, \aj, 136, 2563
\bibitem[Yajima et al.(2019)]{Yajima19} Yajima, Y., Sorai, K., Kuno, N., et al. 2019, \pasj, 71, S13
\bibitem[Yasuda et al.(2020)]{Yasuda20} Yasuda, A., et al. 2020, \pasj, in preparation
\bibitem[Young \& Scoville(1991)]{Young91} Young, J.~S., \& Scoville, N.~Z.\ 1991, \araa, 29, 581
\end{thebibliography}
\end{document}